\documentclass[useAMS,amssymb,usenatbib]{mn2e}
\usepackage{epsfig}
\usepackage{color}
\usepackage{graphicx}
\usepackage{tabularx}
\usepackage{longtable}

\newcommand{\Ms}{\mathrm{M}_\odot}

\title[]{The abundance of (not just) dark matter haloes}

\author[Sawala et al.]  {\parbox{\textwidth}{Till
    Sawala$^{1}$\thanks{E-mail: \texttt{till.sawala@durham.ac.uk}},
    Carlos S.   Frenk$^{1}$, Robert A. Crain$^{2}$, Adrian
    Jenkins$^{1}$, Joop Schaye$^{2}$, Tom Theuns$^{1,3}$ and Jesus
    Zavala$^{4,5}$}\vspace{0.4cm}\\
\parbox{\textwidth}{$^{1}$Institute for Computational Cosmology,  Department of Physics, University of Durham, South Road, Durham DH1  3LE, UK \\ 
$^{2}$Leiden Observatory, Leiden University, Postbus  9513, 2300 RA Leiden, the Netherlands \\ 
$^{3}$Department of  Physics, University of Antwerp, Campus Groenenborger,  Groenenborgerlaan 171, B-2020 Antwerp, Belgium\\
$^{4}$Department of Physics and Astronomy, University of Waterloo, Waterloo, Ontario, N2L 3G1, Canada \\
$^{5}$Perimeter Institute for Theoretical Physics, 31 Caroline St. N., Waterloo, ON, N2L 2Y5, Canada}}

\begin{document}

\date{Accepted 2013 8 February. Received 2013 25 January; in original
  form 2012 27 June}

\pagerange{\pageref{firstpage}--\pageref{lastpage}} \pubyear{2012}

\maketitle

\label{firstpage}

\begin{abstract}
  We study the effect of baryons on the abundance of structures and
  substructures in a $\Lambda$CDM cosmology, using a pair of high
  resolution cosmological simulations from the GIMIC project. Both
  simulations use identical initial conditions, but while one contains
  only dark matter, the other also includes baryons. We find that gas
  pressure, reionisation, supernova feedback, stripping, and truncated
  accretion systematically reduce the total mass and the abundance of
  structures below $\sim10^{12}\Ms$ compared to the pure dark matter
  simulation. Taking this into account and adopting an appropriate
  detection threshold lowers the abundance of observed galaxies with
  maximum circular velocities $v_{\rm{max}} < 100$~kms$^{-1}$,
  significantly reducing the reported discrepancy between $\Lambda$CDM
  and the measured HI velocity function of the ALFALFA survey. We also
  show that the stellar-to-total mass ratios of galaxies with stellar
  masses of $\sim 10^5-10^7\Ms$ inferred from abundance matching of
  the (sub)halo mass function to the observed galaxy mass function
  increase by a factor of $\sim 2$. In addition, we find that an
  important fraction of low-mass subhaloes are completely devoid of
  stars.  Accounting for the presence of {\it dark} subhaloes below
  $10^{10}\Ms$ further reduces the abundance of {\it observable}
  objects, and leads to an additional increase in the inferred
  stellar-to-total mass ratio by factors of 2-10 for galaxies in
  haloes of $10^9-10^{10}\Ms$.  This largely reconciles the abundance
  matching results with the kinematics of individual dwarf galaxies in
  $\Lambda$CDM. We propose approximate corrections to the masses of
  objects derived from pure dark matter calculations to account for
  baryonic effects.
\end{abstract}


\section{Introduction} \label{introduction}

The large-scale evolution of the Universe is determined by gravity
which drives hierarchical structure formation in an expanding
space. While the nature of dark matter remains an open question at a
fundamental level, gravity makes no distinction between baryonic and
dark matter. Consequently, numerical simulations of cosmic structure
formation are generally purely gravitational, or ``Dark Matter Only''
(DMO), which greatly reduces the computational cost and complexity
compared to gas-dynamical simulations. Over the past decades, DMO
simulations have played a key part in establishing the currently
preferred $\Lambda$CDM paradigm \citep[e.g.][]{Davis-1985, Frenk-1988}
and the hierarchical formation scenario \citep[e.g.][]{White-1978,
  White-Frenk-1991}, explaining the observed abundance and correlation
of structures at different redshifts and over many mass and length
scales.

On dwarf galaxy scales (haloes up to $10^{11}\Ms$ or
$v_{\rm{max}}<100$~kms$^{-1}$, or galaxies with stellar masses up to
$\sim10^9\Ms$), however, cold dark matter simulations appear to
overpredict the amount of structure, to a degree that appears
incompatible with observations: the abundance of haloes with a maximum
circular velocity $v_{\rm{max}} \sim 35$~kms$^{-1}$ exceeds the
abundance measured in the ALFALFA HI survey by a factor of $\sim10$
\citep[e.g.][]{Zavala-2009, Papastergis-2011}; the simulations fail to
match the dynamics of satellites, (e.g. \citealt{Parry-2009,
  Strigari-2010, Boylan-Kolchin-2011, Boylan-Kolchin-2012}, but see
\citealt{Wang-2012}); and the stellar-to-total mass ratios inferred
from the abundances do not match those of individual dwarf galaxies
\citep[e.g.][]{Sawala-Matter, Ferrero-2012}. In this work we show that
the inclusion of baryons and their relevant astrophysical processes
changes the abundance of (sub)haloes sufficiently to reconcile CDM
with observations.

While the observable matter in the Universe traces the underlying
total mass distribution, galaxies are ``biased'' with respect to the
dark matter \citep{Kaiser-1984, Davis-1985}, and the stellar-to-total
mass ratio varies greatly with galaxy mass, evolutionary stage, and
environment. Linking simulated structures to observations of galaxies
must therefore be done indirectly, via methods such as (sub)halo
abundance matching \citep[e.g.][]{Vale-2006, Conroy-2009, Moster-2009,
  Guo-2010}, halo occupation distribution models
\citep[e.g.][]{Benson-2000, Seljak-2000, Scoccimarro-2001,
  Berlind-2002}, or semi-analytical models
\citep[e.g.][]{White-Frenk-1991, Kauffmann-1993, Cole-1994,
  Somerville-1999, Croton-2006, Bower-2006}. While these approaches
vary greatly in complexity, aim and predictive power, they share the
fundamental assumption that the total mass distribution in the
Universe is governed by gravity alone, and can be represented by dark
matter.

In this work, we investigate the impact of baryon physics on structure
formation on smaller scales. We find that baryons systematically
affect the masses and abundances of objects in three ways:
\begin{itemize}
\item Interstellar gas is expelled from haloes in simulations that
  include supernova feedback, while intergalactic gas is prevented
  from accreting in the presence of ionising radiation.
\item Ram-pressure stripping preferentially removes gas from low-mass
  satellite galaxies, and satellites are more easily disrupted.
\item The shallower potential well caused by the loss of baryons
  subsequently leads to diminished accretion, both of baryons and of
  dark matter.
\end{itemize}

Of these, the first two effects can substantially reduce the mass of
baryons in small haloes, and lower the total mass of an object by up
to the universal baryon fraction, compared to the corresponding DMO
simulation. The third effect can reduce an object's mass even further,
as haloes with a lower mass and shallower potential accrete less
material. All baryon effects show a strong mass dependence, and
stripping also depends on environment.

To quantify the influence of baryons, we use a set of simulations of
structure formation in identical cosmological volumes: a pure dark
matter simulation, and one which includes a gas physics model that
incorporates hydrodynamics, radiative heating and cooling, a cosmic UV
background, star formation, chemical evolution and supernova
feedback. This simulation, part of the ``Galaxies-Intergalactic Medium
Interaction Calculation'' \citep[``GIMIC'',][]{Crain-2009}, has
sufficient dynamic range and volume to resolve objects with total mass
between $10^9-10^{14} \Ms$, from dwarf galaxies to massive groups and
clusters. It has been used previously to study many aspects of galaxy
formation, such as environmental dependencies \citep{Crain-2009}, disk
galaxies \citep{Sales-2012} and their X-ray coronae
\citep{Crain-2010}, the origin of the baryonic Tully Fisher relation
\citep{McCarthy-2012a}, and the assembly of stellar haloes
\citep{Font-2011, McCarthy-2012b}.

Here, our focus is not on the formation of galaxies, but on the
formation of their dark matter hosts, and in particular, on the
abundance of haloes and subhaloes of different masses and circular
velocities, in different environments, and over time. Matching
individual objects across both simulations also enables us to
understand the mechanisms by which baryons affect the abundance of
structures, and to derive an approximate expression that allows us to
correct for baryonic effects in generic DMO simulations.

The effect of baryons on the structure of haloes has been studied
extensively, both using analytical models
\citep[e.g.][]{Blumenthal-1986, Sellwood-2005} and in numerical
simulations \cite[e.g.][]{Abadi-2010, Duffy-2010, Bryan-2012}. Whereas
dissipation universally increases the concentration of haloes
\citep[e.g.][]{Mo-2010, Gnedin-2011}, a non-adiabatic response to
rapid outflows induced by supernovae can erase the central cusp in the
profiles of low-mass haloes \citep{Navarro-1996, Governato-2010,
  Pontzen-2012}, potentially resolving the discrepancy between cusped
profiles predicted by CDM, and the cored profiles indicated by some
observations (e.g. \citealt{Walker-2011}, but see
\citealt{Strigari-2010, Wolf-2012}).

Comparisons of the masses and abundances of haloes in DMO and gas
simulations have also been performed previously, by
\cite{Weinberg-2008, Rudd-2008, Dolag-2009} and \cite{Cui-2012}, but
these have focused on scales of $10^{12}\Ms$ and above.
\cite{Semboloni-2011} and \cite{VanDaalen-2011} have investigated the
effects of baryons on estimates of cosmological parameters, for
$k=10h/$Mpc and larger. On these scales, \citep{Dolag-2009}
concluded that baryons only have a minor effect on halo masses, with
adiabatic contraction balancing feedback, while \cite{Rudd-2008} and
\citep{Cui-2012} both found an increase in the mass of haloes, with
\cite{Cui-2012} reporting a net increase of $1-2\%$ in $\rm{M_{200}}$
for haloes of $10^{13.5}\Ms$ and above, and \cite{Rudd-2008} report an
increase in the cumulative halo mass function of $\sim10\%$ above
$10^{12} h^{-1} \Ms$. An opposite effect of a reduction in the matter
power spectrum was found by \cite{VanDaalen-2011}, in simulations
which include AGN feedback. \cite{Simha-2012} have compared
hydrodynamic simulations to results of (sub)halo abundance matching,
and found the results to be consistent.

By comparison to these works, our study is based on much higher
resolution simulations, which allows us to investigate the effect of
baryons in lower mass haloes. In showing that the effect of baryons is
highly mass-dependent, we find that while our results are consistent
with previous results for objects of $10^{12}\Ms$ and
above\footnote{All masses and distances in our own work are expressed
  in physical units.}, on smaller scales, baryons significantly reduce
the mass of haloes and subhaloes, with consequences for
abundance matching.


\begin{figure*} 
    \begin{tabular}{c@{}c@{}c@{}c@{}c@{}}
      \hspace{-2mm}\includegraphics*[trim = 10mm  12mm 17mm 15mm, clip, width = 44mm]{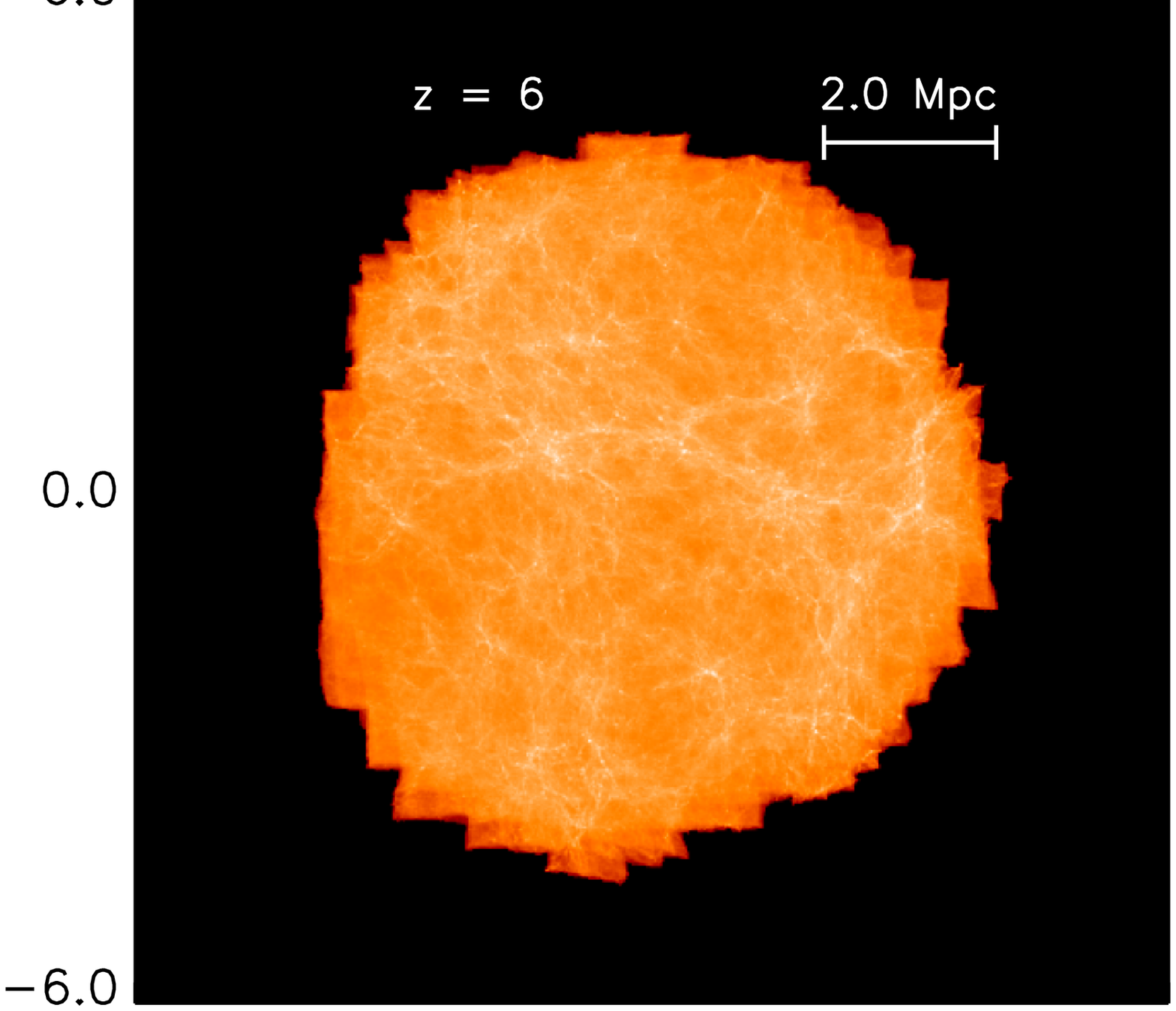} &
      \includegraphics*[trim = 10mm 12mm 17mm 15mm, clip, width = 44mm]{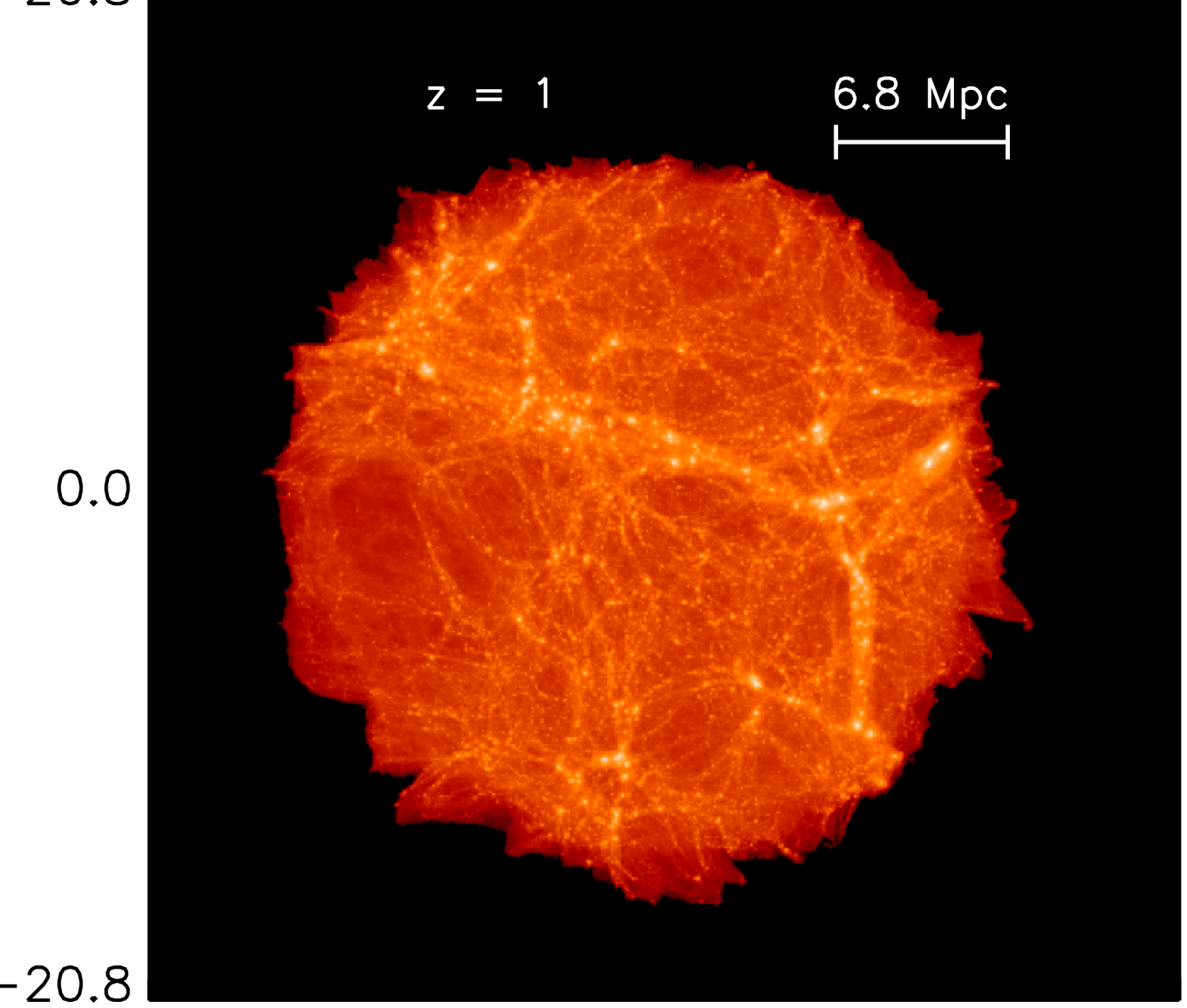} &
      \includegraphics*[trim = 10mm 12mm 17mm 15mm, clip, width = 44mm]{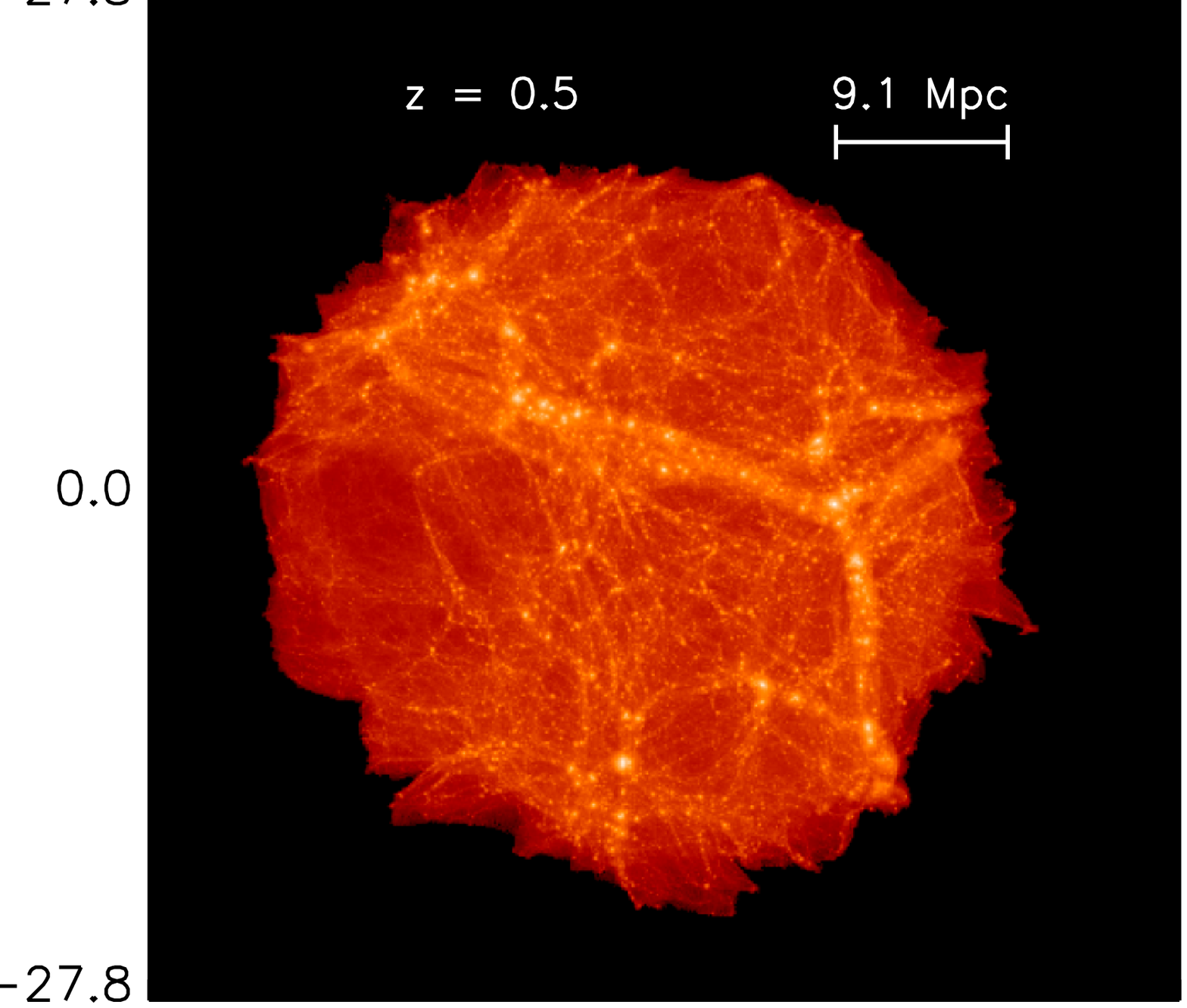} &
      \includegraphics*[trim = 10mm 12mm 17mm 15mm, clip, width = 44mm]{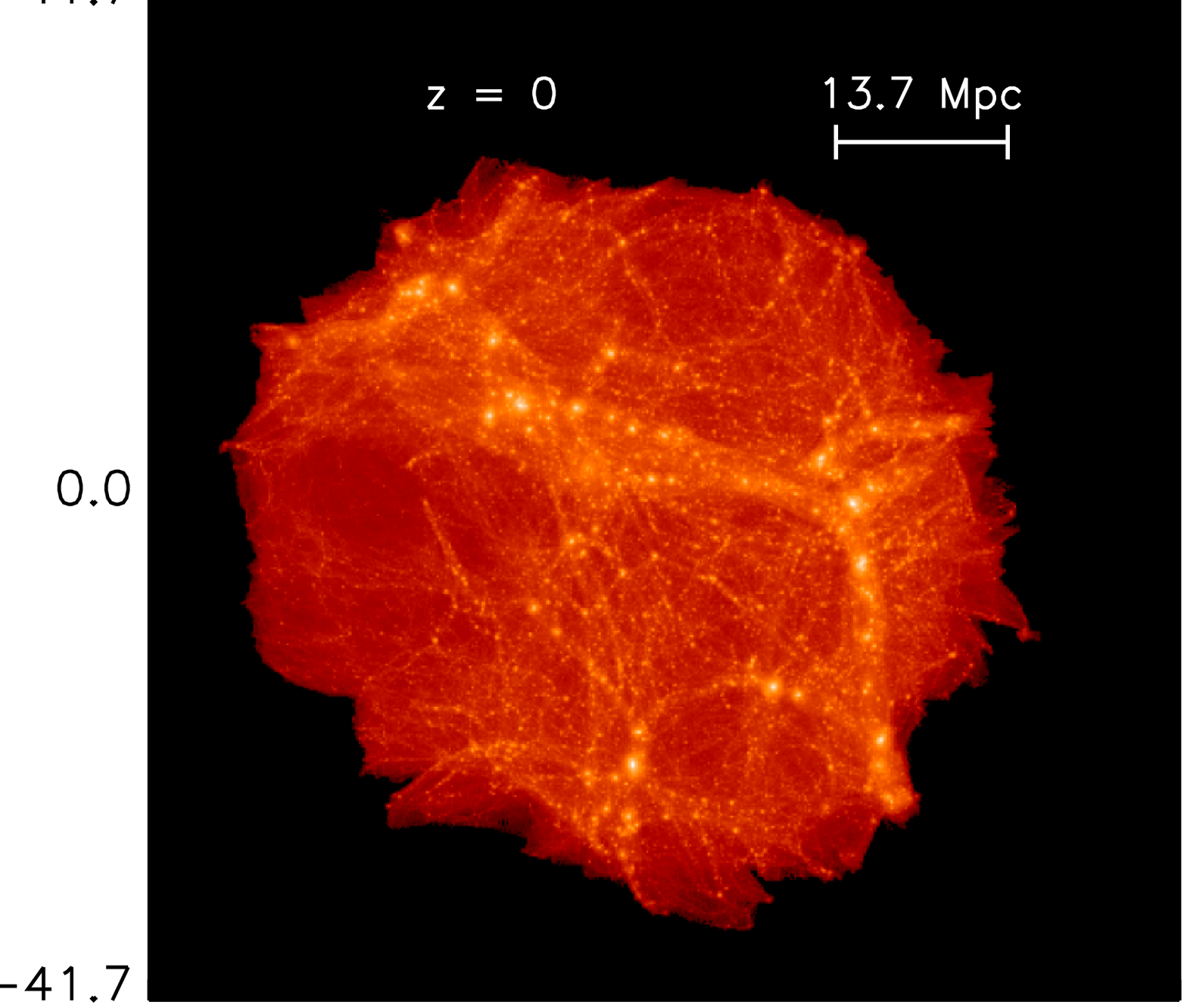} \\
      \hspace{-2mm}\includegraphics*[trim = 10mm 12mm 17mm 15mm, clip, width = 44mm]{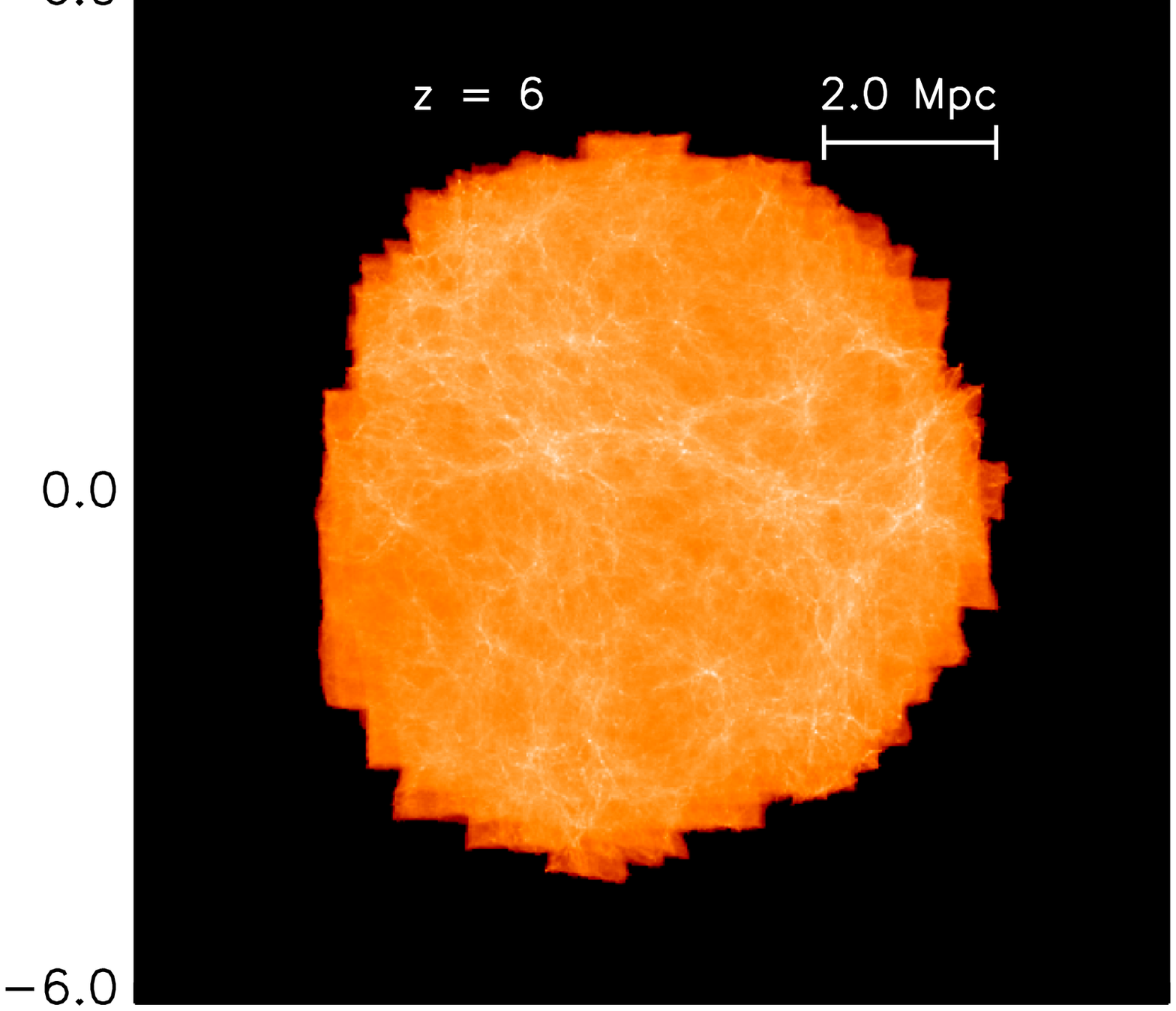} &
      \includegraphics*[trim = 10mm 12mm 17mm 15mm, clip, width = 44mm]{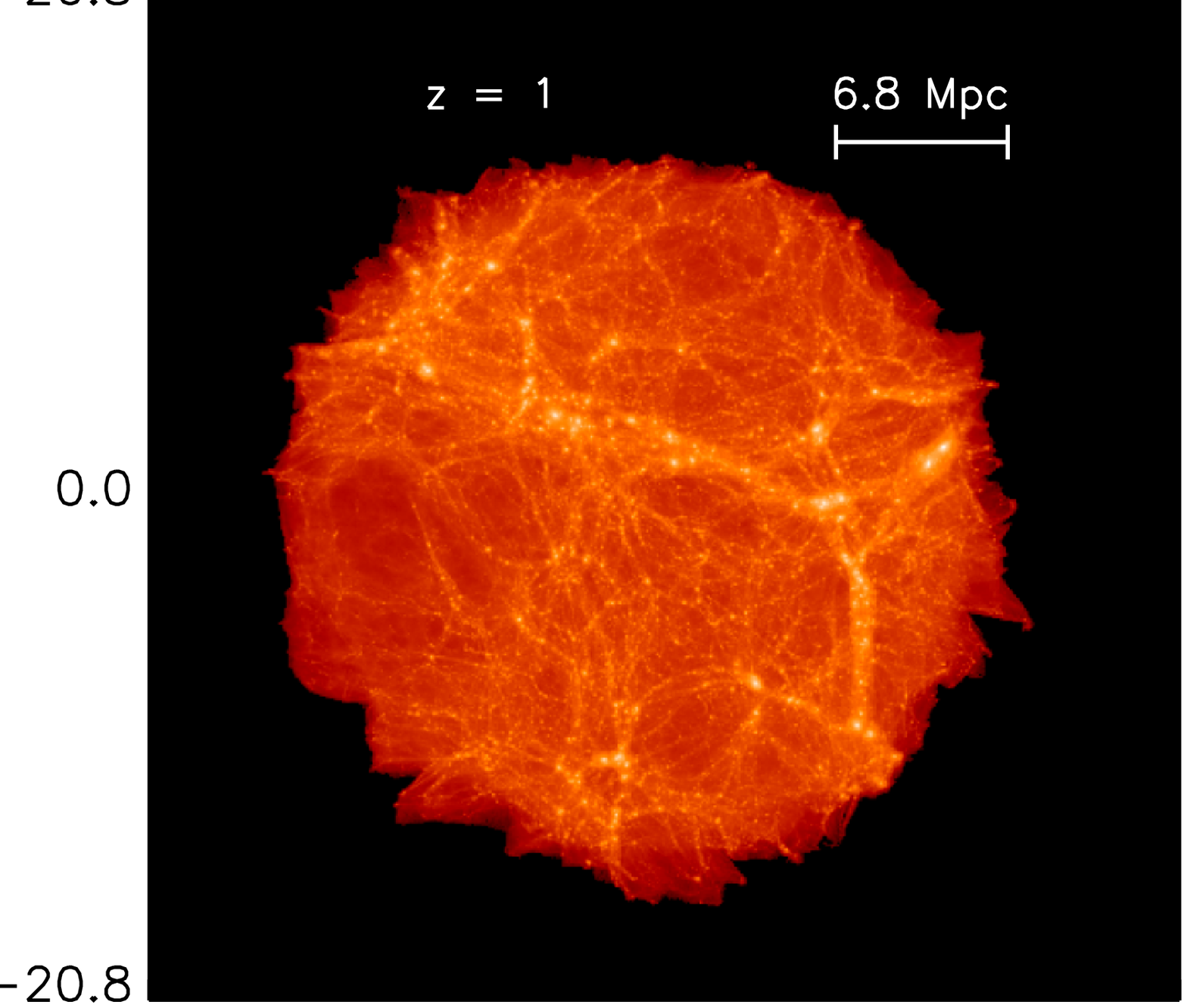} &
      \includegraphics*[trim = 10mm 12mm 17mm 15mm, clip, width = 44mm]{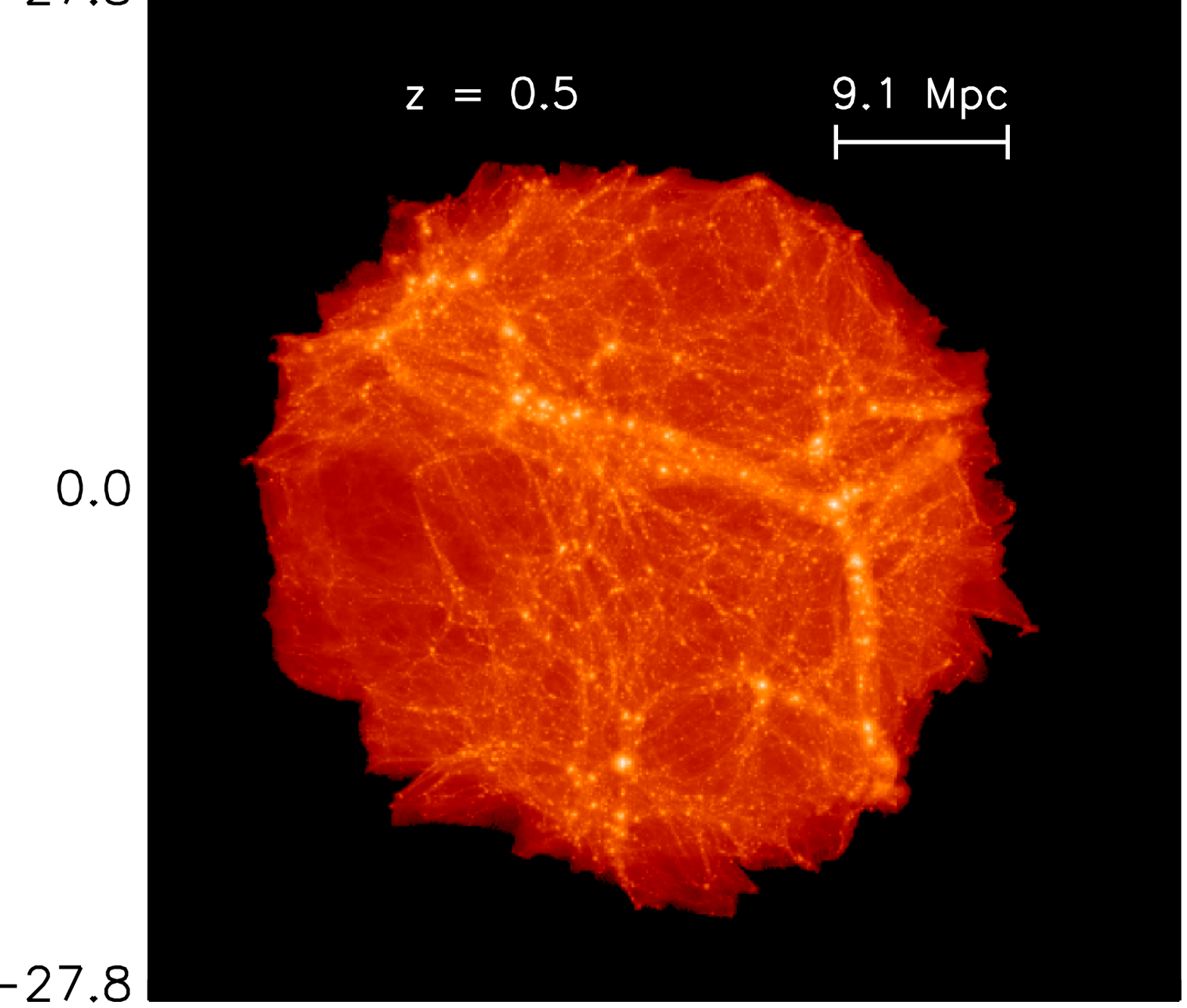} &
      \includegraphics*[trim = 10mm 12mm 17mm 15mm, clip, width = 44mm]{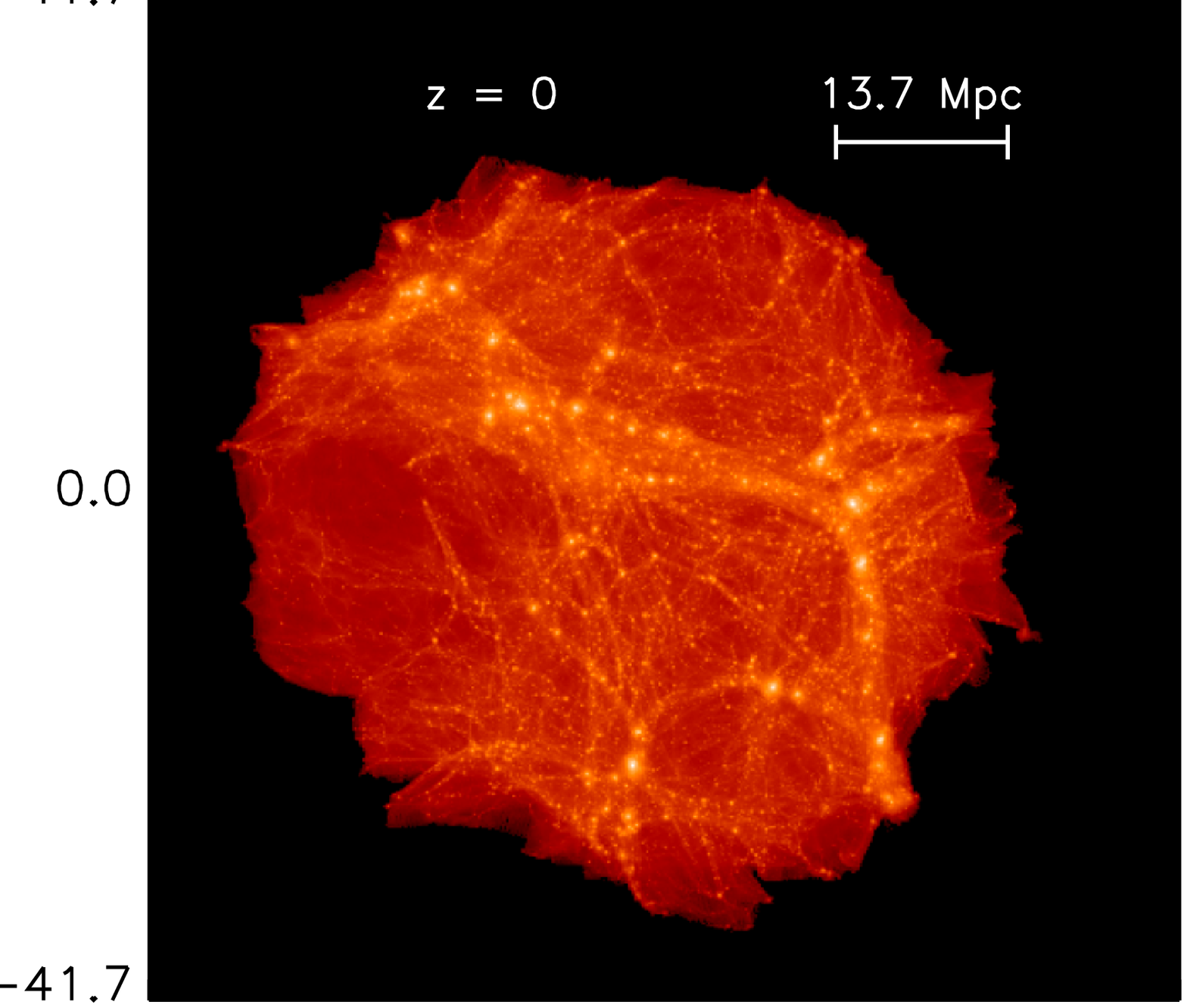} 
    \end{tabular}
  \caption{Evolution of a slice of $16 h^{-1}$~Mpc in depth through
    the total mass distribution within the high-resolution regions of
    the DMO simulation (top) and GIMIC simulation (bottom), at $z=6$,
    1, 0.5, and 0 (left to right). Shown in each panel is a scale bar
    of constant {\it comoving} length $10 h^{-1}$~Mpc, and its
    corresponding length in physical coordinates. On large scales, the
    mass distributions in the DMO and GIMIC simulations appear
    identical.}
  \label{fig:amoeba}
\end{figure*}

This paper is organised as follows: in Section~\ref{sec:methods}, we
describe the simulations, and outline the procedures employed to
identify and link substructures across the two
simulations. Section~\ref{sec:results} contains our main results: in
Section~\ref{sec:origin}, we compare individual structures in both
simulations, and we give an analytic expression for the change in the
median subhalo mass in Section~\ref{sec:analytic-preview}. In
Section~\ref{sec:statistics-mf}, we describe the effect of baryons on
the mass functions of haloes and subhaloes, while in
Section~\ref{sec:statistics-vf}, we examine the $v_{\rm{max}}$
function, and compare our results to observational HI data. In
Section~\ref{sec:satellites}, we show the effect of baryons on the
satellite mass functions of groups and
clusters. Section~\ref{sec:dark} focuses on {\it dark} subhaloes,
which do not contain any gas or stars. In
Section~\ref{sec:applications}, we describe how the baryonic effects
change the results of abundance matching, and compare the inferred
stellar-to-total mass ratios to observations.  We summarise our
results in Section~\ref{sec:summary}. Parameters for the analytical
mass correction are given in Appendix~\ref{sec:correction}, and
numerical convergence is discussed in Appendix~\ref{sec:resolution}.

\section{Methods} \label{sec:methods}

\subsection{The Simulations}\label{sec:setup}
Our comparison is based on a set of N-Body simulations of the same
cosmological volume: a hydrodynamical simulation (``GIMIC''),
performed by the Virgo Consortium and described in detail by
\cite{Crain-2009}, and a matching dark matter only simulation
(``DMO'') where the total mass is composed of dark matter. The
simulations refine a spherical Lagrangian volume (labelled $0\sigma$
in \citealt{Crain-2009}) of radius $\sim 18h^{-1}$~Mpc (a volume of
$\sim 63$~Gpc$^3$ at $z=0$) and median density, selected from the
Millennium Simulation \citep{Springel-Millennium}. Within the zoom
region, additional small scale modes are added using the zoom method
described by \cite{Power-2003}, while the external large scale density
field is represented by lower resolution dark matter particles.

The simulations were performed using the TreePM-SPH code {\sc
  P-Gadget-3}, an extension of the publicly available code {Gadget-2}
\citep{Springel-2005}. The cosmological parameters are identical to
those of the Millennium simulations; density parameters
$\Omega_m=0.25$, $\Omega_\Lambda=0.75$, $\Omega_b=0.045$, Hubble
parameter $h=0.73$, power spectrum normalisation $\sigma_8=0.9$, and
spectral index $n=1$. The evolution of the dark matter distribution in
the full, high-resolution region in both simulations is shown in
Fig.~\ref{fig:amoeba}. By construction, the simulated volume is
approximately spherical at $z=1.5$, and becomes more irregular at both
higher and lower redshifts. On scales of megaparsecs and above,
baryons do not significantly affect structure formation, and no
differences can be seen in the density distribution shown in
Fig.~\ref{fig:amoeba}.

\begin{figure}
  \includegraphics*[trim = 16mm  3mm 5mm 12mm, clip, width = \columnwidth]{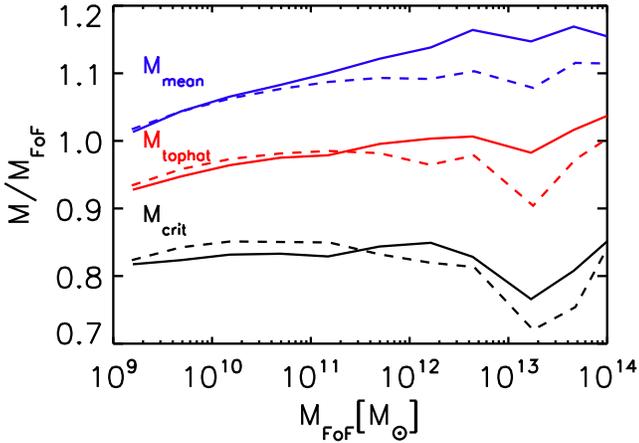}
  \caption{Comparison of halo mass estimators $\rm{M_{200,mean}}$
    (blue), $\rm{M_{tophat}}$ (red), and $\rm{M_{200,crit}}$ (black),
    as a function of $\rm{M}_{FoF}$ for all haloes at $z=0$. Solid
    lines show the results of the GIMIC simulation, dashed lines show
    results of the DMO simulation. As expected, for all objects,
    $\rm{M_{200,mean}}$ gives the highest mass per halo, followed by
    $\rm{M_{tophat}}$ and $\rm{M_{200,crit}}$. Importantly, there is
    no systematic bias in mass estimators between the GIMIC or the DMO
    simulation.}
  \label{fig:mass-definition}
\end{figure}

\begin{figure}
  \begin{center}
  \includegraphics*[width = \columnwidth]{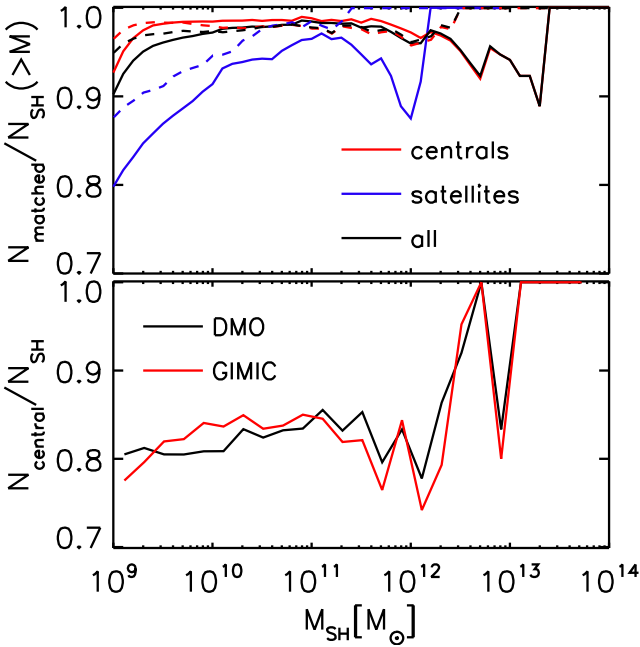} 
  \end{center}
  \caption{Top panel: cumulative fraction of subhaloes in the DMO
    simulation that are uniquely matched to subhaloes in the GIMIC
    simulation at $z=0$ (solid) and $z=1$ (dashed), for centrals
    (red), satellites (blue), and all subhaloes (black). Bottom panel:
    fraction of all subhaloes at $z=0$ that are central in the DMO
    simulation (black), and in the GIMIC simulation (red).  In
    general, the fraction of subhaloes that are matched increases with
    mass, and it is slightly lower at $z=0$ compared to $z=1$,
    particularly for satellites. The lower matching fraction among
    satellites and its decrease over time is attributable to a higher
    rate of mergers, and a higher probability for divergent evolution
    between the two simulations. Overall, more than 90\% of subhaloes
    above $10^9 \Ms$ are matched at $z=0$. In both simulations,
    centrals account for $\sim80\%$ of subhaloes in the mass range
    $10^9$-$10^{12}\Ms$.}
  \label{fig:matching-fraction}
\end{figure}

\subsubsection{Initial Conditions}
Both simulations evolve the same number, $2.75 \times10^8$, of dark
matter particles, from $z=127$ to $z=0$, and are analysed at 58
co-temporal snapshots. In the GIMIC simulation, the high-resolution
region additionally contains an equal number of gas particles that can
be converted to star particles.

The initial particle masses are $m_g=1.98\times 10^6\Ms$ and $m_{DM} =
9.05\times 10^6\Ms$ for gas and dark matter, respectively in the GIMIC
simulation, and $m_{DM} = 1.98+9.05 = 11.03 \times 10^6 \Ms$ in the
DMO simulation. In the GIMIC simulation, gas particles can be
converted into star particles of equal mass, and mass can be exchanged
between gas and star particles in the form of stellar winds and
supernovae. The gravitational softening scale for all particle types
in the high resolution region is initially fixed in comoving
coordinates, and chosen so that it reaches a value of $0.5 h^{-1}$~kpc
Plummer equivalent in physical coordinates at $z=3$, whereupon it
remains fixed in physical coordinates, contracting in comoving
coordinates as $1/a$. To test convergence, both simulations were also
repeated at lower resolution, with all particle masses increased by a
factor of eight, and the softening increased by a factor of
two. Unless we specifically refer to the lower resolution simulations
for the purpose of comparison, all results are based on the
high-resolution simulations. We discuss convergence in
Appendix~\ref{sec:resolution}, and consider the abundance of
(sub)haloes to be unaffected by resolution down to $10^9 \Ms$.

\subsubsection{Physical Model}
While the DMO simulation includes only gravitational interactions, the
GIMIC simulation also includes astrophysical processes, such as gas
dynamics, photo-ionisation, radiative element-by-element cooling, star
formation and supernova feedback. Here, we give a brief outline of the
most important aspects of the physics model; a more complete
description can be found in \cite{Crain-2009} and references therein.

\begin{itemize}
\item Radiative cooling is computed assuming the optically thin
  limit and ionisation equilibrium, depending on redshift, local
  temperature and density, and chemical composition. The IGM is heated
  uniformly, with a redshift-dependent UV/X-ray background as given by
  \cite{Haardt-2001}, with hydrogen and helium-II being reionised at
  $z\sim9$ and $z\sim3.5$, respectively.

\item Following \cite{Schaye-2008}, star formation is implemented by
  imposing an effective equation of state for gas particles above a
  density threshold of $n_H=0.1 \rm{cm}^{-3}$, that makes the Jeans
  mass independent of the gas density. In this regime, gas particles
  can be transformed into star particles at a pressure dependent rate
  that reproduces a local Kennicutt-Schmidt law, with a star formation
  rate column density of $\dot{\Sigma} = 1.5 \times 10^4
  \Ms\rm{yr}^{-1}\rm{kpc}^{-2} \Sigma_g^{1.4}$, where $\Sigma_g$ is
  the gas column density in $\Ms\rm{pc}^{-2}$. Each star particle is
  assumed to represent a single stellar population, with an IMF based
  on \cite{Chabrier-2003}. As described in \cite{Wiersma-2009b},
  stellar evolution follows the abundances of 11 metals, which are
  released over time to the surrounding gas particles by supernovae
  type II, type Ia, and AGB stars.

\item Following \cite{DallaVecchia-2008}, supernova feedback is
  imparted on the surrounding gas particles in the form of kinetic
  energy, with each star in the range of $6-100 \Ms$ providing
  $10^{51}$ ergs, sufficient to accelerate an average of 4 neighbours
  to a wind velocity of 600~kms$^{-1}$.
\end{itemize}

Whilst some parts of the astrophysical model are still subject to
considerable uncertainty, both in terms of the underlying physics and
its numerical implementation, it has been shown to reproduce the
cosmic star formation rate \citep{Crain-2009, Schaye-2010}, and the
properties of individual galaxies in some detail (\citealt{Font-2011},
\citealt{McCarthy-2012a}). By comparison, the results presented in
this paper are quite general.

The GIMIC simulation does not include AGN.  This omission is the most
likely cause of a too high stellar and baryon fraction in the most
massive haloes. Because the total abundance of these massive haloes is
low, the effect on the cumulative mass function is negligible for less
massive haloes.

\subsection{Analysis}
We compare the populations of haloes and subhaloes in the two
simulations, as well as individual objects which are matched across
both. In Section~\ref{sec:identification}, we describe the
identification of structures and substructures, and the definition of
different subsets used in the analysis. In Section~\ref{sec:linking},
we describe the procedure and criteria used for matching substructures
across the two simulations.

\subsubsection{Identification of Substructures} \label{sec:identification}
In each snapshot, structures are identified in a two step
process. First, dark matter {\it haloes} are found using the
Friend-of-Friend (FoF) algorithm, with a linking length of 0.2
\citep{Davis-1985}. In the GIMIC simulation, star particles and gas
particles are then attached to the nearest dark matter particle, and
inherit its FoF association. In a second step, self-bound
substructures within haloes are identified using the {\sc subfind}
algorithm of \cite{Springel-2001}, with the extension of
\cite{Dolag-2009} to account for baryons. {\sc Subfind} computes the
density field within each FoF halo using an SPH-interpolation. A set
of $N > 20$ particles (of any type) is considered as a potential
substructure if it exceeds the local smoothed density estimate. It is
then subjected to unbinding, whereby all particles whose combined
kinetic and internal energy exceeds their gravitational binding energy
to the substructure are iteratively removed, until the substructure
either vanishes, falling below the 20 particle threshold, or is
identified as a genuine self-bound subhalo.


\begin{figure*}
  \begin{center}
      \includegraphics*[trim = 27mm 25mm 9mm 5mm, clip, height = .32\textwidth]{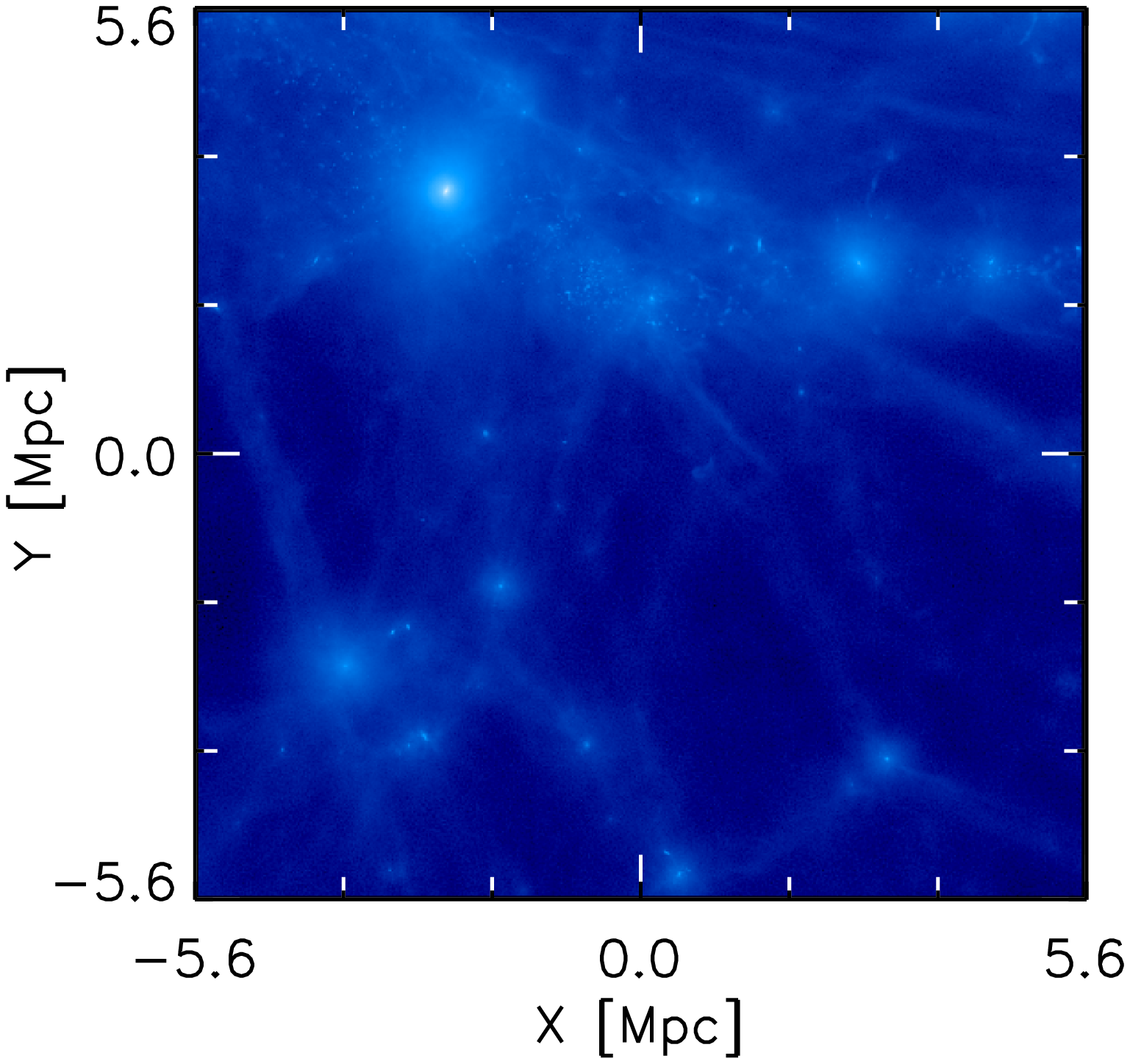}
      \includegraphics*[trim = 27mm 25mm 9mm 5mm, clip, height = .32\textwidth]{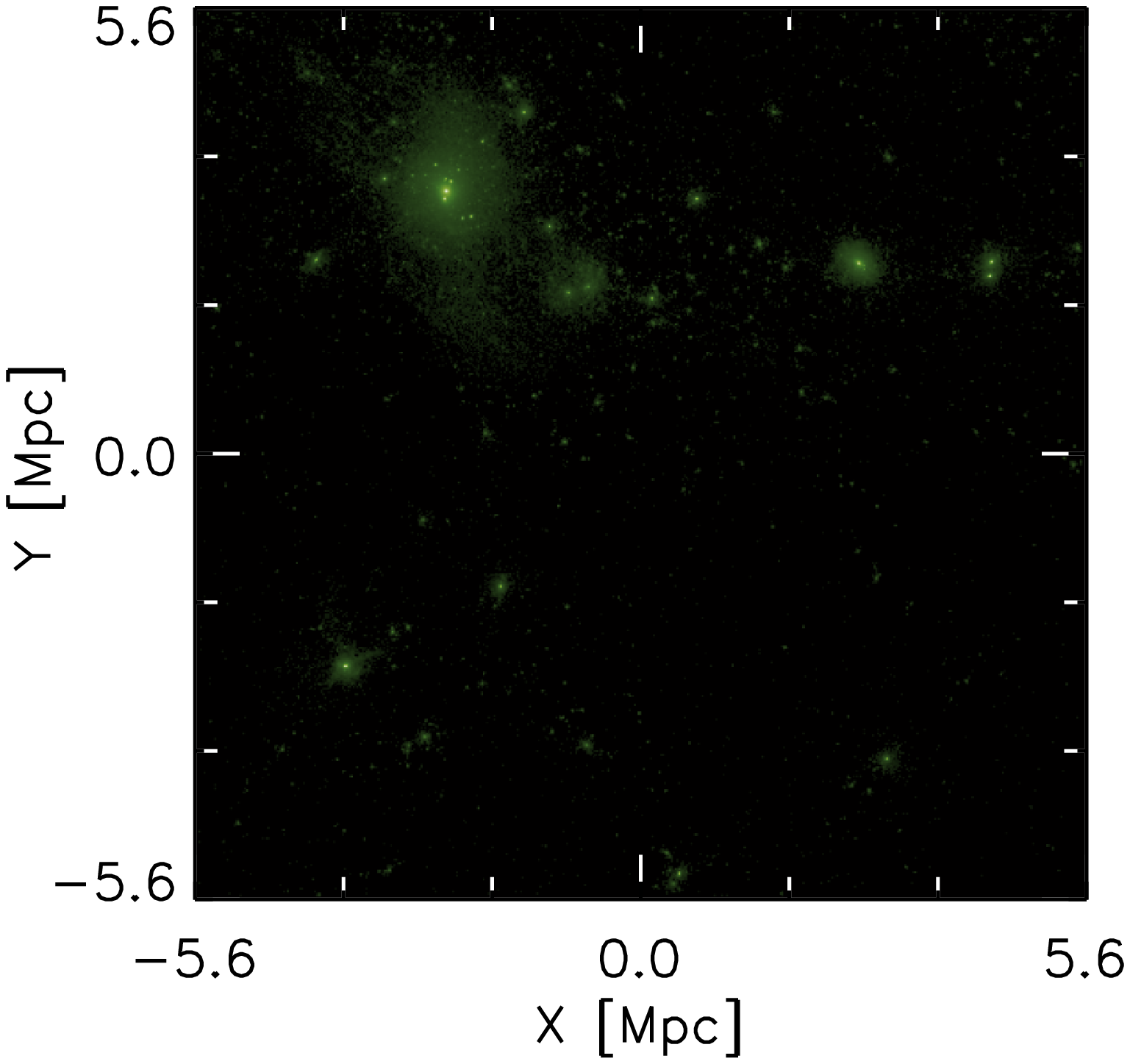}
      \includegraphics*[trim = 27mm 25mm 9mm 5mm, clip, height = .32\textwidth]{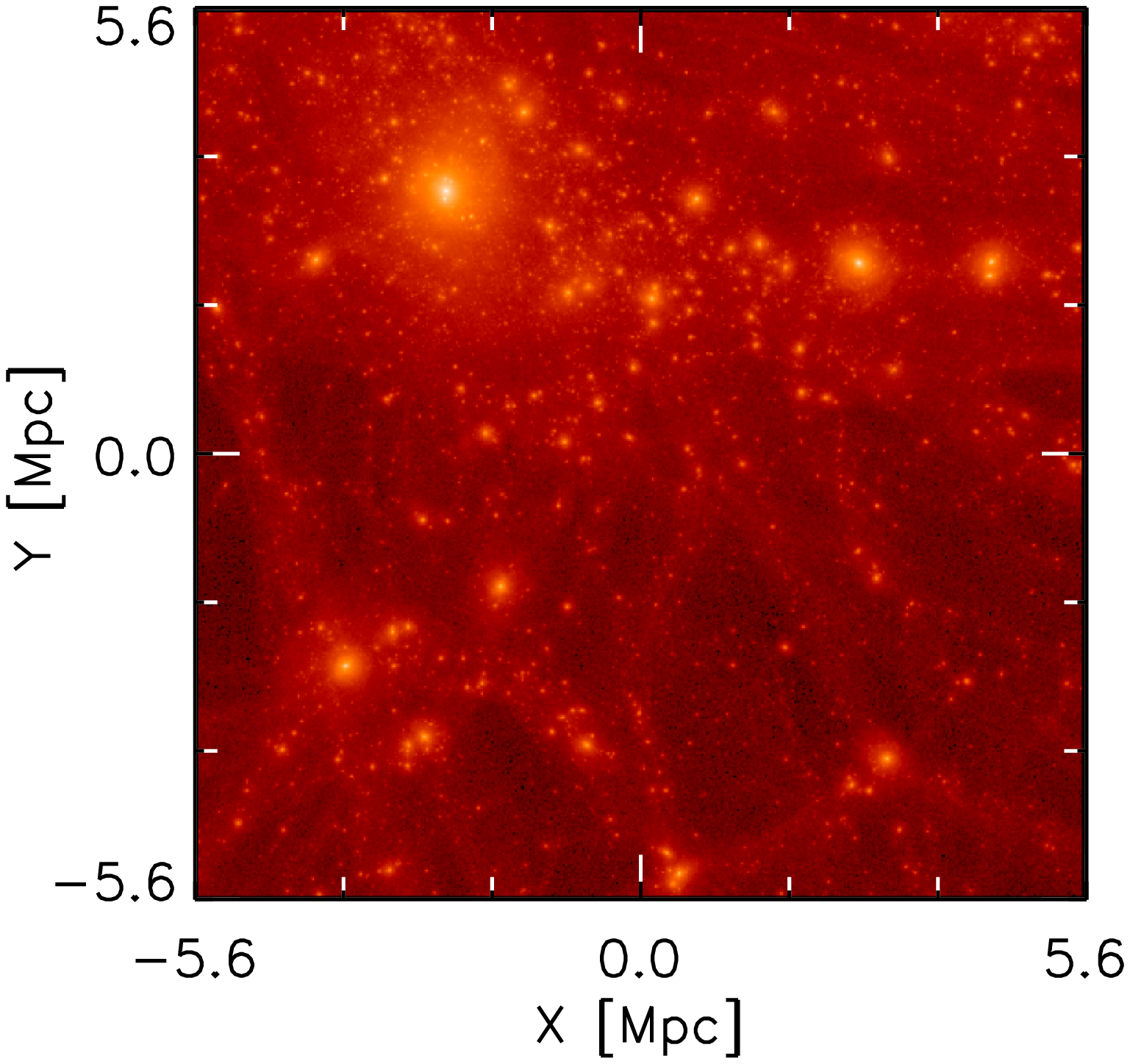}
  \end{center}
  \caption{Projected density distributions of gas (left), stars
    (centre) and dark matter (right) in the GIMIC simulation, from a
    cubic region of side length $l=11.2$~Mpc at $z=0$. On large
    scales, stars and gas follow the dark matter distribution, but on
    galaxy scales, the distributions differ, leading to a difference
    in the total mass distributions between the GIMIC and the DMO
    simulations.}
  \label{fig:high-res-components}
\end{figure*}

\subsubsection{Mass Estimators} \label{sec:mass-definition}
Having identified haloes and sub-haloes, their masses can be measured
in several ways. We define the FoF-mass, $\rm{M_{FoF}}$, of a halo as
the sum of the masses of all particles belonging to the FoF-group. In
structure formation theory, it is customary to characterise a halo by
$\rm{M_{200}}$, which describes an object with an overdensity of 200,
near the threshold for a dark matter halo to collapse and
virialise. Different definitions exist for $\rm{M_{200}}$ and
$\rm{r_{200}}$, the radius which encloses a spherical volume of
corresponding overdensity: $\rm{M_{200,crit}}$ is defined such that
the mean overdensity within $\rm{r_{200}}$ is 200 $\times$ the
critical density; $\rm{M_{200,mean}}$ such that the overdensity within
$\rm{r_{200}}$ is 200 $\times$ the mean background density;
$\rm{M_{tophat}}$ is the mass in a spherical volume whose overdensity
equals the value at virialisation in the top-hat collapse model. In
Fig.~\ref{fig:mass-definition}, we show a comparison of the three mass
estimators relative to the FoF-mass for all haloes at $z=0$. We find
that $\rm{M_{tophat}}$ most closely matches $\rm{M_{FoF}}$, and that
the spread between the three estimators increases with halo mass in
the GIMIC simulation, from $\sim 20\%$ at $10^9 \Ms$ to $\sim 30\%$ at
$10^{14} \Ms$. Importantly for our purpose, below $10^{12}\Ms$ the
ratio between each of the three halo mass estimators and the FoF-mass
is very similar in both simulations, and much smaller than the
measured baryonic effect on the halo mass. We use $\rm{M_{FoF}}$ as
our halo mass definition throughout the rest of the paper, as it is
independent of the concentration of the halo. The fact that our
results are very similar for the different mass estimators indicates
that a measured halo mass difference is a true physical effect.

For subhaloes, the mass $\rm{M_{SH}}$ is defined as the total mass of
particles bound to the subhalo. For {\it isolated} subhaloes,
i.e. central subhaloes with no satellites, we find that the subhalo
mass is typically very close to the halo mass, indicating that most
particles belonging to the halo are also gravitationally bound to
it. We have also examined individual subhaloes, and compared density
profiles of the gravitationally bound particles to those of all
particles in the vicinity. We find no noticeable difference between
the two simulations in the ratio of bound to unbound particles for
each subhalo, indicating that any subhalo mass difference that we find
between the two simulations is a genuine physical effect. Examples of
density profiles, including a comparison of the bound and unbound
particles, are shown in Fig.~\ref{fig:mass-profiles-components}.

We discuss the convergence of the simulations in terms of the halo-
and the subhalo mass functions in Appendix~\ref{sec:resolution}, and
show that the principal effects of the baryon physics are independent
of the definition of the objects, and are not affected by resolution.

\subsubsection{Matching Substructures Across
  Simulations} \label{sec:linking} In order to understand how the
differences between the DMO and the GIMIC simulations arise, we also
compare the evolution of individual objects. We use the fact that dark
matter particles of identical position in the initial
conditions\footnote{A DM particle in the DMO simulation is replaced by
  a pair of DM and gas particles with the same centre of mass in the
  initial conditions of the GIMIC simulation.} have identical IDs in
order to match substructures across both simulations. Specifically, at
each snapshot and for every subhalo in the DMO simulation, we consider
its 5 most bound particles, and require that at least 3 of them belong
to a single subhalo in the GIMIC simulation. This ensures that each
subhalo in the DMO simulation is matched to at most one subhalo in the
GIMIC simulation. Because mergers can occur at different times in the
two simulations (and in some cases occur only in one and not in the
other), we also impose a threshold on the mass ratio, matching only
objects whose mass ratio is less than 4:1.

As shown in Fig.~\ref{fig:matching-fraction}, this method results in
$>90\%$ of subhaloes above $10^9\Ms$ in the DMO simulation being
matched to unique counterparts in the GIMIC simulation, including
$>95\%$ of centrals, and $>80\%$ of satellites. The reduction of the
matching fraction for present-day satellites compared to central
subhaloes is largely due to their more chaotic orbits, and the
associated increase in the likelihood of diverging merger histories of
individual objects in the two simulations. In addition, low-mass
satellites are more likely to be disrupted in the GIMIC simulation
compared to the DMO simulation.

In total, at $z=0$, the DMO simulation contains 214,676 haloes and
254,305 subhaloes; the GIMIC simulation contains 148,765 haloes and
162,866 subhaloes. Including only subhaloes with a total mass above
$10^9\Ms$, the DMO simulation has 86,137 subhaloes, 69,699 centrals
and 16,438 satellites, while the GIMIC simulation has 62,265
subhaloes; 50,077 centrals and 12,188 satellites. Of the central
subhaloes in the DMO simulation with masses above $10^9\Ms$, $92.6\%$
are matched to subhaloes in GIMIC, $97\%$ of which are matched to
centrals and $3\%$ to satellites. In the same mass range, $80.0\%$ of
satellites in the DMO simulation are matched to subhaloes in GIMIC,
$70\%$ of which are matched to satellites and $30\%$ to centrals.

We also compared the matching success rate for subhaloes in the
central parts and near the boundary of the high-resolution region (see
Section~\ref{sec:setup}), but found no significant edge-effects. While
the criteria for matching subhaloes appear somewhat arbitrary, we note
that the matching of objects across the two simulations only serves to
understand the origin of the differences, but that statistical
quantities, including mass functions and satellite mass functions, are
independent of the matching rate and criteria.

\section{Effect of Baryons on Structures}\label{sec:results}
In this Section, we describe the effects that the inclusion of baryons
have on the formation and evolution of structure mainly in statistical
terms. We compare the evolution of individual objects across the two
simulations in Section~\ref{sec:origin}. The effect on the overall
abundance of haloes and subhaloes is discussed in
Section~\ref{sec:statistics-mf}. Section~\ref{sec:statistics-vf} shows
the effect on the $v_{\rm{max}}$ function, and compares our results to
measurements of the ALFALFA HI survey in the Local Volume. In
Section~\ref{sec:satellites}, we focus on the mass functions of
individual groups and clusters. We highlight the importance of {\it
  dark} subhaloes in Section~\ref{sec:dark}, which will be relevant to
the discussion of abundance matching in
Section~\ref{sec:applications}.

From the similarity in the overall mass distribution of the dark
matter only (``DMO'') simulation and its counterpart with baryons
(``GIMIC'') shown in Fig.~\ref{fig:amoeba}, it is clear that the large
scale evolution of structures is nearly identical. On scales of
several Mpc, gravity remains the only relevant force even in the
hydrodynamical simulation, and the gas distribution closely follows
that of the dark matter. Fig.~\ref{fig:high-res-components} shows the
projected density distribution of gas (left), stars (centre) and dark
matter (right) in a region of $11.2^3$ Mpc$^3$ from the GIMIC
simulation at $z=0$. On scales of several hundred kpc and below, it is
apparent that the gas distribution is much smoother than the dark
matter, a reflection of the fact that most of the interstellar medium
of small haloes has been expelled, while the intergalactic medium is
heated through the UV background, resulting in a large Jeans
length \citep{Theuns-2000}. While the stellar mass is also highly clustered, it is only
significant in the most massive objects.


\subsection{Individual Objects}\label{sec:origin}
\begin{figure*}
  \begin{center}
    \includegraphics*[trim = 15mm  25mm 3mm 6mm, clip, width = .32\textwidth]{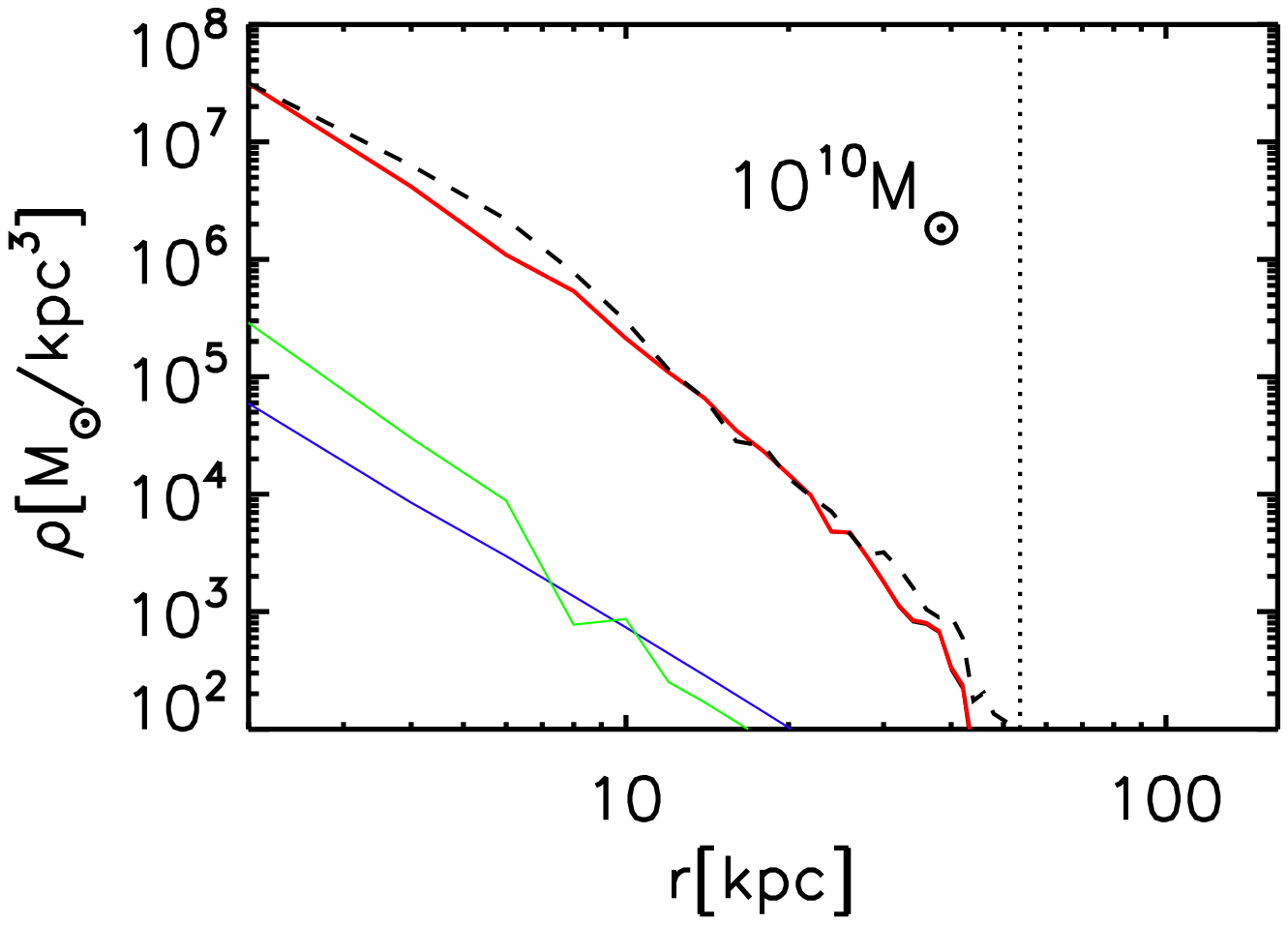}            
    \includegraphics*[trim = 15mm  25mm 3mm 6mm, clip, width = .32\textwidth]{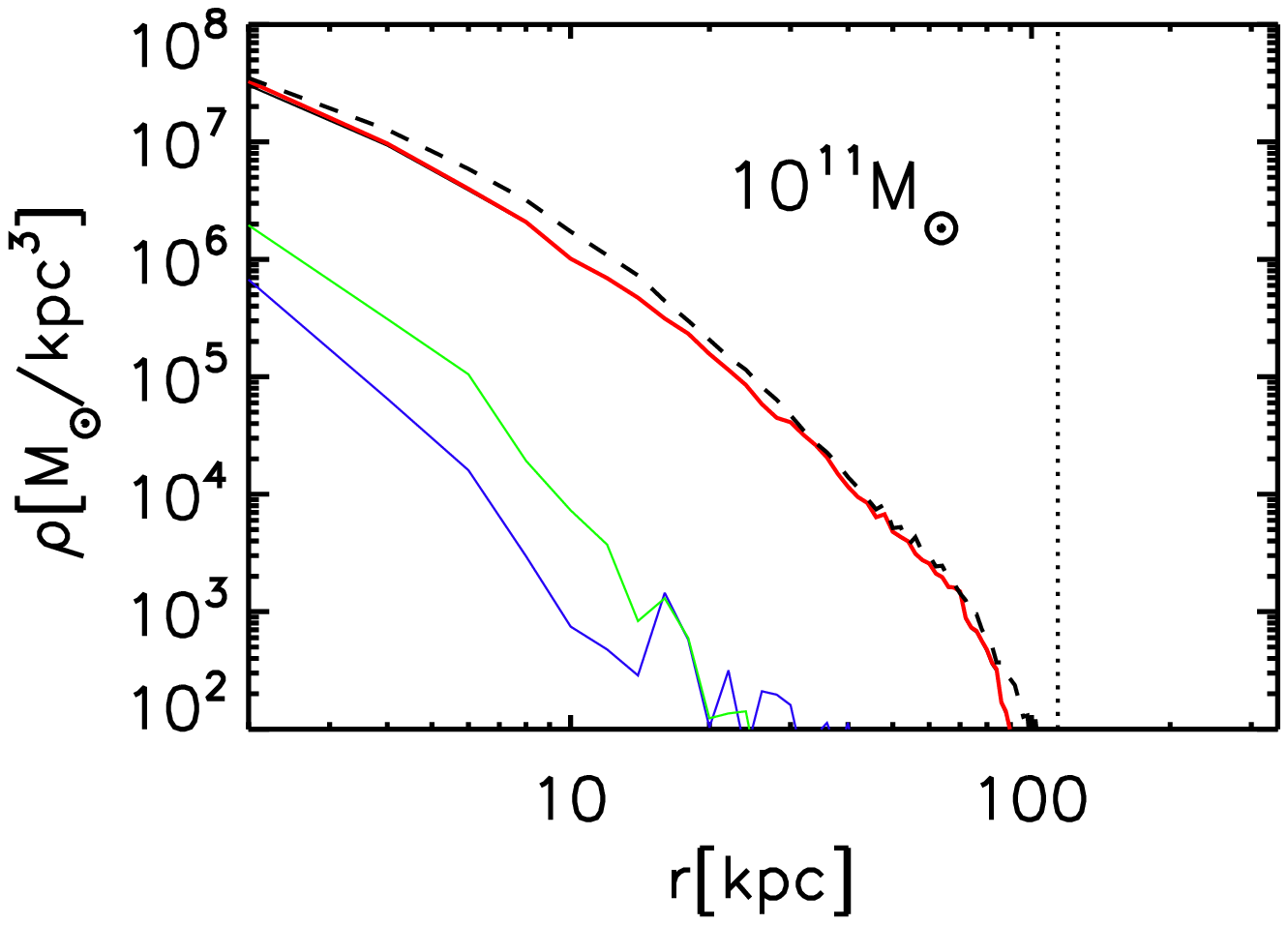}            
    \includegraphics*[trim = 15mm  25mm 3mm 6mm, clip, width = .32\textwidth]{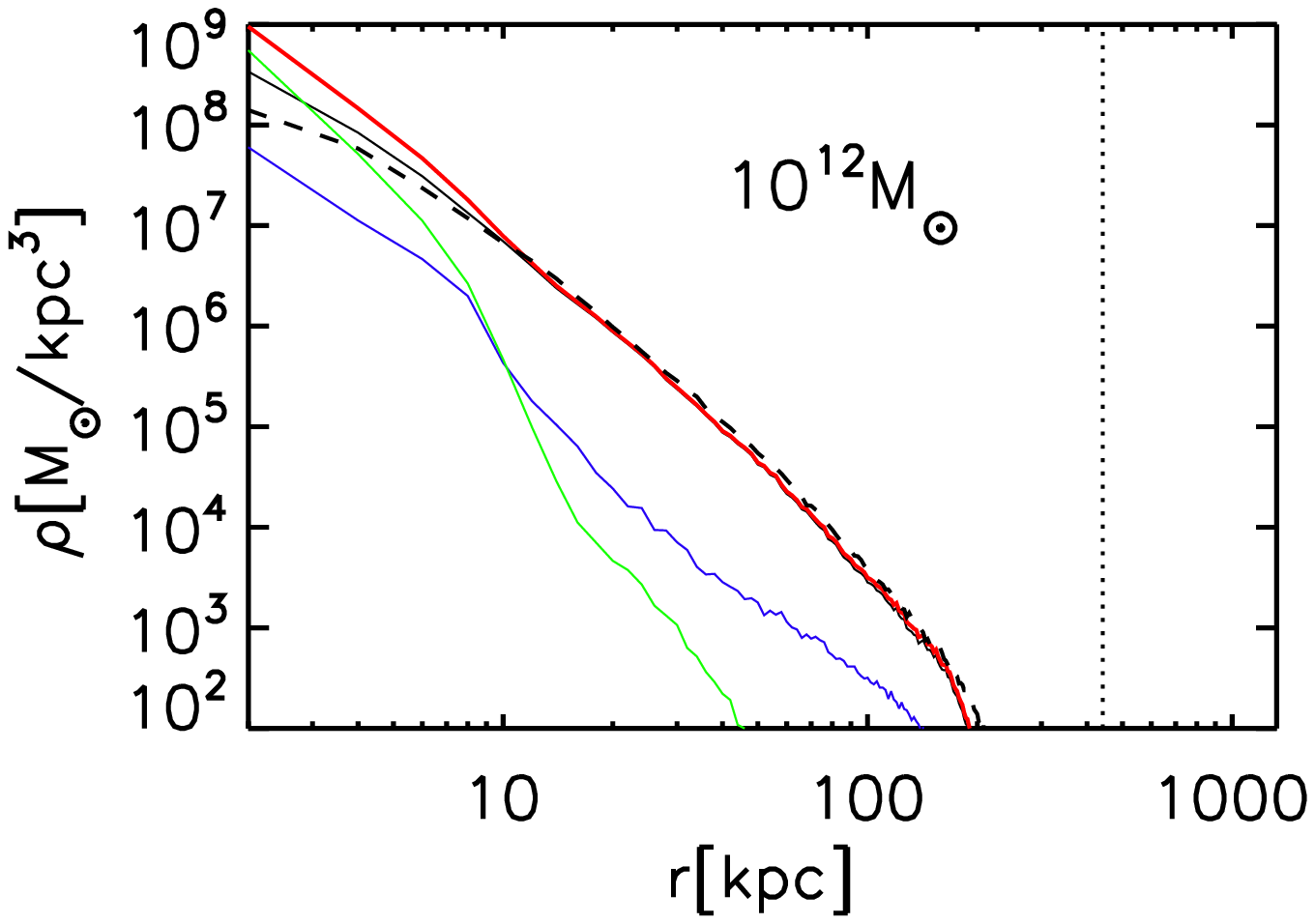} \\
    \includegraphics*[trim = 15mm  5mm 3mm 6mm, clip, width = .32\textwidth]{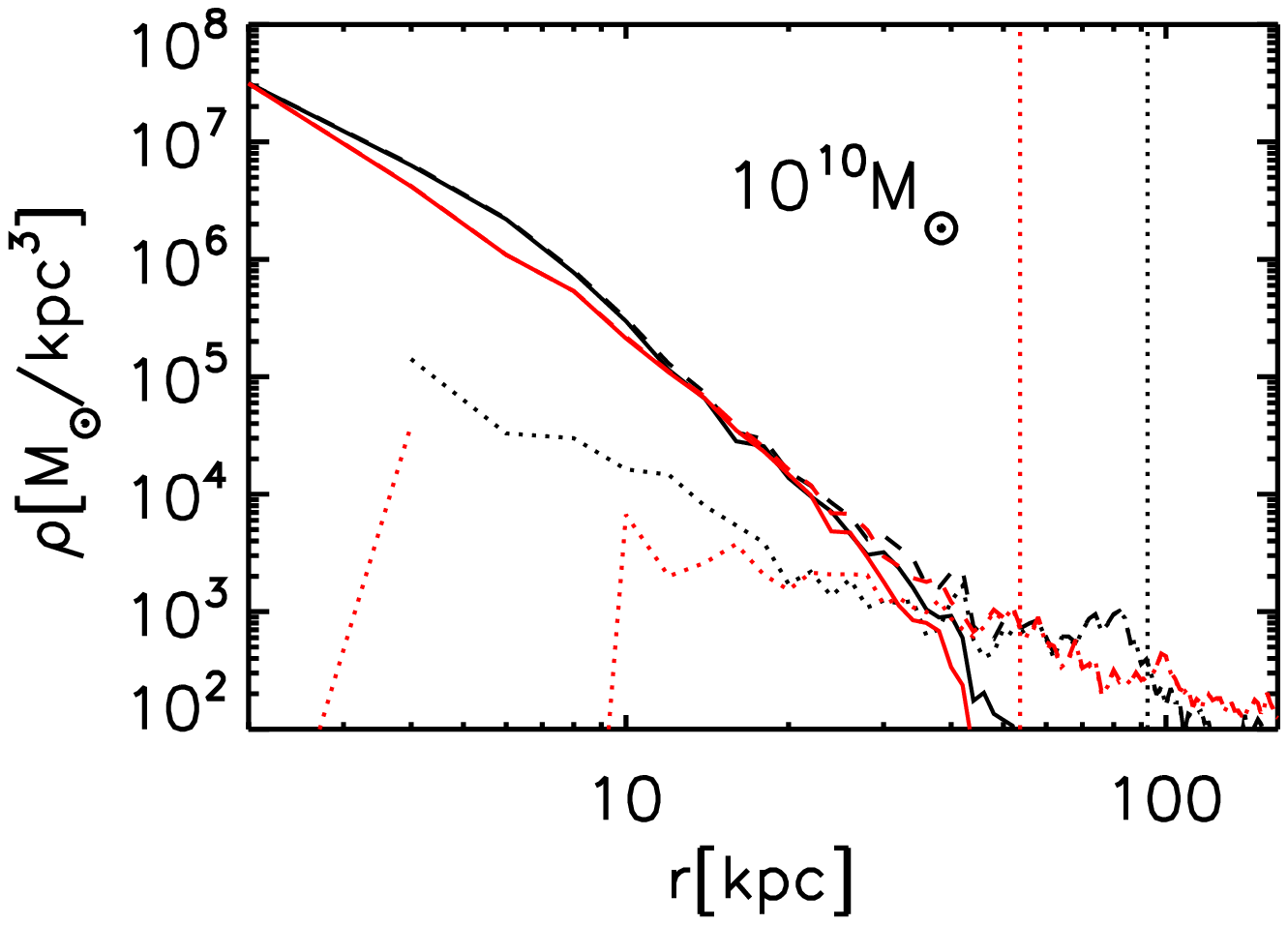}     
    \includegraphics*[trim = 15mm  5mm 3mm 6mm, clip, width = .32\textwidth]{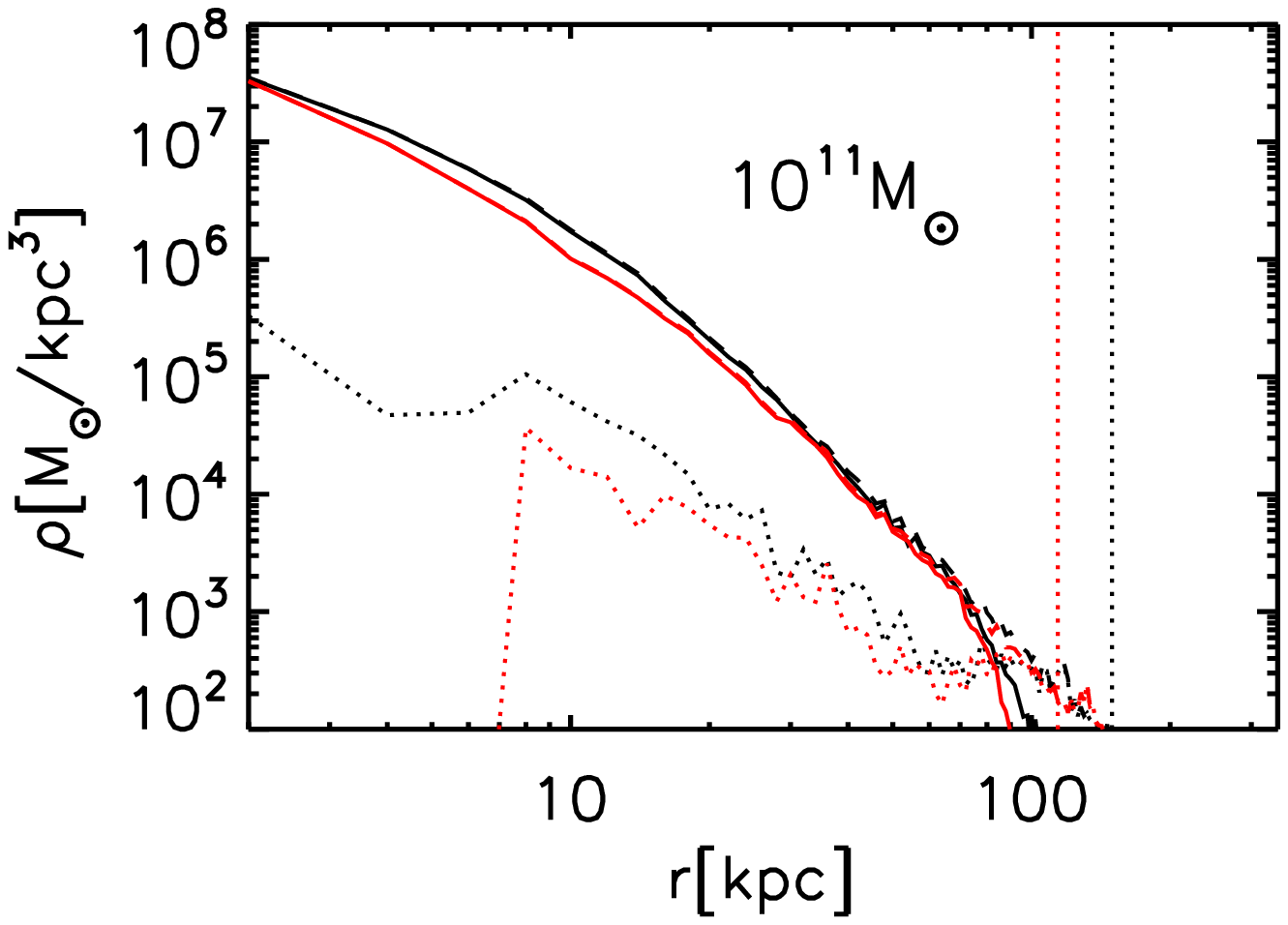}            
    \includegraphics*[trim = 15mm  5mm 3mm 6mm, clip, width = .32\textwidth]{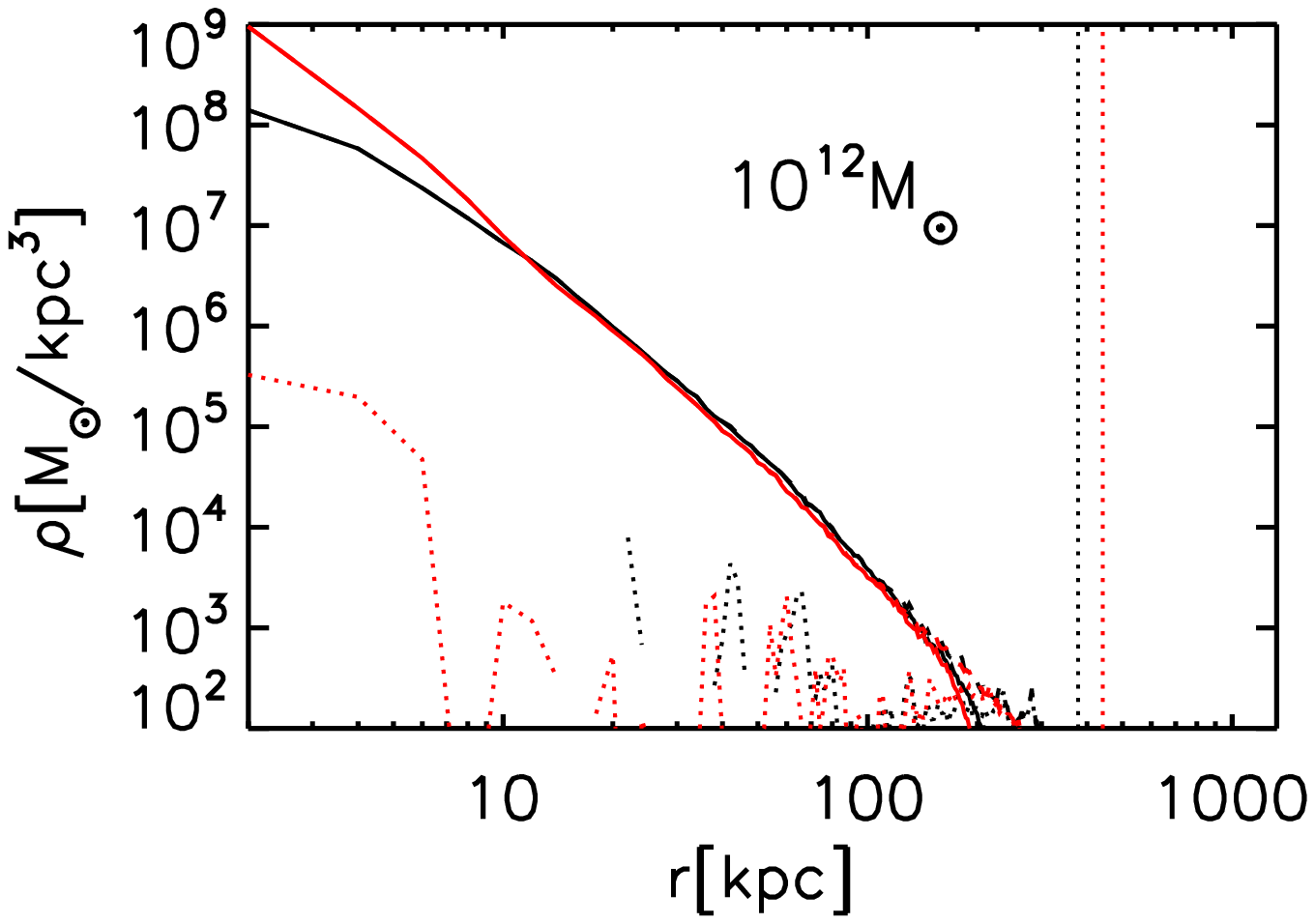} 
  \end{center}
  \caption{Spherical density profiles of 3 pairs of individual
    matched, central subhaloes. In the top row, solid lines show the
    different components in the GIMIC simulation: dark matter (black),
    stars (green), gas (blue), and total mass density (red), each
    including only particles bound to the subhalo. The dashed line
    shows the density of the same object in the DMO simulation, and
    vertical lines indicate the radius of the outermost bound
    particle. The material within $\sim 10^{12}\Ms$ subhaloes has been
    rearranged appreciably in the GIMIC simulation relative to the DMO
    simulation, with baryons accumulating in the centre. The
    difference between the solid and dashed black lines in the inner
    part shows that the dark matter itself is also more concentrated
    in GIMIC, but the total subhalo mass does not change. By contrast,
    $\sim 10^{10}\Ms$ and $\sim 10^{11}\Ms$ subhaloes have lost mass
    in the GIMIC case. Here, the total density is dominated by dark
    matter throughout, and lower than in the DMO simulation. In the
    bottom row, solid lines denote the density profiles computed using
    bound particles, dotted lines show unbound particles in the
    vicinity (not counted towards the subhalo mass), and dashed lines
    show the sum of the densities of bound and unbound particles. In
    all cases, the density of unbound particles is low, and there is
    no significant difference between the two simulations in the ratio
    of bound to unbound particles.}
  \label{fig:mass-profiles-components}
\end{figure*}

\begin{figure*}
  \begin{center}
     \hspace{-.2in} \includegraphics*[trim = 15mm  3mm 3mm 6mm, clip, width = .5\textwidth]{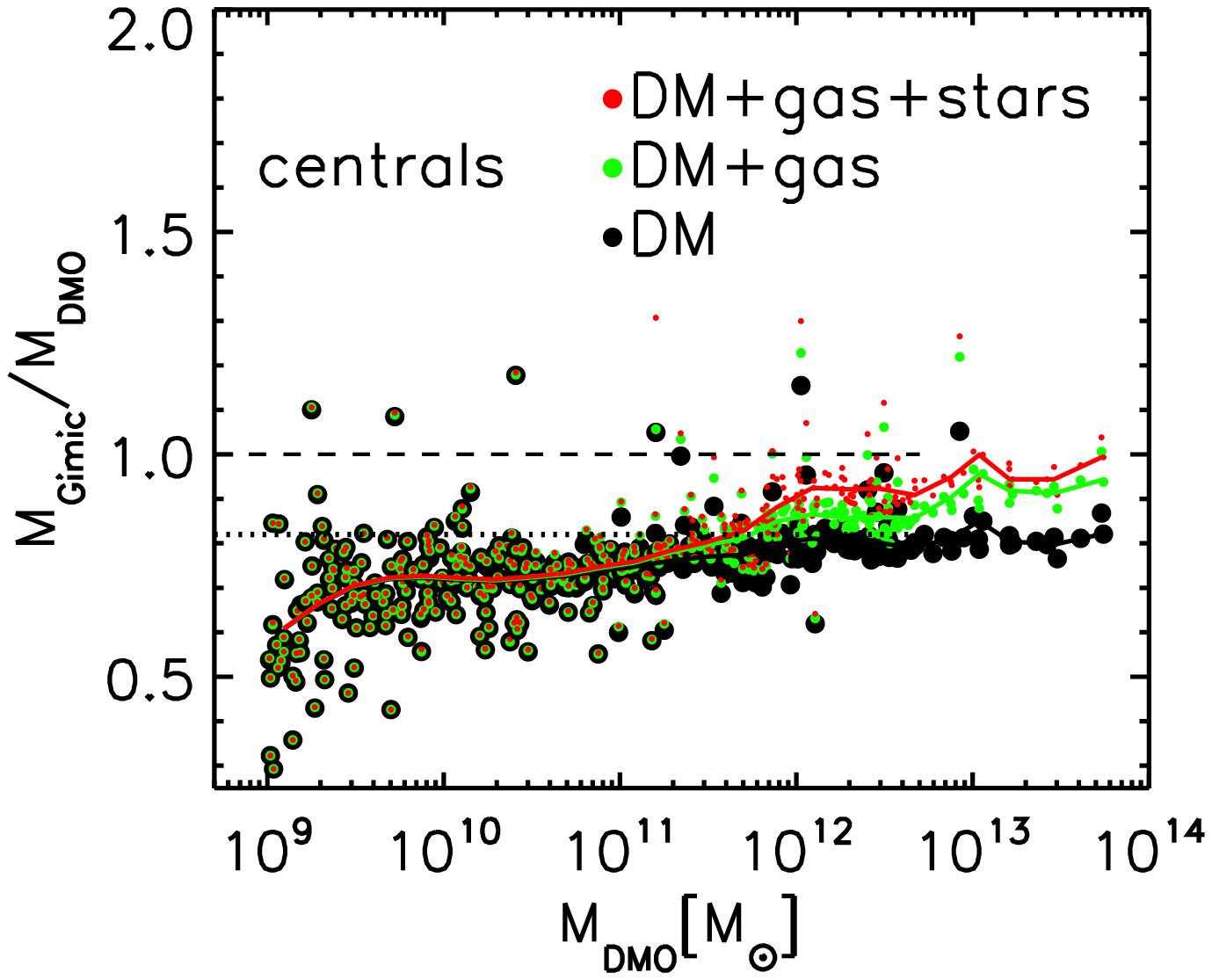}
     \hspace{-.2in} \includegraphics*[trim = 15mm  3mm 3mm 6mm, clip, width = .5\textwidth]{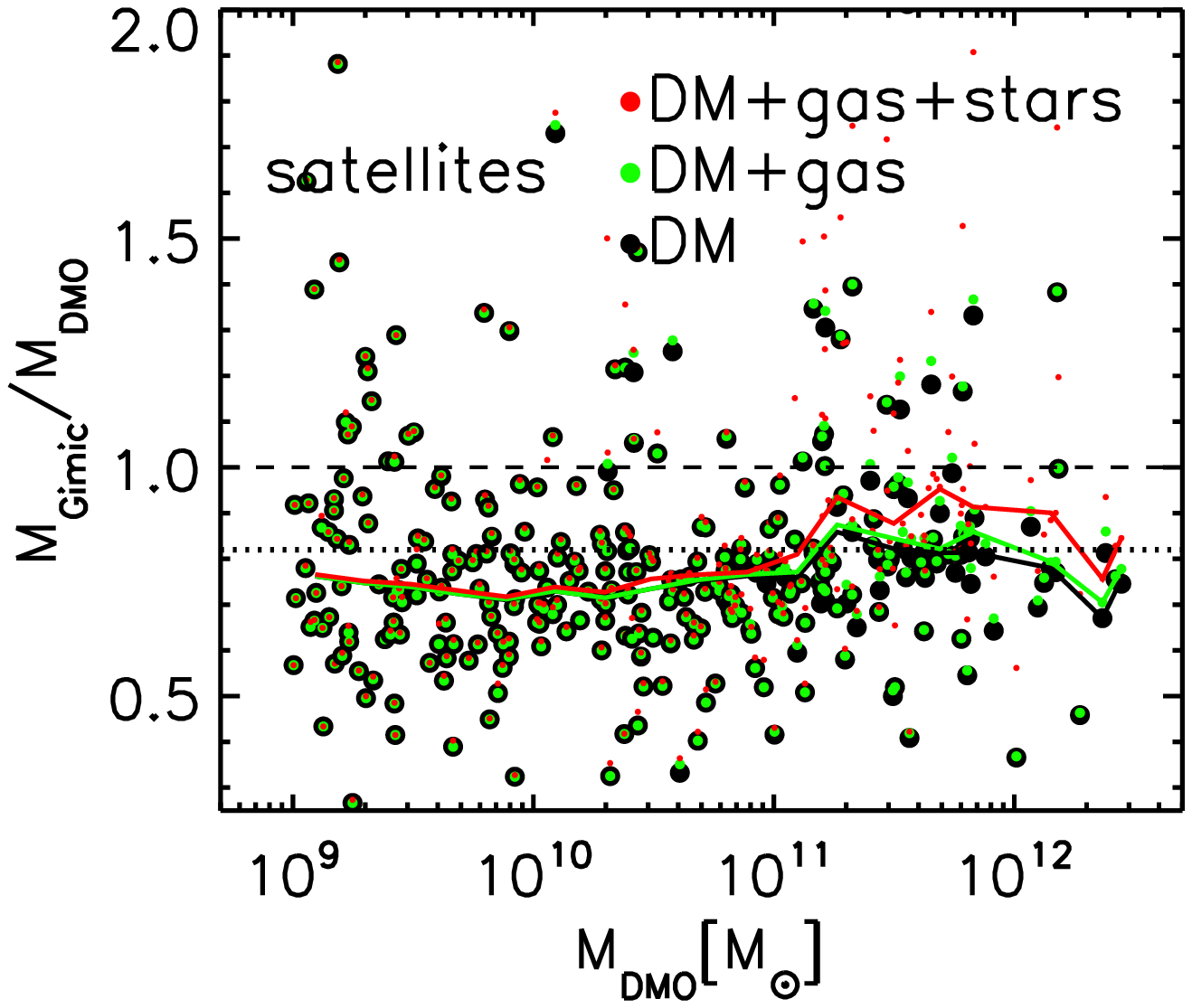}
  \end{center}
  \caption{Ratio of subhalo mass in GIMIC relative to the DMO
    simulation for individual, matched subhaloes at $z=0$, as a
    function of DMO mass, for central subhaloes (left) and satellite
    subhaloes (right). Every subhalo is represented by a set of three
    circles, with the y-values indicating its mass components in the
    GIMIC simulation: only the dark matter (black), dark matter and
    gas (green), and total mass including dark matter, gas and stars
    (red). The solid lines of corresponding colour indicate the median
    value within each bin, from which individual subhaloes are drawn
    at random, so as to show the same number of objects per
    logarithmic mass interval. The black dashed and dotted lines
    denote a 1:1 ratio, and a ratio of
    $\Omega_{DM} / \left(\Omega_{DM}+\Omega_b\right)$, respectively. For lower mass
    subhaloes, the circles lie closer together, indicating a decrease
    in baryon fraction with decreasing subhalo mass, reflected also in
    the convergence of the three curves. The decline of the red curve
    indicates a decrease in total subhalo mass in GIMIC relative to
    the DMO simulation. For satellites, there is a similar trend of
    decreasing baryon fraction and mass ratio for lower mass objects,
    although the scatter among individual subhaloes is much greater.}
  \label{fig:submass-scatter-types}
\end{figure*}

We use pairs of individual matched subhaloes in the DMO and the
GIMIC simulation, to study the effect of baryons on the formation of
haloes and subhaloes. In Fig.~\ref{fig:mass-profiles-components}, we
show the mass profiles in GIMIC at $z=0$, for three typical matched
central subhaloes of $10^{10}$, $10^{11}$ and $10^{12}\Ms$. For each
subhalo, the density in dark matter is shown in black, the density in
stars in green, and the density in gas in blue. As the subhalo mass
increases from left to right, the baryon fraction also
increases. Whereas the total density is dominated by dark matter
throughout for $10^{10}$ and $10^{11}\Ms$ subhaloes, $10^{12}\Ms$
subhaloes contain a significant mass of stars in the centre. For
subhalo masses of $10^{10}$ and $10^{11}\Ms$, the total mass density
in GIMIC is systematically lower than the mass density of the
corresponding object in the DMO simulation, which is plotted as a
black dashed line. Note that for the $10^{12}\Ms$ subhaloes, the
central mass profile is steeper in GIMIC compared to the DMO
simulation as a result of cooling and adiabatic contraction; this is
not seen in the smaller haloes.

Fig.~\ref{fig:submass-scatter-types} compares the masses of many
individual subhaloes across the two simulations, as a function of
their mass in the DMO simulation, subdivided into centrals (left
panel), and satellite (right panel). For the purposes of this
comparison, we only include subhaloes that are of the same type in
both simulations, i.e. satellites matched to satellites, and centrals
matched to centrals, using the criteria discussed in
Section~\ref{sec:linking}. We show a random subset of subhaloes,
normalised so as to have equal numbers per logarithmic mass interval.

Each subhalo is represented by a set of three points that denote the
different composition of the subhalo in the GIMIC simulation: black
circles measure only the dark matter component, green circles also
include the gas, and red circles represent the total mass, consisting
of dark matter, gas and stars.

\begin{figure*}
  \begin{center}
    \hspace{-.2in}\includegraphics*[trim = 0mm 23mm 0mm 9mm, clip,
      width = .5\textwidth]{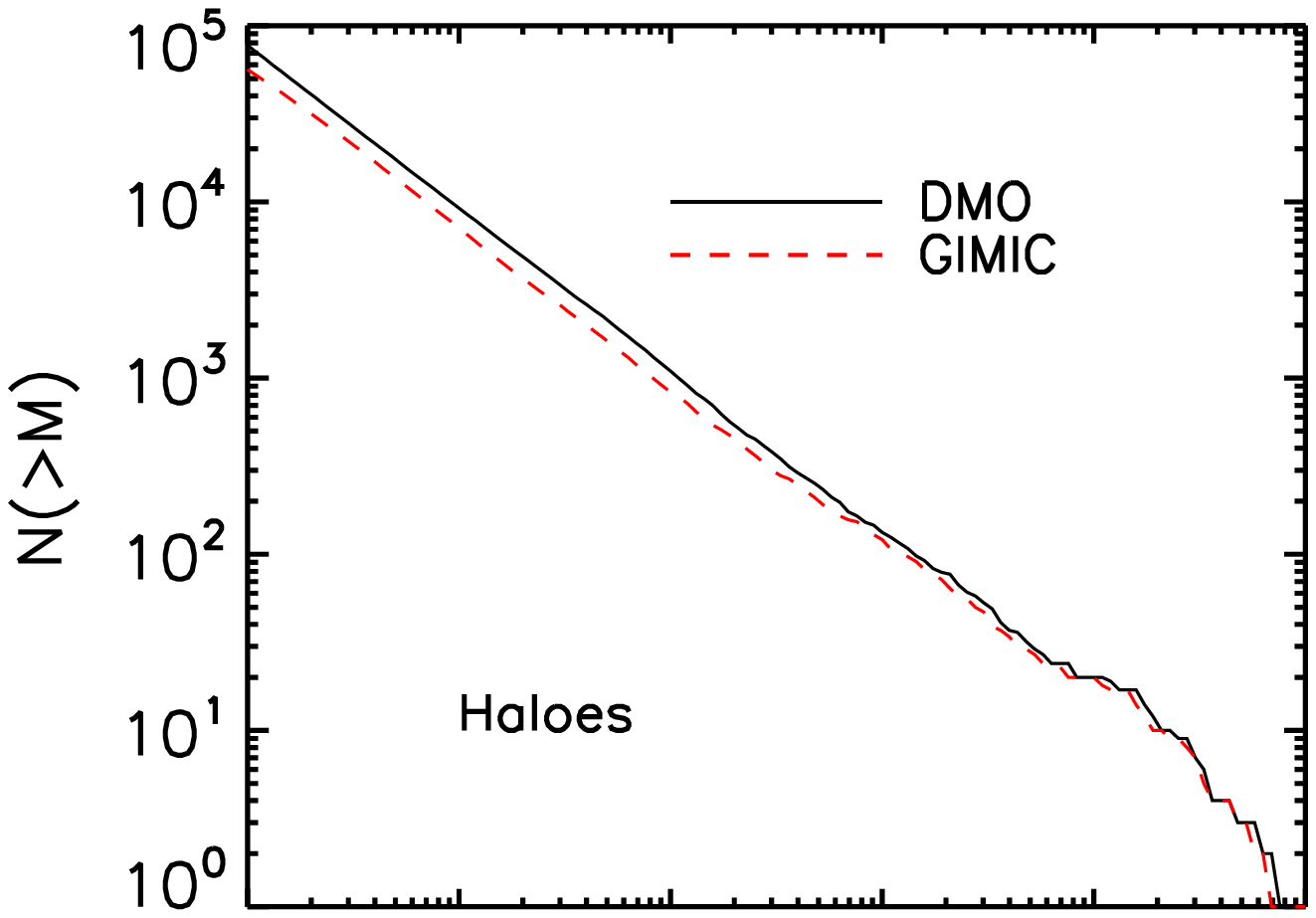}
    \hspace{-.2in}\includegraphics*[trim = 0mm 23mm 0mm 9mm, clip, width = .5\textwidth]{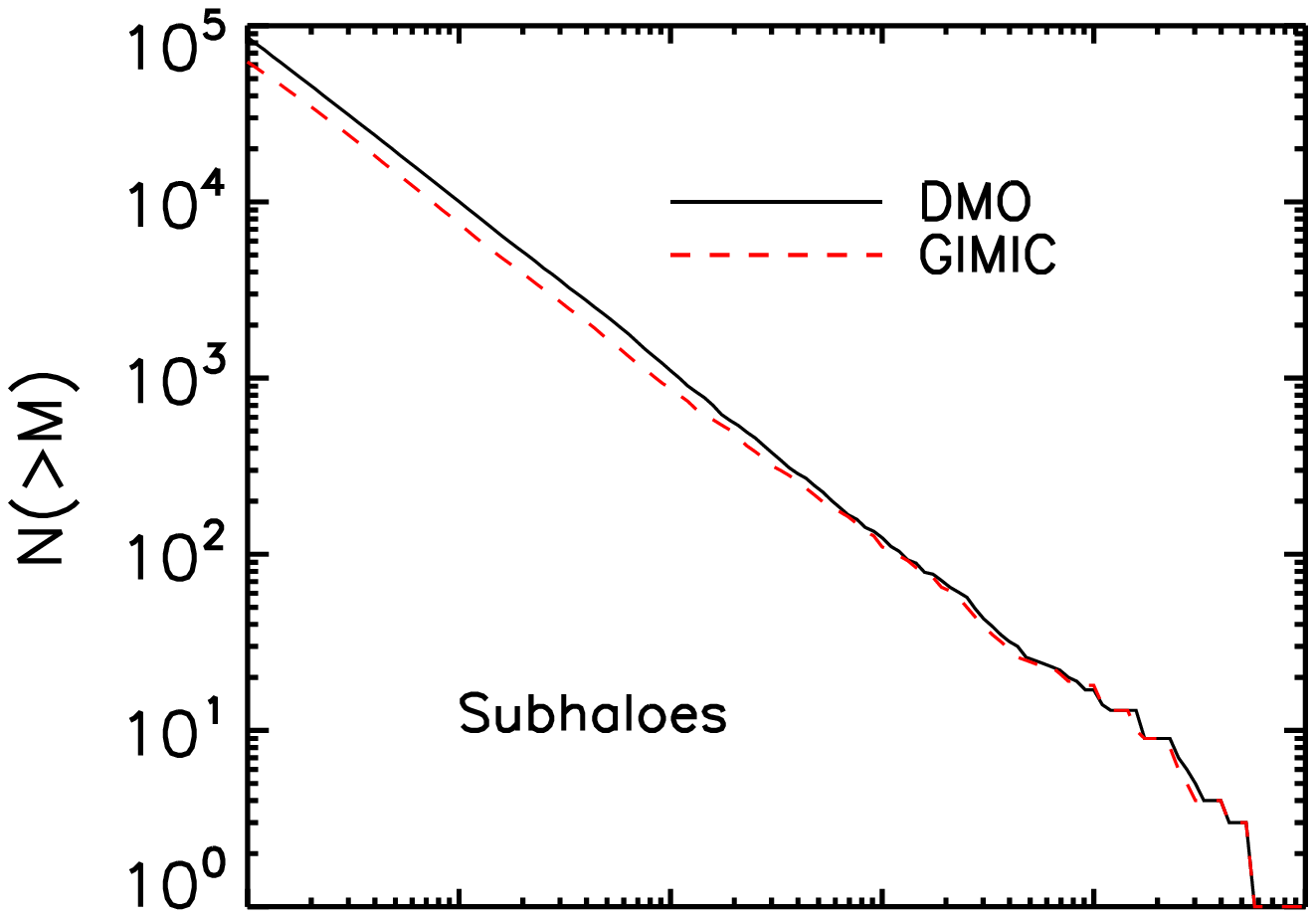} \\
    \hspace{-.2in}\includegraphics*[trim = 0mm 2mm 0mm 12mm, clip, width = .5\textwidth]{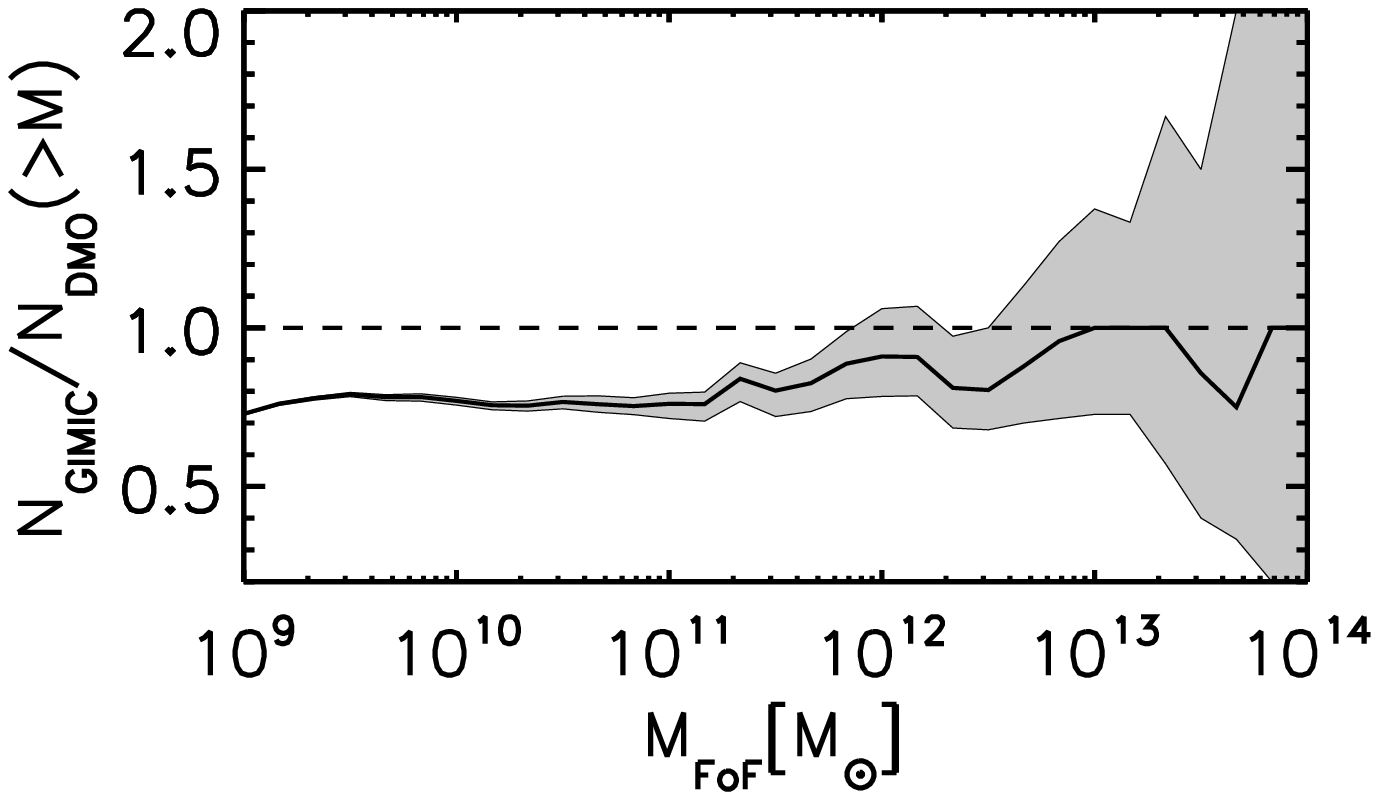}
    \hspace{-.2in}\includegraphics*[trim = 0mm 2mm 0mm 12mm, clip, width = .5\textwidth]{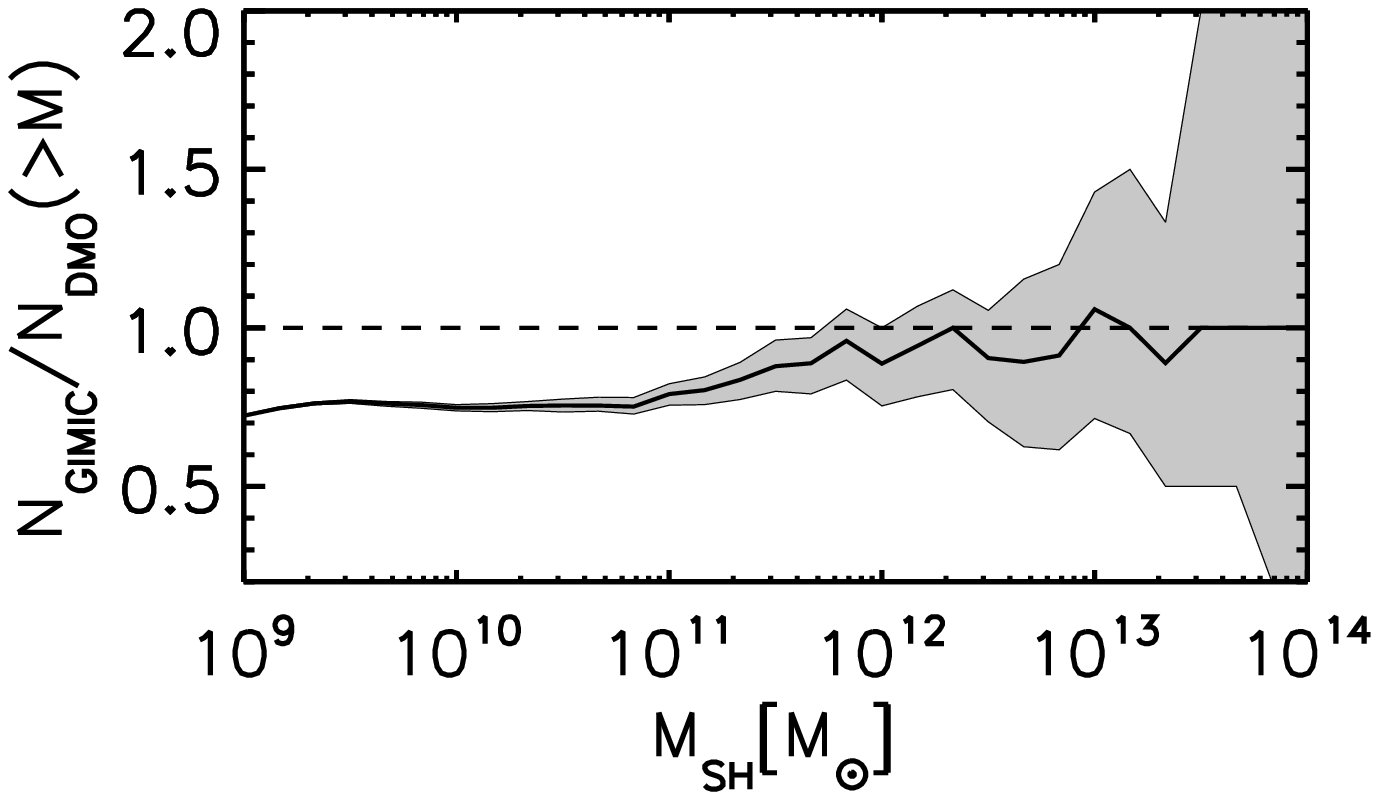}
  \end{center}
  \caption{Comparison of the cumulative halo mass functions (left
    column), and subhalo mass functions (right column) of the GIMIC
    and the DMO simulations. The top row shows the cumulative mass
    functions at redshift $z=0$, with the DMO results shown in black,
    and the GIMIC results overplotted in red. The bottom row shows the
    ratio of the cumulative abundances at $z=0$, with the shaded area
    denoting the statistical uncertainty. The difference in abundance
    depends strongly on mass: Above $\sim10^{12}\Ms$, both the subhalo
    and halo abundances have ratios close to 1, but they drop to
    $\sim0.75$ from $\sim10^{11}\Ms$ and below.}
  \label{fig:HMF}
\end{figure*}

Overplotted on both panels are the median mass ratio of the
corresponding components; black lines for dark matter, green also
including gas, and red including dark matter, gas and stars. As the
baryon fraction decreases with decreasing subhalo mass, circles which
are well separated at high mass become closer, and for lower mass
subhaloes, they mostly overlap. This trend is reflected in the
convergence of the median mass ratios of the three components. At the
high-mass end, on average, subhaloes reach the same total mass in the
GIMIC simulation as in the DMO simulation because their baryon
fraction is close to the universal value. The total mass (red line)
approaches a 1:1 ratio, while the dark matter component (black line)
approaches the value expected from subtracting the universal baryon
fraction. At lower masses, the median total mass in the GIMIC
simulation is dominated by the dark matter component. It is worth
noting that the mass of low-mass subhaloes in the GIMIC simulation is
not only below the mass in the DMO simulation, but even below the
value expected from a simple loss of baryons. This points to the fact
that objects which lose mass during an early stage in their formation
acquire a shallower potential, which also diminishes their subsequent
accretion. Due to the requirement for subhaloes to be self-bound, a
subhalo whose gravitational binding energy is reduced due to outflows
may also lose additional particles that are no longer bound. However,
as we show in~Fig.~\ref{fig:HMF}, we find a very similar reduction in
the masses of FoF-haloes, indicating that any difference in the
attribution of masses by {\sc subfind} is only a secondary effect.

Comparing the central subhaloes in the left panel of
Fig.~\ref{fig:submass-scatter-types} to the satellite subhaloes in the
right panel, it can be seen that the scatter increases substantially
for satellites, and also increases with decreasing mass. As we will
show in Fig.~\ref{fig:baryon-contribution}, the overall gas fraction
is lower for satellites than for centrals as a result of stripping.

\begin{figure}
  \begin{center}
    \includegraphics*[trim = 2mm 30mm 1mm 1mm, clip, width = \columnwidth]{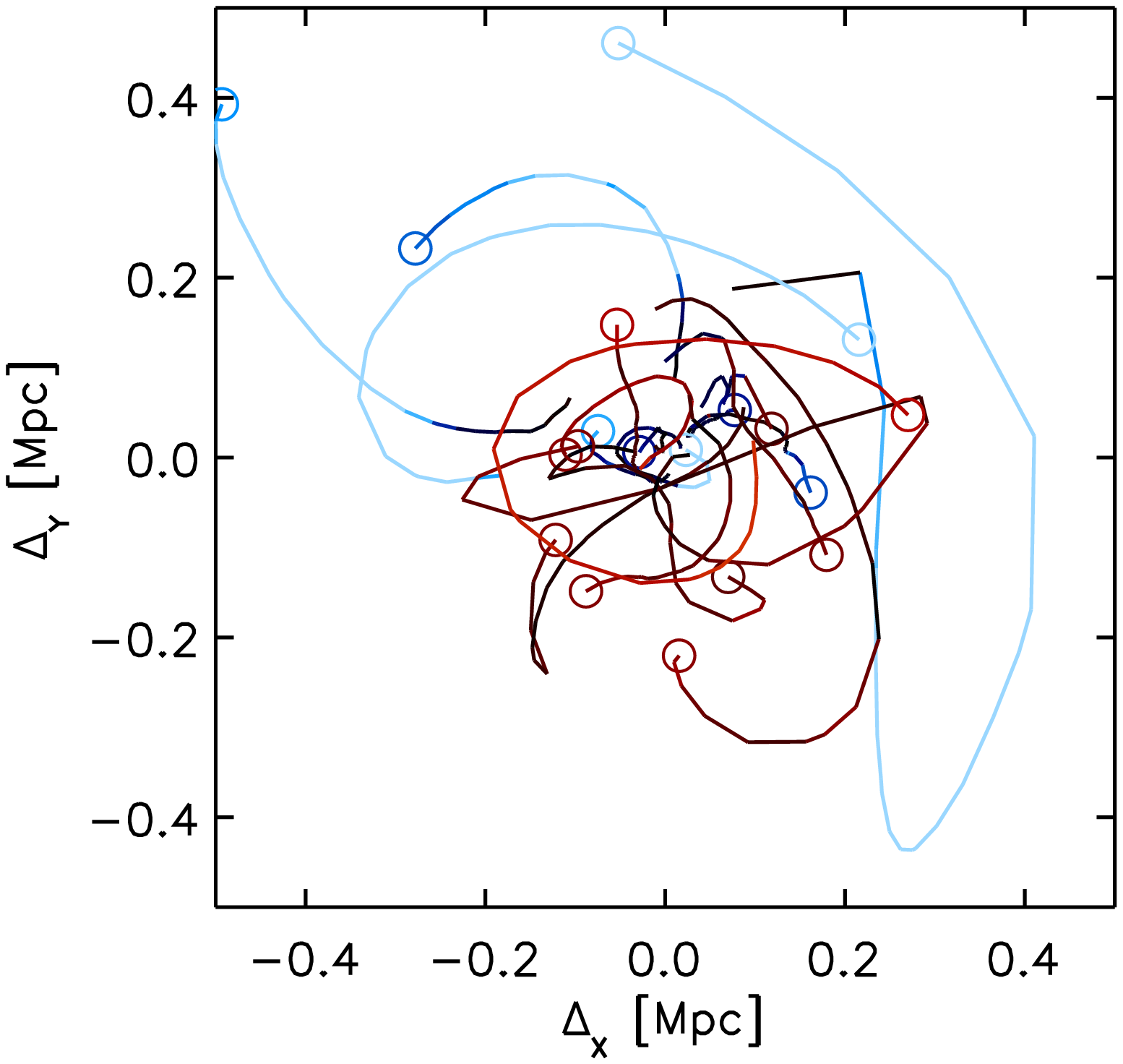} 
    \includegraphics*[trim = 2mm 8mm 1mm 1mm, clip, width = \columnwidth]{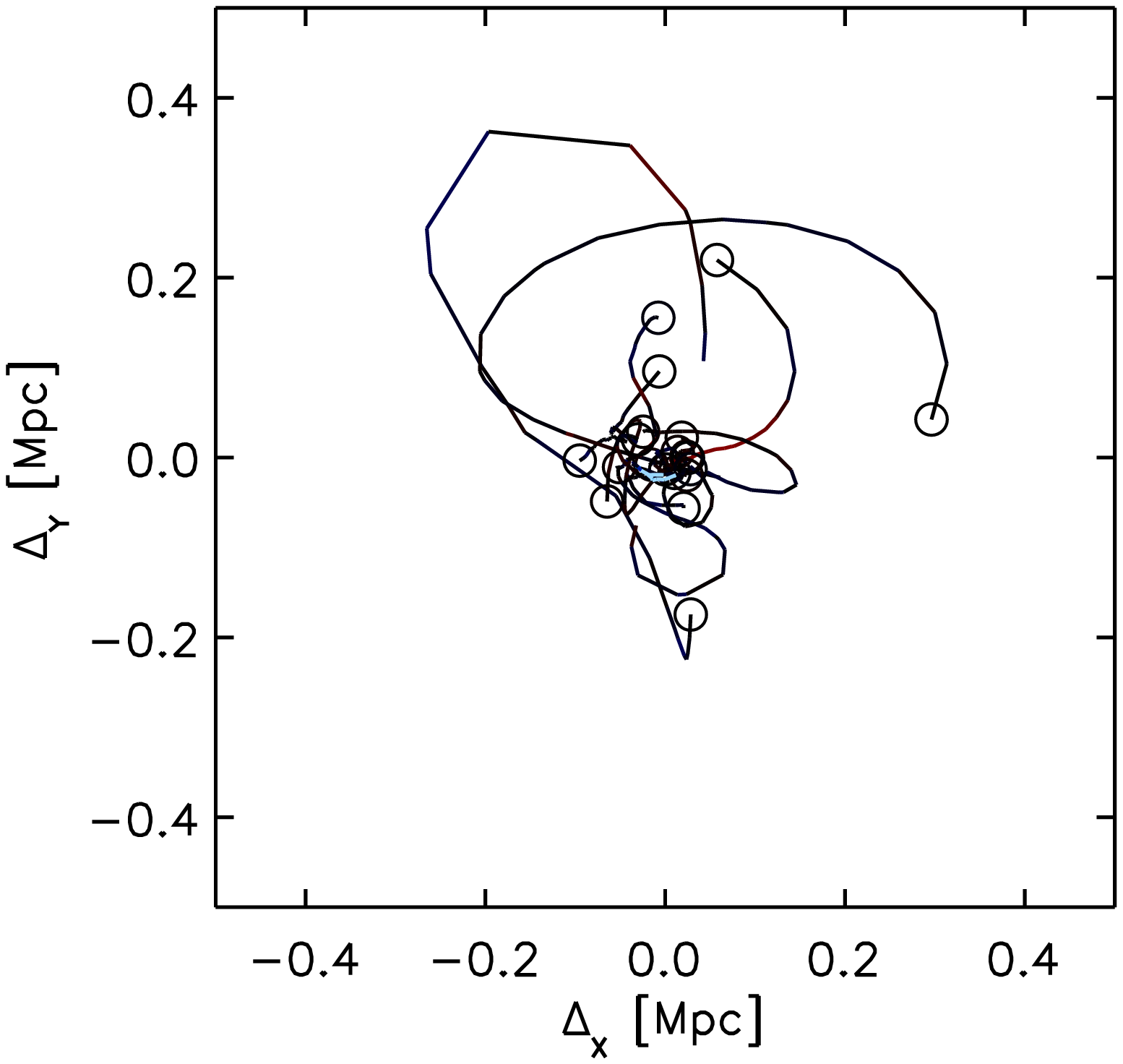} 
   \end{center}
   \caption{Relative displacement between matched pairs of satellite
     subhaloes with final masses of $\sim10^{10}\Ms$, for pairs with
     mass ratios that deviate from the median by more than $2 \sigma$
     (top panel), or by less than 10\% (bottom panel). Circles show
     the present displacement, while traces show the evolution from
     $z=1.5$, coloured according to the mass ratio: black for mass
     ratios close to the median, red (blue) if the mass in GIMIC is
     higher (lower) than the median at the time. While most pairs of
     subhaloes in the bottom panel evolve along similar paths, those
     in the top panel show more divergent histories, linking the
     increased scatter among satellites in
     Fig.~\ref{fig:submass-scatter-types} to their more chaotic
     orbits.}
  \label{fig:traces}
\end{figure}

The main cause for the increased scatter among satellites lies in
their more chaotic orbits, meaning that matched pairs are more likely
to follow divergent evolutionary paths in the two simulations. This
effect is illustrated in Fig. ~\ref{fig:traces}, which shows the
projected displacement between pairs of satellite subhaloes of $\sim
10^{10}\Ms$ in the GIMIC simulation relative to the DMO
simulation. Tracing the evolution since $z=1.5$, it can be seen that
satellite pairs whose final mass ratios deviate from the median by
more than $2 \sigma$ (top panel) are much more likely to evolve along
divergent paths compared to those whose final mass ratios are within
10\% of the median (bottom panel).

\begin{figure}
  \begin{center} 
    \includegraphics*[trim = 10mm 2mm 1mm 10mm, clip, width = \columnwidth]{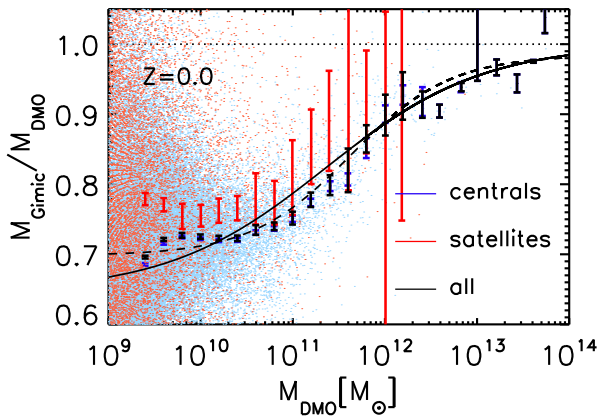} 
   \end{center}
   \caption{Ratio of subhalo masses between the GIMIC and DMO
       simulations for matched subhaloes at $z=0$ compared to analytic
       fits. The blue and red points show the results for centrals and
       satellites, respectively, while the error bars show the
       estimate of the median ratio and its statistical error, for
       centrals (blue), satellites (red) and all subhaloes
       (black). The solid and dashed lines show the best fit to the
       median for all subhaloes, assuming 4 free parameters (dashed
       line, Eq.~\ref{eqn:first-fit}), or 3 free parameters, with the
       upper limit fixed at $b=1$. (solid line, Eq.~\ref{eqn:fit}) }
  \label{fig:fit-comparison}
\end{figure}

Qualitatively, the overall reduction in subhalo mass due to baryonic
effects at $z=0$ is similar for centrals and satellites. However,
despite the considerable scatter, at fixed subhalo mass, the average
mass loss in the GIMIC simulation appears to be somewhat larger for
central subhaloes than for satellites, as is also illustrated in
Fig.~\ref{fig:fit-comparison} and Fig.~\ref{fig:correction-types}
. The offset between the relations seen for satellites and for
centrals may be partly attributed to the fact that while centrals only
experience mass loss due to outflows in the GIMIC simulation, the
masses of satellites are further reduced by stripping in both the DMO
and the GIMIC simulations.

It is worth noting that, despite the fact that every halo contains at
most one central subhalo but potentially many satellites, over $80\%$
of all subhaloes in the simulation volume are central, as was shown in
Fig.~\ref{fig:matching-fraction}. This is easily understood when
recognising that within each halo, by definition, the central subhalo
is more massive than the satellites, so at fixed subhalo mass,
satellites belong to larger haloes than centrals. As a consequence of
the steep halo mass function (see Fig.~\ref{fig:HMF}), the total
subhalo population closely resembles that of centrals, with some
additional scatter introduced by satellites.

\subsection{Analytic Correction}\label{sec:analytic-preview}
Noting from Fig.~\ref{fig:submass-scatter-types} that the change in
mass ratio between the GIMIC and the DMO simulations saturate both at
the high-mass and at the low-mass limit, we parametrise the mean ratio
of the subhalo mass in the GIMIC simulation, $\rm{M_{GIMIC}}$, to that
in the DMO simulation, $\rm{M_{DMO}}$, in the following way:

\begin{equation} \label{eqn:first-fit}
\frac{\rm{M_{GIMIC}}}{\rm{M_{DMO}}} = b \cdot \frac{a/b + \left(\rm{M_{DMO}/M_t}\right)^w}{1 + \left(\rm{M_{DMO}/M_t}\right)^w}
\end{equation} 
With $w > 0$, Eq.~\ref{eqn:first-fit} has two horizontal asymptotes,
$$ \lim_{\rm{M_{DMO}} << \rm{M_t}}\left( \frac{\rm{M_{GIMIC}}}{\rm{M_{DMO}}} \right) = a $$
and 
$$ \lim_{\rm{M_{DMO}} >> \rm{M_t}} \left( \frac{\rm{M_{GIMIC}}}{\rm{M_{DMO}}} \right) = b.$$
The mass ratio takes an intermediate value of $(a+b)/2$ at
$\rm{M_{DMO}} = M_t$, i.e. $\rm{M_t}$ sets the scale for the
transition between the two limits, while the width of the transition
is parametrised by the value of $w$. Assigning equal weight to the
median ratio in each logarithmic mass interval, the best fit to all
subhaloes at $z=0$ yields values of $a=0.69$, $b = 0.98$,
$\rm{M_t}=10^{11.6}\Ms$ and $w=0.79$. It is shown by the dashed line
in Fig.~\ref{fig:fit-comparison}.

The asymptotic behaviour in the low-mass limit signifies an almost
complete loss of baryons at very early times. The fact that the value
of $b$ at $z=0$ is very close to unity indicates that the physical
processes in GIMIC lead to no significant change in mass at the high
mass end. By setting $b=1$, Eq.~\ref{eqn:first-fit} simplifies to:

\begin{equation} \label{eqn:fit}
\frac{\rm{M_{GIMIC}}}{\rm{M_{DMO}}} = \frac{a + \left(\rm{M_{DMO}/M_t}\right)^w}{1 + \left(\rm{M_{DMO}/M_t}\right)^w}
\end{equation} 
with horizontal asymptotes at $a$ and $1$, and an intermediate value
of $(a+1)/2$. A fit to this equation yields values of $a=0.65$,
$\rm{M_t}=10^{11.4}\Ms$ and $w=0.51$. The solid line in
Fig.~\ref{fig:fit-comparison} shows the fit to Eq.~\ref{eqn:fit}, and
can be compared to the dashed line obtained from
Eq.~\ref{eqn:first-fit}.

Because the additional free parameter of Eq.~\ref{eqn:first-fit} does
not change the results significantly, we adopt the simpler
Eq.~\ref{eqn:fit} for the remainder of this work. It is possible that
other physical processes, particularly the effect of AGN, could lead
to a significant effect also at the high mass end, in which case the
more general form of Eq.~\ref{eqn:first-fit} would be more
appropriate. Separate fits for satellites and centrals, as well as for
different redshifts, are discussed in Appendix~\ref{sec:correction}.

\subsection{Global Mass Functions}\label{sec:statistics-mf}
Mass- or multiplicity functions of haloes or subhaloes constitute one
of the most fundamental measures of structure formation. Because dark
matter haloes (or more precisely: subhaloes), are believed to be the
locations of galaxy formation, predicting their abundance also allows
us to relate observations of galaxies to theories about the underlying
mass distribution. We return to this point in
Section~\ref{sec:applications}.

In Fig.~\ref{fig:HMF}, we show the cumulative mass functions of haloes
(left panel) and subhaloes (right panel), for the DMO simulation
(shown in black) and the GIMIC simulation (red). In both cases, the
mass includes the total mass; i.e. the mass in dark matter in the DMO
simulation, and the combined mass of dark matter, gas, and stars in
the GIMIC simulation. The cumulative mass functions are nearly
identical for the most massive objects, but below $\sim 10^{12} \Ms$,
the abundance in the GIMIC simulation is significantly reduced
relative to the DMO simulation, with the difference increasing at
lower masses.

It should be noted that this statistical result is not reliant on the
matching of haloes or subhaloes and, as we show in
Appendix~\ref{sec:resolution}, it is not affected by the resolution of
our simulations. Above the resolution threshold, the decrease in
numbers is only partly due to the erasure of structures, but mostly
due to each object having a lower mass in the GIMIC simulation. The
shape of the mass functions then results in this ``horizontal'' shift
in mass also manifesting itself as a ``vertical'' shift in abundance.

\subsection{Velocity Functions}\label{sec:statistics-vf}
\begin{figure}
  \begin{center}
      \hspace{-.1in} \includegraphics*[trim = 5mm 3mm 5mm 8mm, clip,  width = \columnwidth]{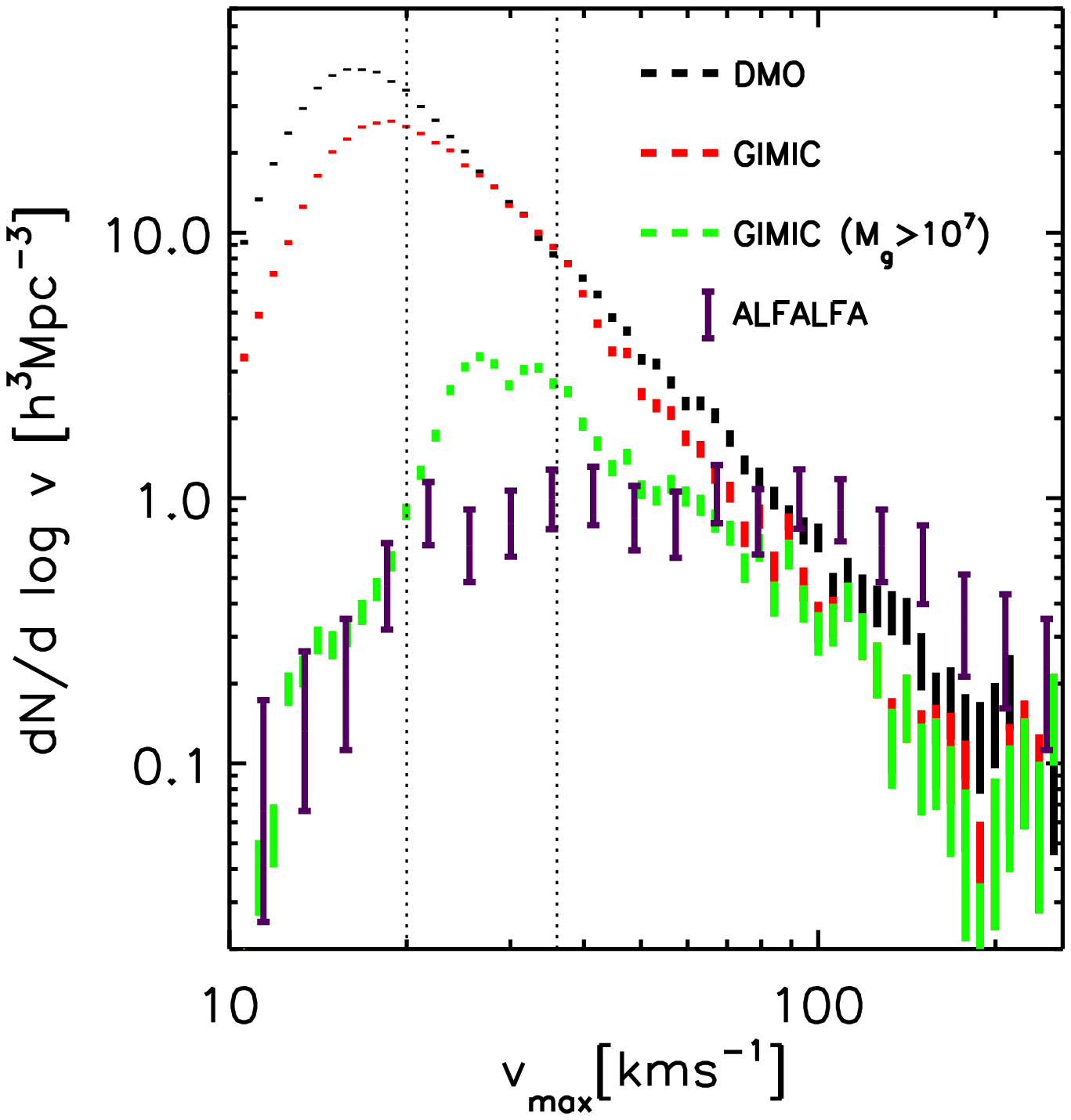}
  \end{center}
  \caption{Differential velocity function of subhaloes in the DMO
    simulation and in the GIMIC simulation, compared to the HI
    velocity function measured in the ALFALFA survey for a detection
    limit of $\rm{M_{HI}}=10^7 h^{-1}\Ms$. The results of the DMO
    simulation are shown in black; for GIMIC, red shows the velocity
    function computed using all subhaloes, and green only includes
    those subhaloes with $\rm{M_g}>10^7h^{-1}\Ms$. The simulated mass
    functions are corrected for the overdensity of the survey volume,
    their width denote the Poisson errors within each bin. The
    vertical lines at 20~kms$^{-1}$ and 35~kms$^{-1}$ indicate the
    resolution limit of our simulations and the completeness limit of
    the ALFALFA survey, respectively. While the DMO simulation
    significantly over-predicts the abundance of subhaloes with
    $v_{max}=35$kms$^{-1}$, accounting for baryonic effects and
    adopting an appropriate detection threshold for the ALFALFA survey
    significantly reduces the reported discrepancies between CDM and
    the measured velocity function, down to the survey limit of
    35~kms$^{-1}$.}
    \label{fig:ALFALFA-comparison}
\end{figure}

The circular velocity, $v_{\rm{c}}(r)$, at a given radius, $r$, is
related to the enclosed mass $m(<r)$ via $v_{\rm{c}}(r)=\sqrt{G
  m(<r)/r}$. Thus, measurements of the velocity profiles of HI gas in
galaxies can provide a measurement of the underlying mass
distribution. In Fig.~\ref{fig:ALFALFA-comparison} we compare the
differential velocity function of subhaloes in the DMO simulation to
results from the GIMIC simulation, and to measurements from the
Arecibo Legacy Fast ALFA (ALFALFA) survey \citep{Giovanelli-2005,
  Giovanelli-2007}. Following constrained simulations of the Local
Volume by \cite{Zavala-2009}, we have normalised the simulated
velocity functions to account for the overdensity in the survey
volume, which is dominated by the Local Supercluster, and whose
velocity function exceeds the universal value. A detailed description
of the normalisation can be found in \cite{Zavala-2009}, but we note
that it has little effect on scales where the DMO results
significantly exceed the observations.

When compared to the ALFALFA observations, it can be seen that the
velocity function of the DMO simulation (black) over-predicts the
abundance of subhaloes of $v_{\rm{max}} \sim 35$kms$^{-1}$ by a factor
of $\sim10$. This discrepancy, which has been pointed out by
\cite{Zavala-2009}, \cite{Trujillo-Gomez-2010} and
\cite{Papastergis-2011}, persists when the observations are compared
to the velocity function computed using all subhaloes in GIMIC
(red). It has been considered a challenge to the cold dark matter
(CDM) paradigm, and evidence in favour of warm dark matter
(WDM). However, allowing for the detection threshold of the survey for
the sample of nearby galaxies analysed by \cite{Zavala-2009},
$\rm{M_{HI}}>10^7h^{-1}\Ms$, and recomputing the GIMIC velocity
function only for subhaloes with $\rm{M_g}>10^7h^{-1}\Ms$, we find
that the resulting velocity function (green symbols) agrees to
within a factor of $\sim3$ with the ALFALFA observations, while also
giving a better fit to the shape of the $v_{\rm{max}}$
function. Considering that the total gas mass, $\rm{M_g}$, is only an
upper limit for $\rm{M_{HI}}$, and that we have not included other
constraints such as the inclination limit of $30°$ for galaxies in the
ALFALFA survey, the remaining difference may be yet be explained
within the CDM framework.

\subsection{Satellite Mass Functions}\label{sec:satellites}
Local Group dwarf galaxies and Milky Way satellites constitute
important test cases for CDM theory: they are among the smallest and
most dark-matter dominated galaxies, and their proximity enables
detailed studies of their abundance and their internal structure.
While the apparent discrepancy between the predicted satellite mass
function and the observed satellite {\it luminosity} function
{\citep[e.g.][]{Klypin-1999, Moore-1999}} may be explained through
inefficient star formation due to astrophysical mechanisms including
photo-ionisation and feedback \citep[e.g.][]{Bullock-2000,
  Benson-2002, Somerville-2002}, the reported discrepancies between
the predicted and measured satellite {\it mass} functions are more
challenging: $\Lambda$CDM simulations such as the Aquarius runs
\citep{Springel-Aquarius} produce a higher number of satellites with
circular velocities above $50$~kms$^{-1}$ than is allowed by
observations \citep{Boylan-Kolchin-2011, Boylan-Kolchin-2012,
  Parry-2012}. It has also been pointed out that the Milky Way appears
special in having two satellites as bright as the Magellanic Clouds
\citep{Busha-2011, Tollerud-2011, Guo-2012}.

\begin{figure}
  \begin{center}
 \hspace{-.2in}\includegraphics*[trim = 0mm 0mm 0mm 9mm, clip, width =  .5\textwidth]{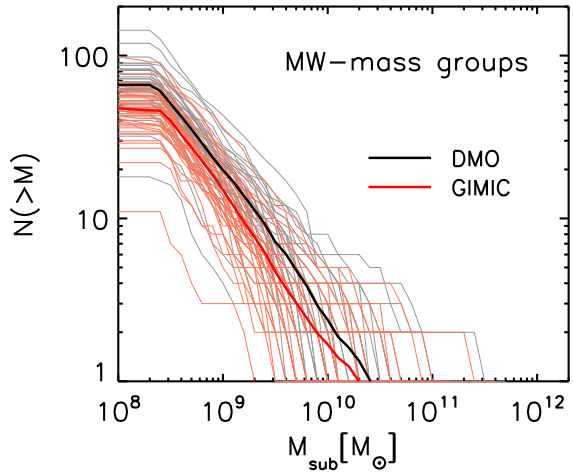} \\
 \hspace{-.2in}\includegraphics*[trim = 0mm 0mm 0mm 9mm, clip, width =  .5\textwidth]{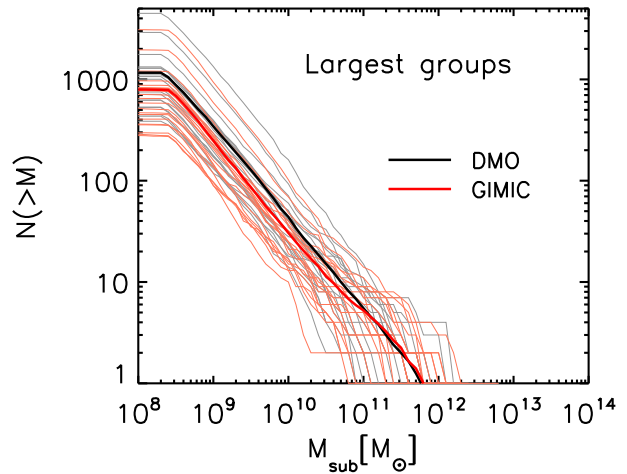} 
   \end{center}
   \caption{Cumulative satellite mass functions for $\sim 50$
     individual, Milky-Way sized ($10^{12} \Ms < M_{200} <
     2.5\times10^{12}\Ms$) haloes (top), and for the 20 most massive
     clusters (bottom), in the DMO simulation (black) and the GIMIC
     simulation (red), at $z=0$. Thin lines show the mass functions of
     individual groups, thick lines show the average abundance for the
     groups in either simulation. For clusters and for Milky Way groups,
     the number of members is reduced in the GIMIC simulation, as
     expected from the overall reduction in subhalo abundance. It is
     also worth noting that there is more than an order of magnitude
     variation in the number of satellites amongst individual groups.}
  \label{fig:satellite-mf-mw}
\end{figure}

In Fig.~\ref{fig:satellite-mf-mw}, we compare the satellite mass
functions of groups of different mass in the DMO simulation (black
lines) and the GIMIC simulation (red lines). In the top panel, in both
simulations, we have selected groups with total mass in the range of
$1-2.5\times10^{12}\Ms$, compatible with current mass estimates of the
Milky Way \citep[e.g.][]{Li-2008, Xue-2008, Gnedin-2010}. Each thin
line shows the satellite mass function of an individual group, while
the two thick lines represent the average cumulative subhalo abundance
over all groups in the mass range in both simulations.

We find that, on average, the MW-mass groups contain fewer satellites
in GIMIC than in the DMO simulation. The reduction in the number of
satellites in haloes of a fixed mass can be understood because
the mass-dependent baryonic effects are stronger in the lower mass
satellites than in their more massive hosts.

In line with previous results by \cite{Di-Cintio-2011}, the average
baryon effects we see in the GIMIC simulation alone are not enough to
sufficiently reduce the number of {\it massive} satellites in order to
explain the results of \cite{Boylan-Kolchin-2011}.  However, in both
the GIMIC and the DMO simulation, we also find a scatter of more than
one order of magnitude in the satellite abundances of individual
groups. In particular, among the sample of $\sim50$~Milky-Way mass
groups, we find several examples with satellite numbers almost an
order of magnitude below the median. Confirming the results of
\cite{Wang-2012}, we therefore find a considerable chance for a low
number of massive satellites in a Milky-Way mass halo, to which baryon
effects contribute further.

In the bottom panel of Fig.~\ref{fig:satellite-mf-mw}, we repeat the
same analysis for the satellite mass functions of the 20 most massive
groups in both simulations; clusters with masses in the range $\sim
10^{13}-10^{14}\Ms$. As expected from the global mass functions shown
in Fig.~\ref{fig:HMF}, at the high-mass end, the average satellite
mass functions of clusters show no systematic difference between the
two simulations, a result noted previously by
\cite{Dolag-2009}. However, at lower masses, clusters also have fewer
satellites in GIMIC than in the DMO simulation. Compared to the
satellite mass function of MW-sized groups shown in the top panel of
Fig.~\ref{fig:satellite-mf-mw}, the satellite mass functions of
individual clusters show less scatter, even though the mass range of
clusters shown is larger than that of MW-sized groups.

\subsection{Dark Subhaloes}\label{sec:dark}
\begin{figure*}
  \begin{center}
      \includegraphics*[trim = 10mm 3mm 9mm 12mm, clip, width = .48\textwidth]{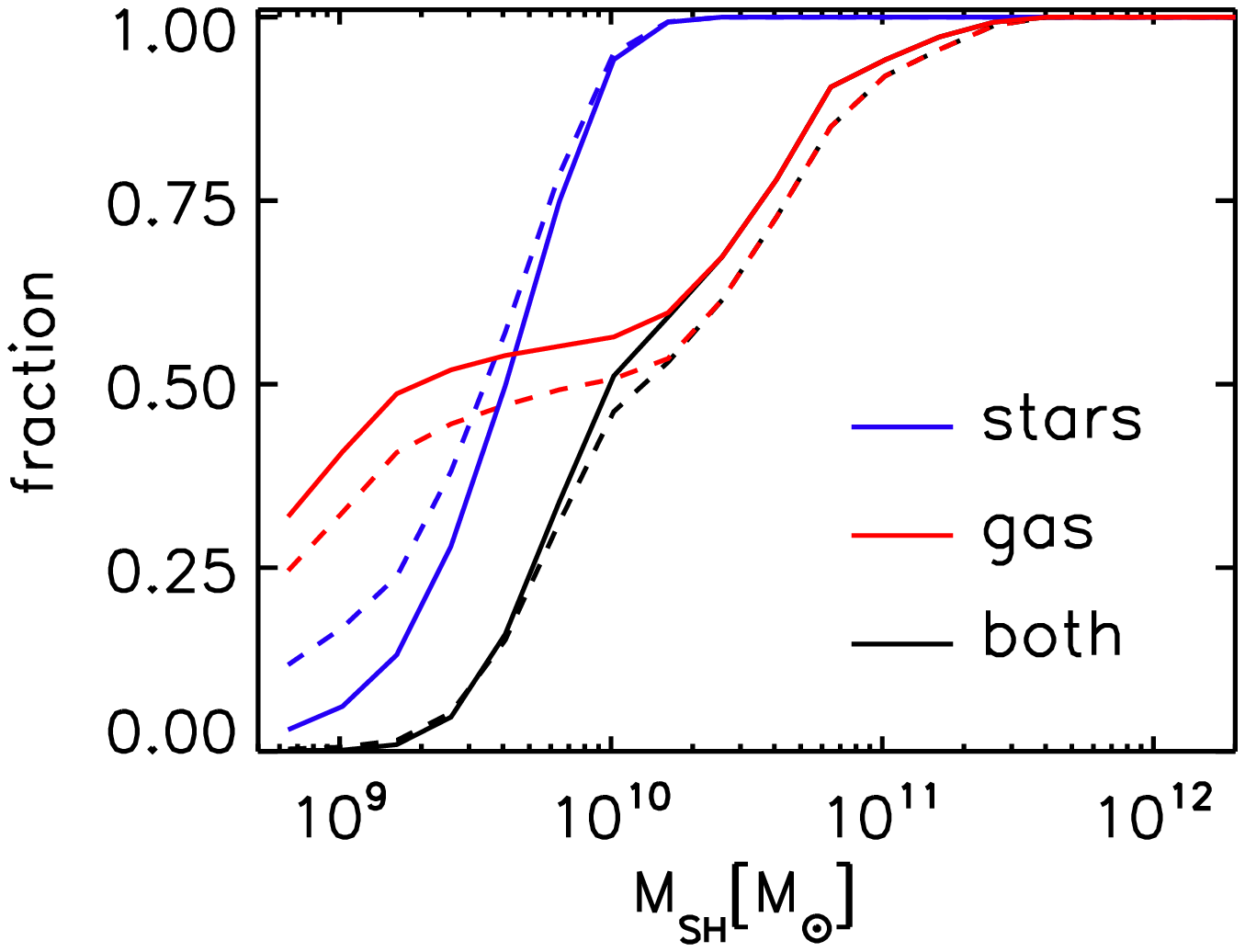}
      \includegraphics*[trim = 10mm 3mm 9mm 12mm, clip, width = .48\textwidth]{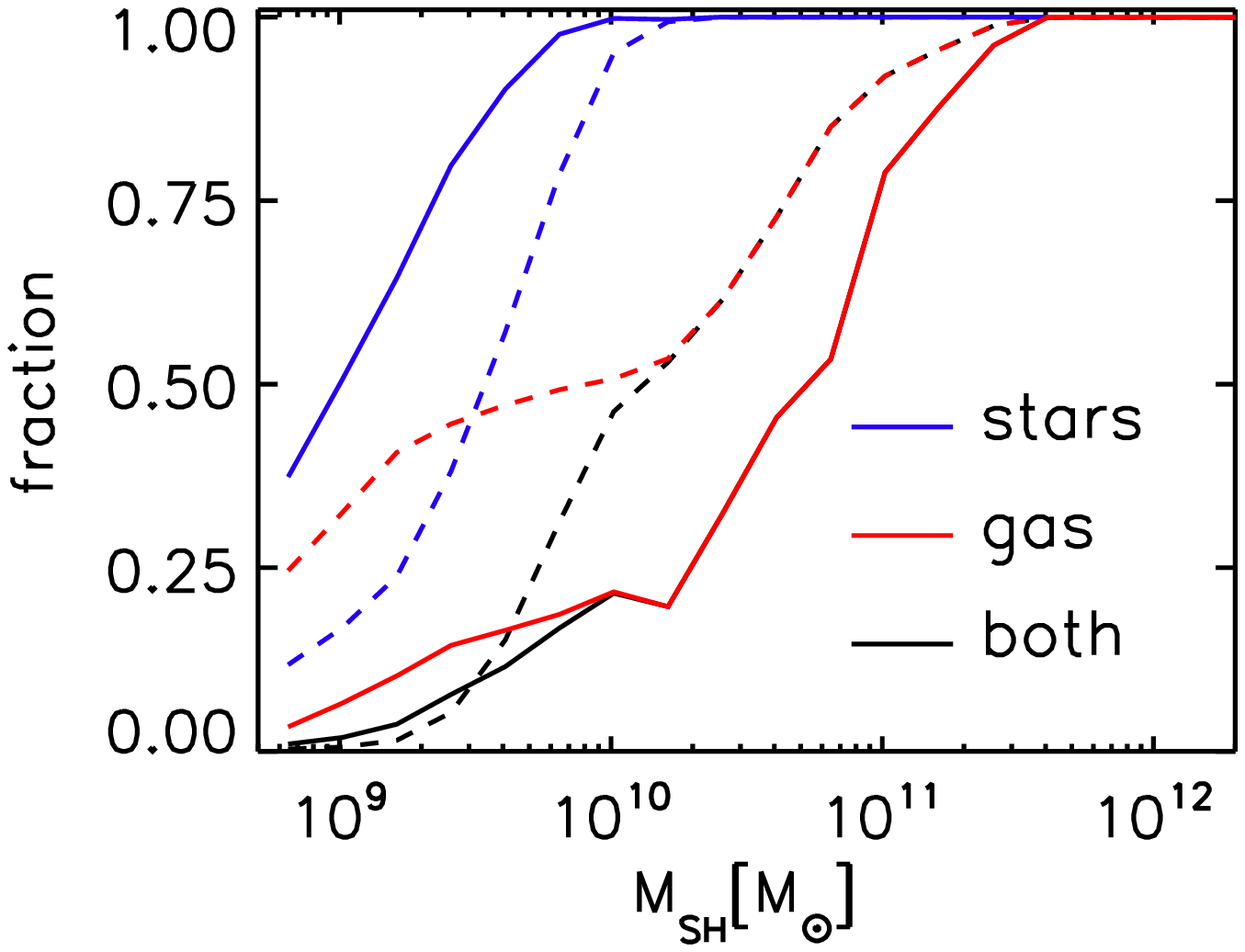} 
  \end{center}
  \caption{Fraction of subhaloes with stars (blue), with gas (red), or
    both (black), at $z=0$. Solid lines show the results for centrals
    (left panel) and satellites (right panel); dashed lines show the
    fraction of all subhaloes for reference and are identical in the
    two panels. For both centrals and satellites, the fraction of
    subhaloes with stars decreases for subhalo masses below $~\sim
    2\times 10^{10}\Ms$, while the fraction of subhaloes with gas
    starts to decrease below $~\sim 3\times 10^{11}\Ms$. The main
    environmental dependence is in the gas content; while the fraction
    of central galaxies with gas remains above $\sim50\%$ down to
    $10^9\Ms$, it drops below $\sim25\%$ among satellites with a total
    mass of $\sim 2\times 10^{10} \Ms$. The decrease in the fraction
    of subhaloes without stars is less steep for satellites than for
    centrals; this is because stars form in more bound objects, which
    are more likely to survive as satellites.}
  \label{fig:baryon-contribution}
\end{figure*}

At $z=0$, the GIMIC simulation contains a substantial number of
low-mass subhaloes which are completely free of stars or gas, i.e.
that they do not contain a single (bound) star or gas particle of mass
$1.98\times 10^6\Ms$. {\it Dark} subhaloes are predicted to exist in
$\Lambda$CDM, as they may explain the ratio of the predicted number of
substructures to the observed number of dwarf galaxies
\cite{Moore-1999}, and because astrophysical processes including
reionisation prevent cooling and star formation in the smallest
systems. Likewise, very low-mass galaxies like dwarf spheroidals
contain stars but no detectable gas, which is understood to have been
lost either through supernova feedback, reionisation (if stars formed
earlier), stripping, or a combination of these effects.

\begin{figure}
\begin{center}
  \includegraphics*[trim = 10mm  0mm 5mm 8mm, clip, width = \columnwidth]{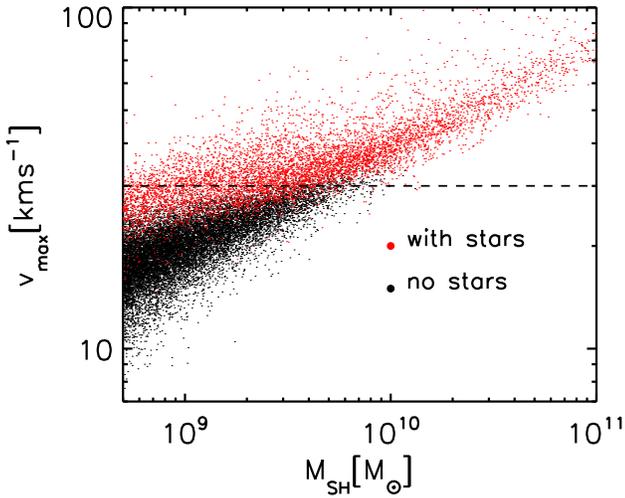}
\end{center}
  \caption{Distribution of maximum circular velocities, $v_{max}$, of
    a random sample of low-mass subhaloes in the GIMIC simulation, as
    a function of subhalo mass. Subhaloes that contain stars are shown
    in red, those without stars are shown in black At masses below
    $\sim10^{10}$, it is apparent that a $v_{max}$ of $\sim
    30$~kms$^{-1}$ splits the two populations: subhaloes with $v_{max}
    > 30$~kms$^{-1}$ have typically been able to form stars, while
    most subhaloes with lower $v_{max}$ have not.}
  \label{fig:m-vmax}
\end{figure}

In Fig.~\ref{fig:baryon-contribution} we show the fraction of low-mass
subhaloes in the GIMIC simulation containing some amount of stars
(blue lines), gas (red lines), or both (black lines), as a function of
total subhalo mass. For both centrals (left panel) and satellites
(right panel), we find that the fraction of {\it dark} subhaloes
increases with decreasing mass below $\sim 10^{10}\Ms$
(i.e. $\sim10^3$ particles). Intriguingly, the fraction of subhaloes
with stars decreases less steeply with mass among satellites compared
to centrals; while $\sim 50 \%$ of satellites with a total mass of
$10^9\Ms$ contain stars, the fraction is below $10\%$ for centrals of
the same mass at $z=0$. This difference can be understood from the
effective threshold in binding energy required for star formation,
apparent from Fig.~\ref{fig:m-vmax}: subhaloes that are more
concentrated, and therefore have a higher maximum circular velocity
$v_{max}$, are both more likely to cool gas efficiently to allow star
formation, and also less likely to be destroyed by stripping. As can
be seen in Fig.~\ref{fig:m-vmax}, a $v_{max}$ threshold of $\sim
30$~kms$^{-1}$, rather than a threshold in total mass, separates
subhaloes that contain stars from those that do not. In preferentially
destroying less bound objects, tidal disruption leads to a higher
fraction of subhaloes with stars among the surviving satellites
relative to the fraction among centrals, which suffer weaker tidal
forces. Our result is in agreement with simulations of
\cite{Crain-2007}, who found that photo-heating can prevent the
collapse of baryons in systems with circular velocities below
$35$~kms$^{-1}$ (at $z=0$), and with the high-resolution simulations
of \cite{Okamoto-2008}, who found that haloes with a circular velocity
of $25$~kms$^{-1}$ can lose half of their baryons due to
photo-heating, and are unable to accrete and cool gas efficiently.

The fraction of subhaloes with gas decreases from $\sim 10^{11}\Ms$
downwards, and decreases much faster for satellites than for
centrals. While stars in satellites are concentrated towards the
centre, gas is easily removed from satellites by tidal and ram
pressure stripping \citep[e.g.][]{McCarthy-2008}. At $10^{10}\Ms$,
where most subhaloes contain stars, the fraction with gas is $\sim
55\%$ among centrals, and only $\sim 20\%$ among satellites.

We also find that the fraction of subhaloes without gas, and the
fraction without stars, are correlated, but not proportional. The
black lines in Fig.~\ref{fig:baryon-contribution} show the fraction of
subhaloes that contain both stars {\it and} gas. It can be seen that
central galaxies with total mass below $\sim 3\times 10^9\Ms$
typically only contain either gas {\it or} stars: subhaloes that
undergo star formation lose their gas due to feedback, while those
that never form stars due to a lack of cooling, can typically retain
their gas.

As subhaloes without stars are not directly observable, their
 {presence} has important consequences for abundance matching, that
will be addressed in Section~\ref{sec:applications}. Of course, the finite
particle mass of $m_g=1.98\times 10^6\Ms$ sets a natural lower limit
to the baryonic mass detectable in our simulation. However, the
occurrence of {\it dark} subhaloes at relatively high subhalo masses,
and the dependence on $v_{max}$ and environment, rather than on
subhalo mass, indicate the physical origin of their existence. As we
will discuss in Appendix~\ref{sec:resolution}, while the gas-free
subhaloes are well converged, predictions for the exact fraction of
{\it dark} subhaloes are still affected by resolution in the GIMIC
simulation.

\section{Abundance Matching}\label{sec:applications}
Abundance matching \citep{Yang-2003, Vale-2006, Moster-2009,
  Guo-2010}, is a simple but powerful method to statistically link
galaxies to dark matter haloes, without making detailed assumptions
about the physics of galaxy formation.  By assuming a monotonic
relationship (with a limited amount of scatter) between an observed
property such as stellar mass and a simulated property such as total
mass, and equating their cumulative abundances, it allows a
determination of quantities like the stellar mass--total mass ratio as
a function of mass. In integral form, the relation (for stellar mass)
can be expressed, in the absence of scatter, as follows:
\begin{equation}\label{eqn:abundance-matching}
\int^{\rm{M_{h,min}}}_{\rm{M_{h,max}}} N_h(m) dm = \int^{\rm{M_{\star,min}}}_{\rm{M_{\star,max}}} N_\star(m) dm,
\end{equation} 
where $N_h(m)$ and $N_\star(m)$ are the subhalo and stellar mass
functions. The upper limits of the subhalo and stellar mass ranges,
$\rm{M_{h,max}}$ and $\rm{M_{\star,max}}$, are fixed, while the
respective lower limits, $\rm{M_{h,min}}$ and $\rm{M_{\star,min}}$,
are chosen such that the relation holds over the entire
interval. Equivalently, abundance matching corresponds to assigning
the most massive observed galaxy to the most massive (sub)halo, and
stepping down towards the minimum galaxy mass or subhalo mass that can
still be matched, which determine $\rm{M_{h,min}}$ and
$\rm{M_{\star,min}}$.

\begin{figure}
  \includegraphics*[trim = 10mm 3mm 5mm 8mm, clip, width =
    \columnwidth]{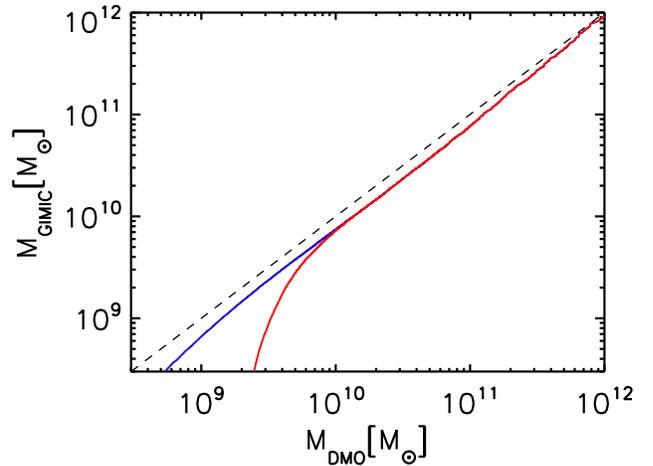}
  \caption{Mass of subhaloes in the GIMIC simulation and in the DMO
    simulation, at {\it equal cumulative subhalo abundance}. The blue
    line includes all subhaloes, while the red line assumes the {\it
      reduced} mass function, with {\it dark} subhaloes removed. The
    dashed line provides a reference for a 1:1 abundance ratio. Below
    $10^{12}\Ms$, the mass attributed to subhaloes of equal abundance
    is considerably lower in the GIMIC simulation compared to the DMO
    simulation, and the occurrence of {\it dark} subhaloes becomes the
    dominant factor below $\sim4\times10^9\Ms$.
  \label{fig:m-equal-abundance}}
\end{figure}

\begin{figure*}
  \begin{center}
    \begin{tabular}{cc}
      \hspace{-.1in} \includegraphics*[trim = 10mm  3mm 5mm 8mm, clip, width = .48 \textwidth]{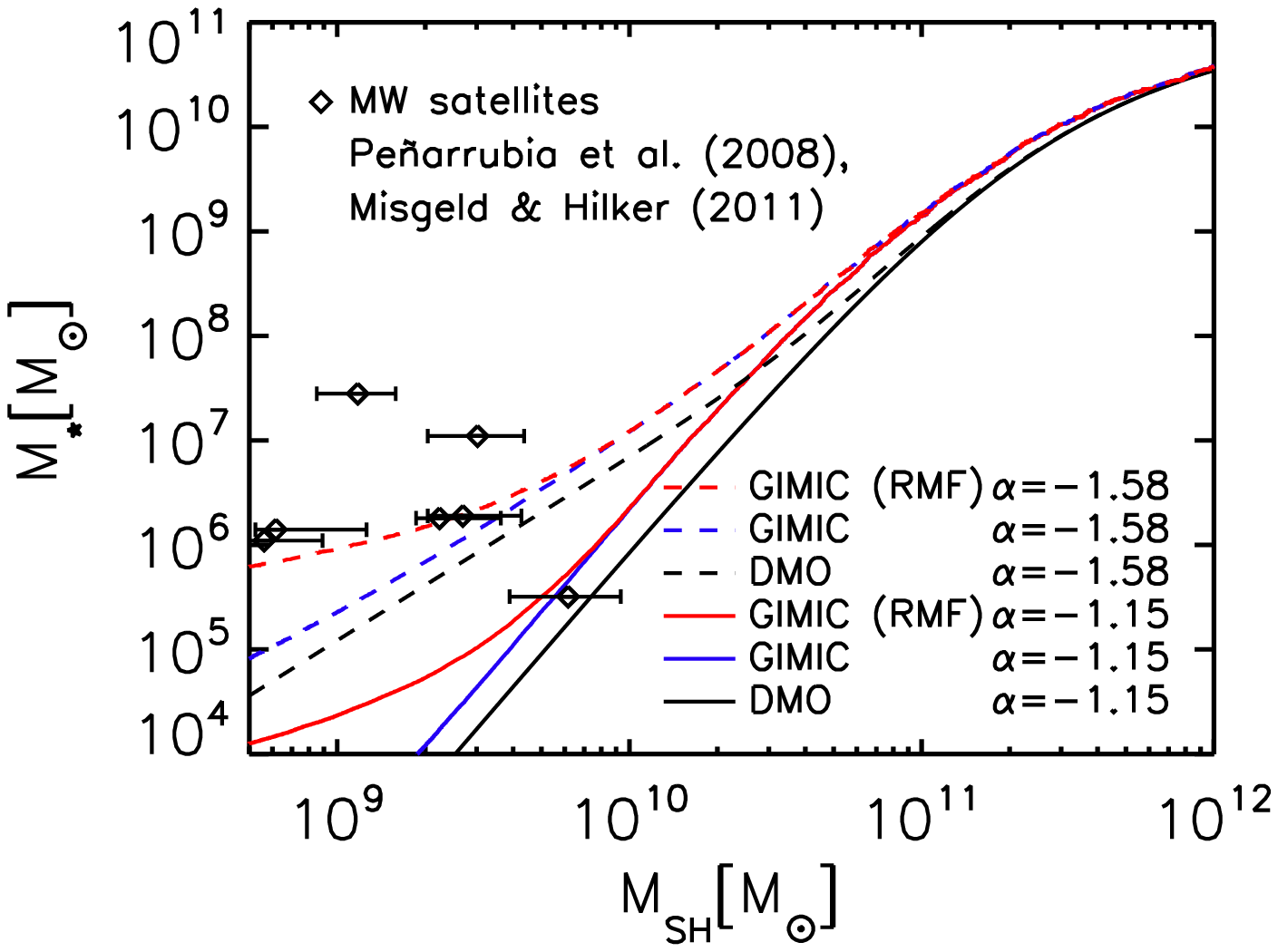} &
      \includegraphics*[trim = 10mm  3mm 5mm 8mm, clip, width = .48 \textwidth]{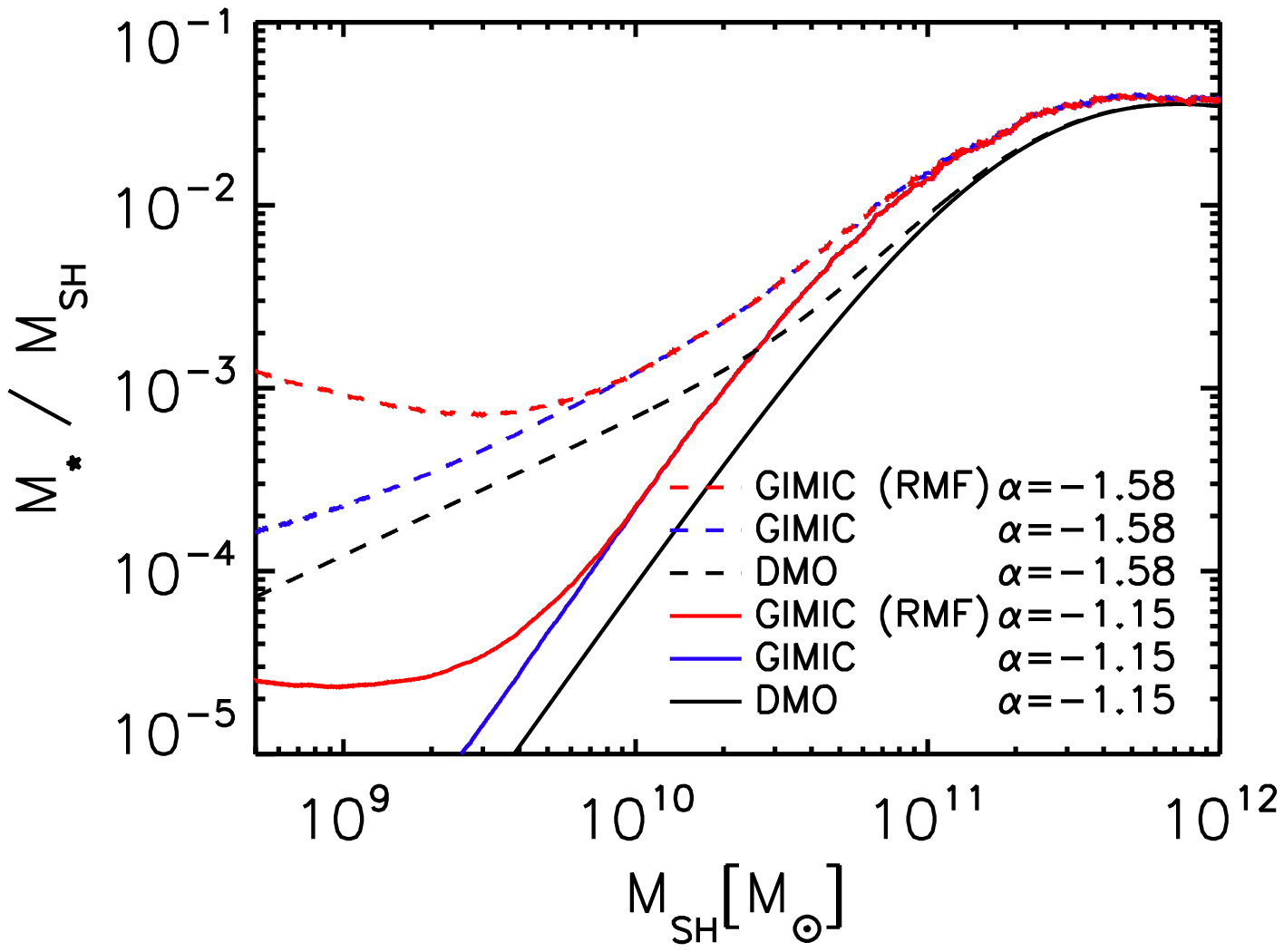}
    \end{tabular}
  \end{center}
  \caption{Stellar mass as a function of subhalo mass (left panel),
    and ratio of stellar mass to subhalo mass (right panel). Black
    lines are adapted from Guo et al. (2010), using the halo mass
    function of the Millennium I and II dark matter only
    simulations. Blue lines are the result of taking into account the
    baryonic effects on the subhalo mass function measured in GIMIC,
    as shown in Fig.~\ref{fig:HMF}. Red lines are obtained using the
    reduced subhalo mass function (RMF) of GIMIC, with {\it dark}
    subhaloes removed. Solid lines correspond to a faint-end slope of
    the stellar mass function of $\alpha=-1.15$ (Li \& White 2009),
    while dashed lines assume $\alpha=-1.58$.  For comparison, we show
    stellar masses \citep{Misgeld-2011} and dynamical masses
    \citep{Penarrubia-2008} of Milky Way satellites. The inclusion of
    baryons doubles the inferred stellar-to-total mass ratio for
    subhaloes below $10^{11}\Ms$, while the effect of the RMF further
    increases the inferred ratios by up to an order of magnitude for
    subhaloes in the range $10^{9}-10^{10}\Ms$. Combined with a steep
    faint-end slope for the stellar mass function, the reported
    discrepancy between abundance matching and measurements of
    individual galaxies is alleviated. The RMF also results in a
    finite minimum stellar-to-total mass ratio in subhaloes below
    $\sim10^9\Ms$.}
  \label{fig:Guo-comparison}
\end{figure*}

\subsection{Abundance matching with baryons}
The effect of baryons on the abundance of subhaloes shown in
Fig.~\ref{fig:HMF} has consequences for the results of abundance
matching. Furthermore, as described in Section~\ref{sec:dark}, an
increasing fraction of low-mass subhaloes does not host any stars.
Because these would not be observable, we propose to modify
Eq.~\ref{eqn:abundance-matching} further, by multiplying the (sub)halo
mass function by a completeness term $0 \le f_{\star}(m) \le 1$, to
give the {\it reduced} subhalo mass function (RMF) of {\it observable}
objects:

\begin{equation}\label{eqn:abundance-matching-effective}
\int^{\rm{M_{h,min}}}_{\rm{M_{h,max}}} f_{\star}(m) N_h(m) dm = \int^{M_{\star,min}}_{M_{\star,max}} N_\star(m) dm,
\end{equation}
Unlike a simple scatter term, dropping the assumption that every
simulated dark matter subhalo with $m > \rm{M_{h,min}}$ hosts a galaxy
has the effect of extending the domain to lower mass subhaloes, and
assigning higher stellar masses to subhaloes above the new limit.

For the purpose of abundance matching, the reduced subhalo mass
function differs depending on the observable: for a stellar mass
survey, the RMF includes only subhaloes that contain stars, while for
an HI survey \citep[e.g.][]{Zavala-2009}, it includes those galaxies
that contain gas, or both stars and gas.

From the ratio of the subhalo mass functions, we compute the subhalo
mass in GIMIC with the same cumulative abundance as a subhalo of a
given mass in the DMO simulation. Fig.~\ref{fig:m-equal-abundance} shows
the relations of equating the DMO subhalo mass function to the total
subhalo mass function in GIMIC (blue line), and to the reduced subhalo
mass function in GIMIC (red line), where subhaloes without stars have
been removed. These relations can be applied directly to abundance
matching results based on a dark matter only simulation, to derive
stellar-to-total mass ratios in a universe that contains baryons.

Applying our results to the abundance matching performed by
\cite{Guo-2010}, we note that the GIMIC simulation uses the same
cosmological parameters as the Millennium simulations studied by
\citeauthor{Guo-2010}. However, while the total abundance of
(sub)haloes depends on the slope and the normalisation of the power
spectrum \citep[e.g.][]{Vale-2006}, within the CDM paradigm, the
effect of baryons should be similar.

In line with other authors, \citeauthor{Guo-2010} have computed the
subhalo mass function as a combination of the present-day mass for
centrals, and the mass before infall for satellites. In adopting their
result for the DMO case which forms the baseline of all of our
abundance matching results, we use the same prescription, but we
consider the ratio of the abundance at $z=0$ when we apply the baryon
effects. As shown in Fig.~\ref{fig:correction-types}, the average mass
ratio shows little evolution between $z=1$ and $z=0$, suggesting that
the difference between applying the correction at infall or at $z=0$
would be negligible.

In Fig.~\ref{fig:Guo-comparison}, we show the effect of baryons on the
results of abundance matching, in terms of the inferred stellar mass
(left panel), and the stellar-to-total mass ratio (right panel). For
reference, black lines show the DMO case; the solid line denotes the
original results of \cite{Guo-2010}, while the dashed line reflects a
change in the faint-end slope of the stellar mass function from
$\alpha=-1.15$ \citep{Li-2009} to $\alpha=-1.58$ \citep{Baldry-2008}.
The set of blue lines shows the corresponding relations, but taking
into account the baryonic effects on the subhalo mass function
obtained from the GIMIC simulation. The set of red lines uses the {\it
  reduced} subhalo mass function (RMF), which includes baryonic
effects, and additionally has the {\it dark} subhaloes removed from the
subhalo mass function.

The effect of baryons on the attributed stellar mass strongly depends
on the subhalo mass. It is insignificant at subhalo masses above
$10^{12} \Ms$, but appreciable at the faint end. Purely due to the
inclusion of baryons, the stellar mass--total mass ratio increases by
a factor of $\sim 1.5$ for subhaloes of $10^{11}\Ms$, and by a factor
of $\sim2.2$ for subhaloes of $10^{10}\Ms$ and below. If the RMF is
considered, stellar masses assigned to subhaloes below $10^{10}\Ms$
increase further, and the mean stellar-to-total mass ratio reaches a
finite minimum at $1:10^3$ - $1:10^5$, depending on the slope of the
stellar mass function.

The apparent discrepancy in the stellar-to-total mass ratios inferred
from abundance matching, and state-of-the-art simulations of dwarf
galaxies \citep[][and references therein]{Sawala-Matter}, as well as
kinematics of individual dwarf galaxies \citep{Ferrero-2012}, have
been described as a challenge to the CDM paradigm. Included in the
left panel of Fig.~\ref{fig:Guo-comparison} are observational
estimates for 7 classical dwarf spheroidals, with stellar masses
adopted from \cite{Misgeld-2011}, and virial masses derived by
\cite{Penarrubia-2008}.

It is worth pointing out that at such low masses, a direct comparison
to observations should be taken with a grain of salt: mass
measurements rely on assumptions about the shape and anisotropy of the
dark matter halo; the halo mass is measured at present rather than at
infall, and the limited sample may be biased towards the objects with
higher stellar mass (indeed, \cite{Wolf-2010} show that despite
stellar masses extending to $\sim 10^4\Ms$, most Milky Way satellites
are compatible with a formation in $\sim 10^9 \Ms$
haloes). Nevertheless, it can be seen that when the CDM subhalo mass
function is corrected to take into account baryonic effects, the
stellar masses inferred from abundance matching become consistent with
observations of individual dwarf galaxies.

\cite{Tikhonov-2009} have likewise argued that the velocity function
of voids in the Local Volume presents a challenge to CDM.  The
observed number of voids is so high that only haloes down to masses of
$6-8 \times10^{9}\Ms$ or $1-2 \times 10^{10}\Ms$ (for values of
$\sigma_8$ of 0.75 and 0.9, respectively) can be populated by galaxies
brighter than $M_B = -12$, but individual dwarf galaxies are observed
with HI rotational velocities below $v_{\rm{max}} \sim 25$ kms$^{-1}$
indicative of halo masses closer to $10^{9}\Ms$. Our results suggest
that this discrepancy may also be resolved by baryonic processes. As
shown in Fig.~\ref{fig:m-equal-abundance}, the mass limits derived by
\cite{Tikhonov-2009} based on the abundance of haloes in dark matter
only simulations would have to be reduced in a universe where baryons
affect the masses of low mass haloes. In addition, the appearance of
{\it dark} subhaloes discussed in Section~\ref{sec:dark} instead of a
sharp threshold in (sub)halo mass can reconcile the observation of a
population of galaxies in lower mass haloes with theoretical
predictions.

\section{Discussion} \label{sec:summary} We have compared a set of
high-resolution cosmological N-body simulations with and without
baryons to study the effect that baryonic processes have on structure
formation in a $\Lambda$CDM universe. While both simulations agree
well on large scales, objects below $\sim10^{12}\Ms$ have
systematically lower masses in the GIMIC simulation that includes
baryons compared to its ``Dark Matter Only'' (DMO)
counterpart. Consequently, the cumulative abundance of haloes and
subhaloes at every mass below $\sim10^{12}\Ms$ is reduced in the GIMIC
simulation. Given that the Universe includes baryonic processes such
as gas pressure, cooling, reionisation, star formation and supernova
feedback, and assuming that they are represented at least
approximately in the GIMIC simulation, it follows that DMO simulations
{\it overpredict} the true abundance of substructures, by $\sim10\%$
at $10^{11.5}\Ms$, $\sim20\%$ at $10^{11}\Ms$, and $\sim 30\%$ at
$10^{10}\Ms$.

The difference in halo and subhalo abundance has consequences for all
analyses that take their mass functions as a starting point. For
example, the stellar-to-total mass ratio of dwarf galaxies with
stellar masses of $\sim10^5-10^9\Ms$ inferred from subhalo abundance
matching doubles after taking into account the lower mass of
subhaloes, resulting from baryonic effects. If, in addition, the {\it
  reduced} mass function accounts for the fact that not all subhaloes
contain stars, stellar-to-total mass ratios increase further,
particularly for galaxies with stellar masses below $\sim10^6\Ms$,
i.e. the classical dwarf spheroidals. It appears that both effects are
needed in order to reconcile abundance matching results with kinematic
estimates of mass-to-light ratios of dwarf galaxies
\citep{Ferrero-2012}, or current high-resolution simulations
\citep[e.g.][]{Sawala-Matter}.

We conclude that the discrepancy between dwarf galaxy abundances and
kinematics, and the results of cold dark matter (CDM) simulations
paired with abundance matching, may be due to the inability of dark
matter only simulations to capture the physical effects relevant on
small scales, aggravated by the assumption that a visible galaxy forms
inside every single dark matter (sub)halo. Likewise, taking into
account baryonic effects and excluding subhaloes that fall below the
observational detection threshold significantly reduces the reported
discrepancy between the $\Lambda$CDM paradigm, and the velocity
function measured in the ALFALFA HI survey \citep{Zavala-2009,
  Papastergis-2011}.

For haloes more massive than $\sim10^{12}\Ms$, our results corroborate
those of previous works \citep[e.g.][]{Dolag-2009, Cui-2012}, who
concluded that the net effect of gas loss via stripping and feedback
is largely counterbalanced by the increased concentration of the
stellar component. We cannot exclude the possibility, however, that as
suggested by \cite{VanDaalen-2011}, efficient AGN feedback would also
change the results on these scales; indeed \cite{McCarthy-2012b} have
reported overcooling in galaxies with stellar masses above
$10^{11.5}\Ms$ in the GIMIC simulation. While there are also
differences in the physics model, most notably the inclusion of
photo-ionisation in the GIMIC simulation that suppresses star
formation in small objects, the most significant difference compared
to previous works is the increased resolution, which has allowed us to
study objects more than two orders of magnitude lower in mass. As we
show, the baryonic results are strongly mass dependent. At low masses,
re-ionisation, feedback and stripping outweigh adiabatic contraction,
leading to a net mass loss. The GIMIC results also confirm the results
of high-resolution resimulations of individual dark matter haloes with
baryons \citep{Sawala-Matter}, which showed a similar reduction in
mass due to efficient outflows from $\sim10^{10}\Ms$ haloes.

Quantitatively, our results will likely be refined by future
simulations at still higher resolution and with more complete physics
models. In particular, our result for the fraction of {\it dark}
subhaloes straddles the resolution limit of our simulations. However,
the requirement of a {\it reduced} subhalo mass function to match the
faint end of observable galaxies, of taking into account the gas
fraction when comparing to HI surveys, and of including baryon physics
on the underlying mass distribution of (sub)haloes, appear to be
largely model-independent. To the precision achievable with our
simulations, the failures previously attributed to the CDM paradigm
appear to be largely attributable to a neglect of baryons in dark
matter only simulations.

Overall, we believe that the GIMIC simulation, which has already been
demonstrated to successfully reproduce many aspects of galaxy
formation (\citealt{Crain-2010} \citealt{Font-2011},
\citealt{McCarthy-2012a}, \citealt{McCarthy-2012b}), is a much closer
approximation to the observable Universe than a dark matter only
model. With current and upcoming surveys such as ALFALFA, GAMA and
GAIA pushing the observations to even fainter limits, simulations must
not only increase in resolution, but will also have to take baryons
into account, if they are to resolve the scales required to
distinguish alternative dark matter models, or to model galaxy
formation on sub-Milky-Way scales.

While it seems somewhat unsatisfactory to modify apparently
``assumption-free'' and elegant methods like abundance matching and to
replace dark matter simulations with complex, and in many ways
uncertain astrophysical simulations, it was clear from the outset that
a ``dark matter only universe'' itself was merely an assumption of
convenience. On small scales, it falls short.

\section*{Acknowledgements}

We are indebted to Ian McCarthy, who developed and performed some of
the original GIMIC simulations. We are very grateful to Q.~Guo for
providing the abundance matching data. T.~S. acknowledges a fellowship
by the European Commission's Framework Programme 7, through the Marie
Curie Initial Training Network Cosmo-Comp
(PITN-GA-2009-238356). C.~S.~F. acknowledges a Royal Society Research
Merit Award and ERC Advanced Investigator Grant COSMIWAY. Some
simulations used in this paper were performed on the Cosmology Machine
supercomputer at the ICC, which is part of the DiRAC Facility jointly
funded by STFC, the Large Facilities Capital Fund of BIS, and Durham
University. This work was supported in part by an STFC rolling grant
to the ICC and by the National Science Foundation under Grant No. NSF
PHY11-25915. J.~Z. is supported by the University of Waterloo and the
Perimeter Institute for Theoretical Physics. Research at Perimeter
Institute is supported by the Government of Canada through Industry
Canada and by the Province of Ontario through the Ministry of Research
\& Innovation. J.~Z. acknowledges financial support by a CITA National
Fellowship.

\bibliographystyle{mn2e}
\bibliography{gimic}

\appendix

\section{Mass Correction for Centrals and Satellites}\label{sec:correction}
\begin{figure*}
  \begin{center}
    \begin{tabular}{ccc}
      \hspace{-.3in} \includegraphics*[trim = 10mm 2mm 1mm 10mm, clip, width = .5\textwidth]{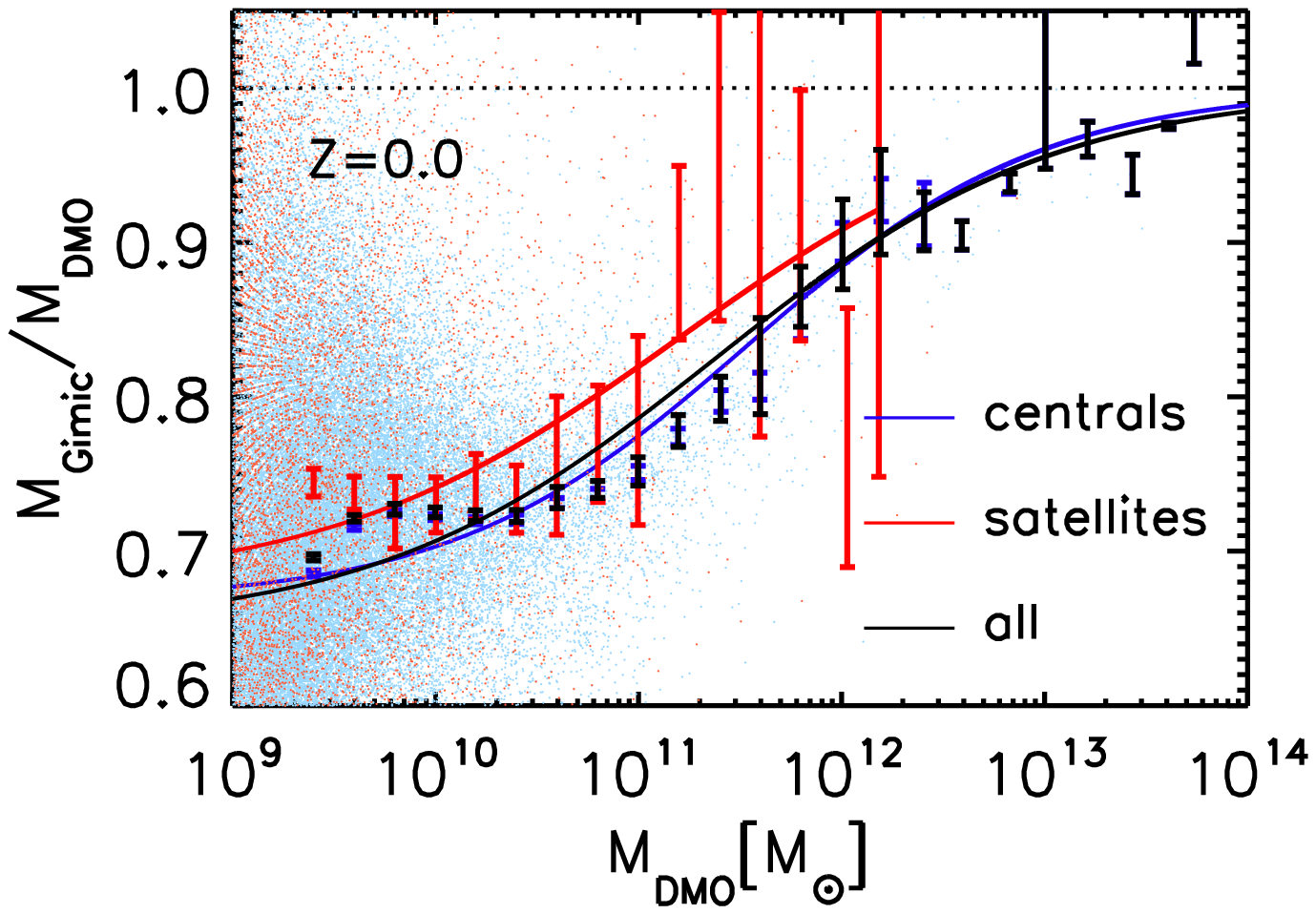}
      \hspace{-.3in} \includegraphics*[trim = 10mm 2mm 1mm 10mm, clip, width =  .5\textwidth]{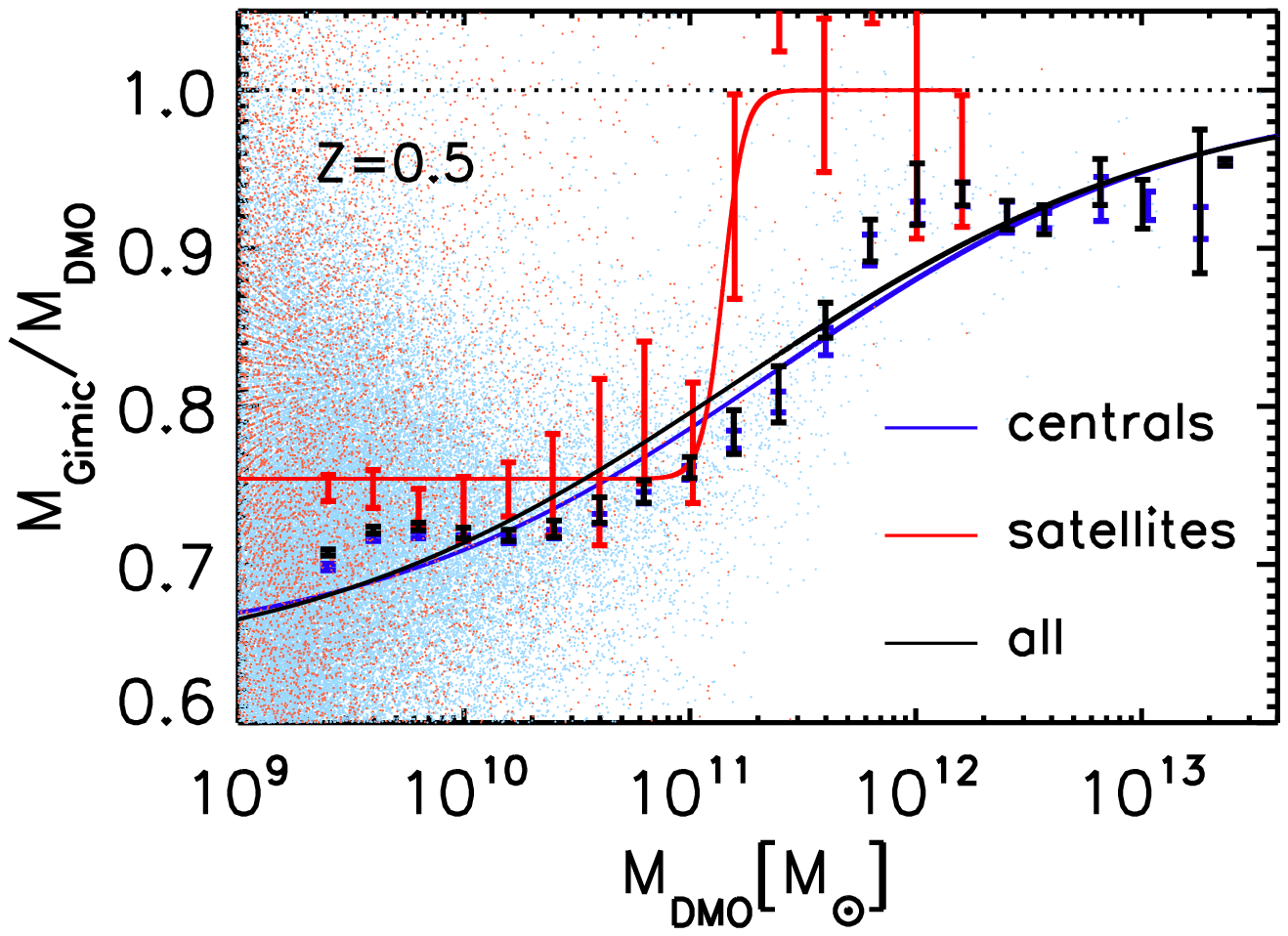} \\
      \hspace{-.3in} \includegraphics*[trim = 10mm 2mm 1mm 10mm, clip, width = .5\textwidth]{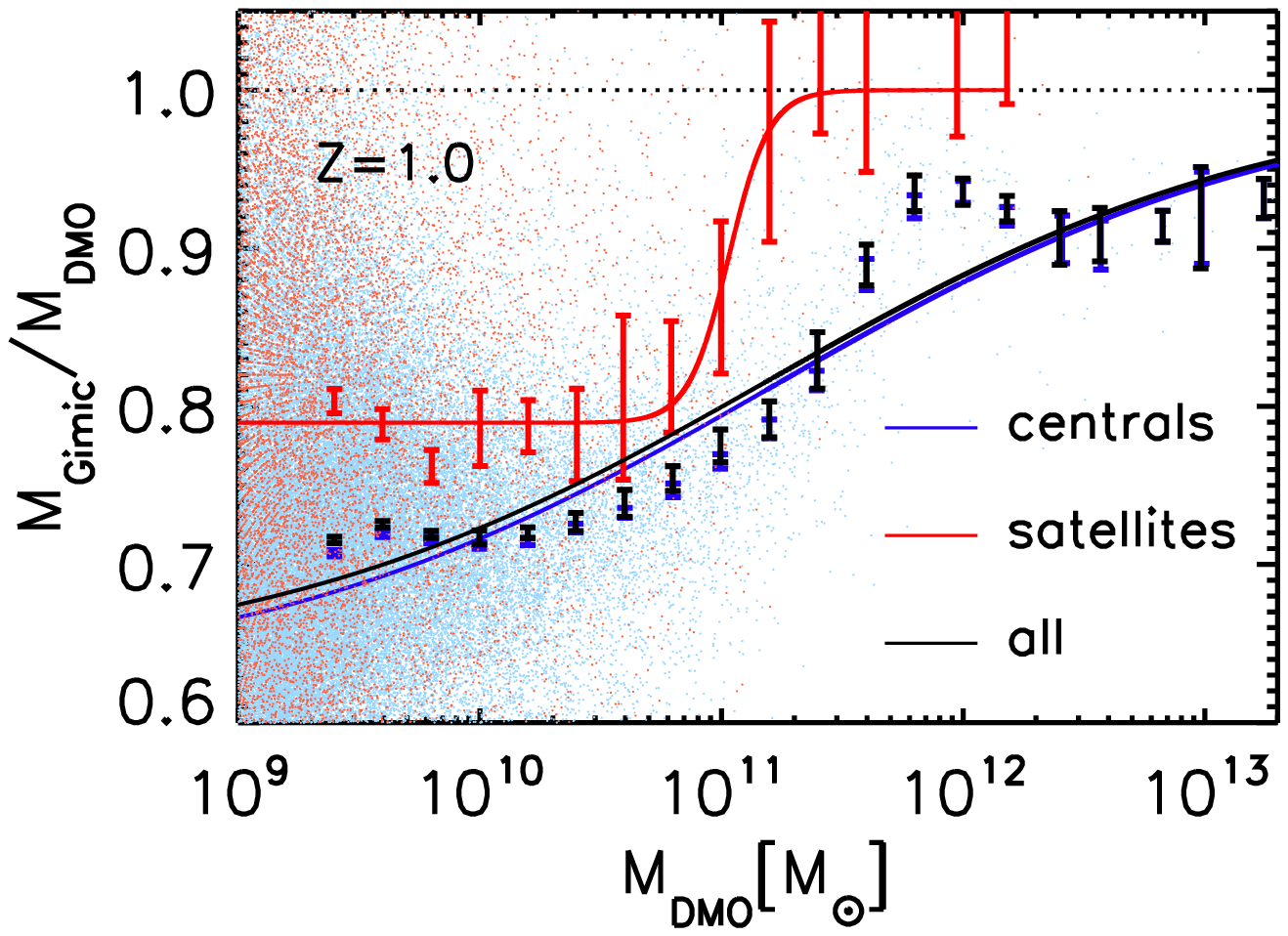}
      \hspace{-.3in} \includegraphics*[trim = 10mm 2mm 1mm 10mm, clip, width = .5\textwidth]{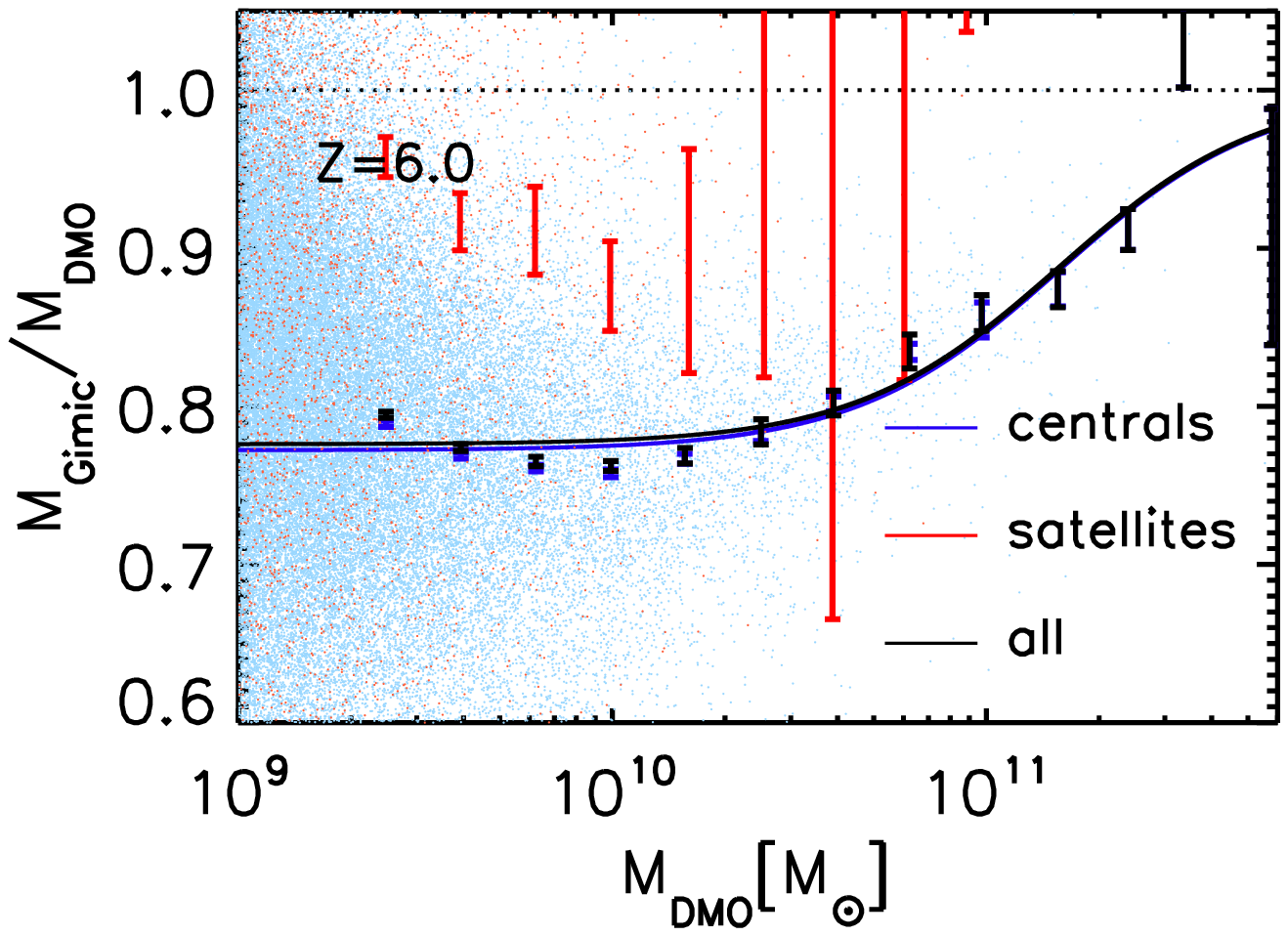}    
\end{tabular}
  \end{center}
  \caption{Ratio of subhalo masses between the GIMIC and DMO
      simulations for matched pairs of subhaloes at different
      redshift. The error bars indicate the estimate of the median and
      its error for centrals (blue), satellites (red) and all
      subhaloes (black), while the lines show the corresponding best
      fits to Eqn.~\ref{eqn:fit}, with coefficients listed in Table~\ref{table:coeffs}.}
  \label{fig:correction-types}
\end{figure*}

In~Section~\ref{sec:analytic-preview}, we parametrised the average
relative change of a subhalo's mass from the DMO to the GIMIC
simulation in Eq.~\ref{eqn:fit}:
$$
\frac{\rm{M_{GIMIC}}}{\rm{M_{DMO}}} = \frac{a + \left(\rm{M_{DMO}/M_t}\right)^w}{1 + \left(\rm{M_{DMO}/M_t}\right)^w}
$$ In Fig.~\ref{fig:correction-types}, we show the ratio of subhalo
  masses for matched pairs of satellites, centrals, and the
  combination of all subhaloes at four different redshifts, from $z=6$
  to $z=0$. Allowing $a$, $\rm{M_t}$ and $w$ to vary freely, we fit
  Eq.~\ref{eqn:fit} to the different populations of subhaloes, giving
  equal weight to the median ratio within each logarithmic mass
  bin. It can be seen that for central subhaloes, the lower limit does
  not evolve strongly with redshift from $z=1$, although the mass
  $\rm{M_t}$ at which the mass ratio reaches the intermediate value of
  $(a+1)/2$ rises from $10^{11.2} \Ms$ at $z=1$ to $10^{11.6} \Ms$ at
  $z=0$. At $z=6$, the reduction in subhalo mass is smaller.

We noted in Section \ref{sec:origin} that at fixed subhalo mass
  at $z=0$, the average difference in mass between the GIMIC
  simulation and the DMO simulation is slightly less for satellites
  than for centrals, which may be attributed to the fact that
  satellites also experience tidal effects, which are similar in both
  simulations. A pair of isolated subhaloes that has evolved to a
  given mass ratio before infall, and whose mass is subsequently
  reduced in equal proportion by tidal effects in both simulations,
  would evolve to the same mass ratio, but for a lower total
  mass. Fig.~\ref{fig:correction-types} shows that a difference also
  exists at higher redshifts, where the decrease in mass ratio appears
  at higher masses for satellites than for centrals, although at high
  redshift, the number of massive satellites is low, and the
  associated statistical uncertainty is large. It is worth noting that
  the fraction of satellites increases with time, and that more than
  half of the satellites at $z=0$ were still (isolated) centrals up to
  $z=0.5$, partly explaining the convergence of the two curves over
  time. At all times, the average reduction in mass for the total
  population of subhaloes closely resembles that for the centrals.

Table~\ref{table:coeffs} contains the coefficients of the different
fits, and may be useful for constructing subhalo mass functions based
on a DMO catalogue.

\begin{table}
  \begin{center}
    \begin{tabular}{lllllll} \hline \hline              
      $z$ & type & a & $log_{10}M_t$ & w \\ \hline

      0.0 & all & 0.65 & 11.37 & 0.51 \\
      & centrals & 0.66 & 11.54 & 0.59 \\
      & satellites & 0.68 & 11.20 & 0.50 \\ \hline

      0.5 & all & 0.63 & 11.20 & 0.65 \\
      & centrals & 0.64 & 11.38 & 0.47 \\
      & satellites & 0.75 & 11.14 & 9 \\ \hline

      1.0 & all & 0.63 & 11.20 &  0.41\\
      & centrals & 0.62 & 11.20 & 0.40 \\
      & satellites & 0.78 & 11.07 & 5 \\ \hline

      6.0 & all & 0.77 & 11.18 & 1.95 \\
      & centrals & 0.77 & 11.20 & 1.58 \\
      & satellites & -- & -- & -- \\ 

    \end{tabular}
  \end{center}

\caption{Coefficients of Eq.~\ref{eqn:fit} for fits to the median mass
  ratios of individual, matched subhaloes in the GIMIC relative to the
  DMO simulation. For each redshift, the rows show the result of
  different sets of subhaloes, as defined in Section~\ref{sec:linking}. Note
  that at high redshift, scatter dominates the mass ratio for
  satellites, but the total population remains dominated by
  centrals. \label{table:coeffs}}

\end{table}

\section{Convergence}\label{sec:resolution}

\begin{figure*}
  \begin{center}
    \hspace{-.3in} \includegraphics*[trim = 0mm  0mm 0mm 5mm, clip, width = .5\textwidth]{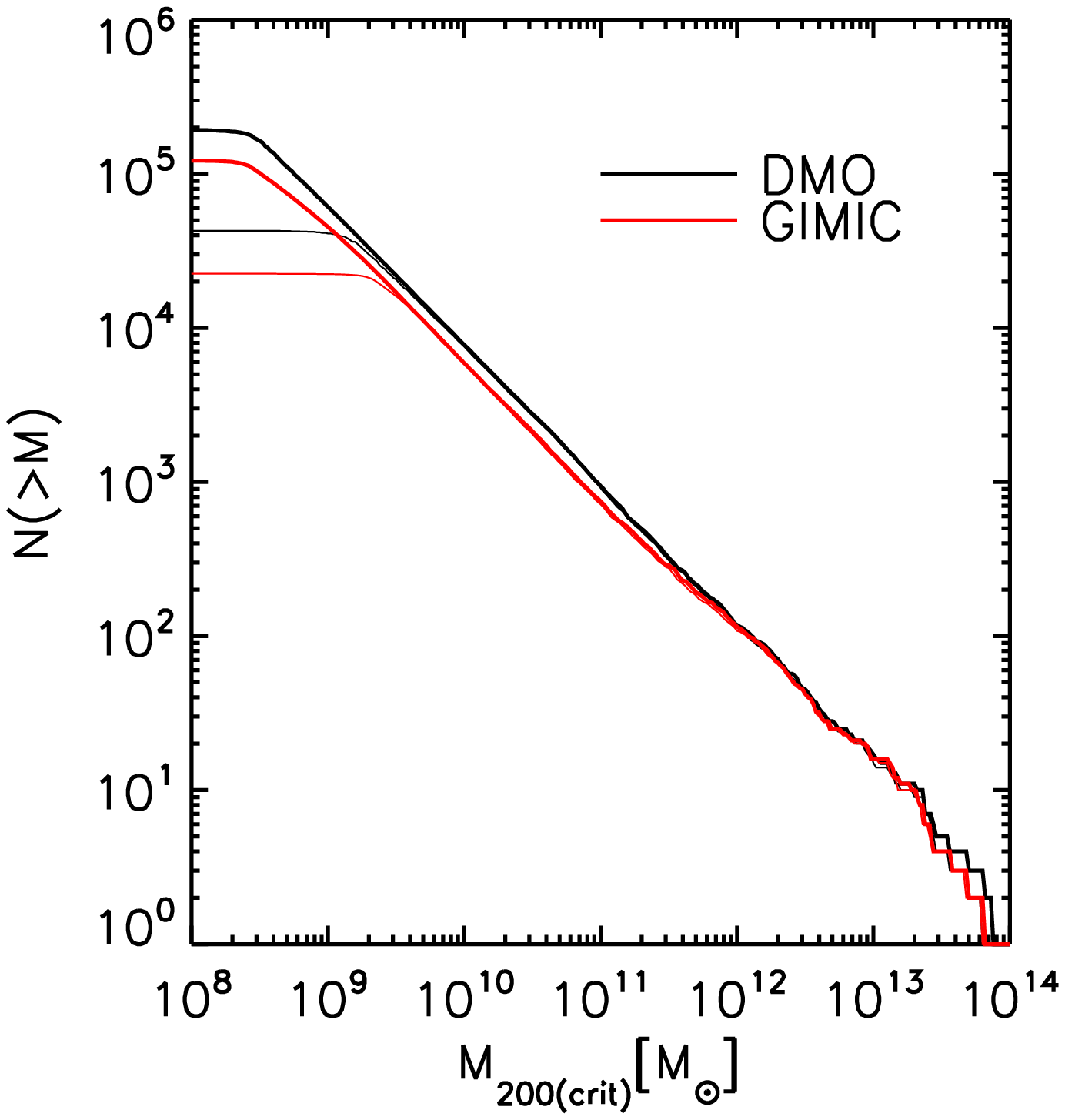} 
    \hspace{-.4in} \includegraphics*[trim = 0mm  0mm 0mm 5mm, clip, width = .5\textwidth]{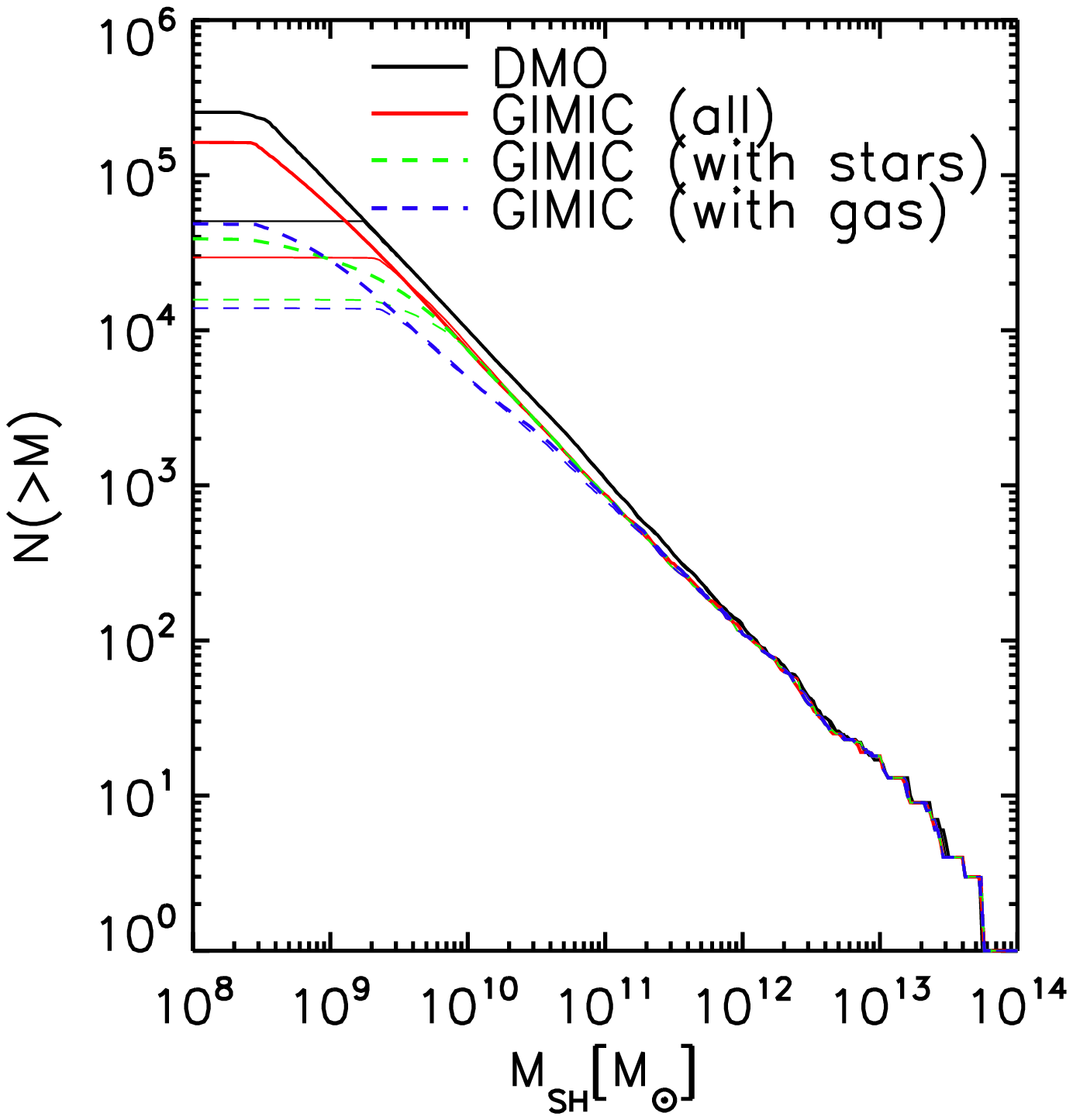}
  \end{center}
  
  \caption{Cumulative mass function of haloes (left panel) and
    subhaloes (right panel) at $z=0$ in the high-resolution (thick
    lines) and intermediate resolution (thin lines) realisations. In
    both panels, results from the DMO simulations are shown in black,
    while those of the GIMIC simulations are shown in red. Also
    plotted in the right panel are the {\it reduced} subhalo mass
    functions for subhaloes with stars (green), and with gas
    (blue). It can be seen that the total number of haloes and
    subhaloes are converged to the resolution limit, both in the DMO
    and the GIMIC simulations. While the reduced mass functions for
    subhaloes with gas are also converged, the reduced mass functions
    of subhaloes with stars begin to diverge at higher masses,
    indicating some resolution dependence.}
  \label{fig:resolution-comparison}
\end{figure*}

As described in Section~\ref{sec:methods}, the GIMIC and DMO
simulations are each performed at two different resolutions, with
particle masses that differ by a factor of 8. In
Fig.~\ref{fig:resolution-comparison}, we compare the cumulative mass
functions in both sets of simulations, with thin and thick lines
denoting the low- and high- resolution results respectively. The left
panel shows the halo mass function, while the right panel shows the
subhalo mass function, both at $z=0$. In both panels, black lines
denote the DMO simulations and red lines the total mass functions of
the GIMIC simulations. While the total number of both haloes and
subhaloes decreases by an expected factor of $\sim 5$ for the lower
resolution runs, it can be seen that the mass functions are well
converged, up to the resolution limit. Hence, the baryonic effects on
the total number of haloes and subhaloes as a function of mass, as
measured in the GIMIC simulation, are largely independent of
resolution, at least for masses of $\sim10^9 \Ms$ and above.

In the right panel of Fig.~\ref{fig:resolution-comparison}, we also
plot the {\it reduced} cumulative mass functions of subhaloes that
contain stars (green), or gas (blue). While the reduced mass functions
for subhaloes with gas are well converged, the reduced mass functions
of subhaloes with stars begin to diverge above the absolute resolution
limit, indicating some resolution dependence of the star formation
threshold. In addition to model dependence, this lack of convergence
implies some uncertainty in the effect on stellar-to-total mass ratios
discussed in Section~\ref{sec:applications}, but does not change our
finding that a significant fraction of {\it dark} subhaloes are
present in a universe with baryons, which has a significant effect on
abundance matching at the low-mass end.

\label{lastpage}
\end{document}